\documentclass[a4, 12pt]{article}
    \usepackage[round]{natbib}
    \usepackage{fullpage}
    \usepackage{geometry}
    \usepackage{amssymb}
    \usepackage{float}
    \usepackage{amsmath}
    \usepackage{mathtools}
    \usepackage{dsfont}
    \usepackage{enumerate}
    \usepackage{bm}
    \usepackage{adjustbox}
    \usepackage{longtable}
    \usepackage{threeparttablex}
    \usepackage{booktabs}
    \usepackage{dcolumn} 
    \newcolumntype{d}[1]{D{.}{.}{#1}}

    \usepackage[labelfont=bf, font=footnotesize, labelsep=period]{caption}
    \usepackage{subcaption}

    \usepackage{graphicx}
    \graphicspath{{./Graphs/}}
    \usepackage[pagebackref=true]{hyperref}
    \renewcommand*{\backref}[1]{}
    \renewcommand*{\backrefalt}[4]{%
        \ifcase #1 (Not cited.)%
        \or        [Cited on page~#2.]%
        \else      [Cited on pages~#2.]%
        \fi}
    \usepackage{url}
        
    \usepackage[utf8]{inputenc}
    \makeatletter
    \def\input@path{{./Tables/}}
    \makeatother

    \usepackage{scalerel,stackengine}
    \stackMath
    \newcommand\reallywidehat[1]{%
    \savestack{\tmpbox}{\stretchto{%
    \scaleto{%
        \scalerel*[\widthof{\ensuremath{#1}}]{\kern.1pt\mathchar"0362\kern.1pt}%
        {\rule{0ex}{\textheight}}
    }{\textheight}%
    }{2.4ex}}%
    \stackon[-6.9pt]{#1}{\tmpbox}%
    }
    \usepackage{titlesec}
    \titleformat*{\section}{\large\bfseries}
    \titleformat*{\subsection}{\normalsize\itshape}
    \titleformat*{\paragraph}{\normalsize\bfseries}

    \usepackage{wrapfig}  
    \usepackage{verbatim}  
    \usepackage{lscape}  

    \usepackage[table,xcdraw]{xcolor}

    \usepackage{tikz}
        \usetikzlibrary{calc}
        \usetikzlibrary{trees}
        \usetikzlibrary{decorations.pathreplacing,angles,quotes}
        \usetikzlibrary{positioning, matrix}
        \usepackage{pgfplotstable} 

    \usepackage{titling}

\usepackage{algpseudocode,algorithm,algorithmicx}

\algrenewcommand\algorithmicrequire{\textbf{Precondition:}}
\algrenewcommand\algorithmicensure{\textbf{Postcondition:}}

\begin{document}

\title{Conceiving Naturally After IVF: \\
    the effect of assisted reproduction on obstetric interventions and child
    health at birth.}

\author{Fabio I. Martinenghi, Xian Zhang, \\
    Luk Rombauts and Georgina M. Chambers}
\date{}
\maketitle

\section{Introduction}
It is estimated that 439,039 babies were born via assisted reproduction
technology (ART) in 2018 alone
\citep{Chambers21}. That same year, almost 5\% of all the
children born in Australia were conceived via in-vitro fertilisation
\citep[IVF, a form of ART; ][]{Newman20},
while estimates for the US and Europe are at 1.6\% and 4.5\%, respectively,
and growing \citep{DeGeyter2018, Sunderam2018}.
In the context of this significant contribution of ART to global
fertility, pregnancies conceived via ART are
currently considered high-risk
\citep[see][for a recent review of the literature]{Berntsen2019}.
However, the causal evidence supporting this claim remains scant.

Indeed, while the characteristics of couples who use ART
strongly correlate with poorer outcomes at birth\footnote{
    This is not only true for multiple pregnancies, but also for singleton
    pregnancies.
    Singleton pregnancies are pregnancies where the mother is expecting only
    one child.
    Multiple pregnancies, such as twins and triplets, increase risks of
    adverse outcomes at birth for the mother and the children.
    They are common among IVF pregnancies although largely preventable
    by transferring a single embryo per each cycle.
    The Reproductive Technology Accreditation Committee of
    the Fertility Society of Australia and New Zealand requires to
    ``minimise the incidence of multiple pregnancies'', which includes
    recommending a single embryo transfer per cycle,
    ensuring that no more than two embryos are transferred per cycle,
    and informing the patient of the risks associated with double transfers
    \citep{RTAC21}.
}, both for mother (e.g. caesarean sections) and the baby
(e.g. pre-term births),
it remains unclear whether ART treatment independently increases the
risk of such adverse outcomes.
These characteristics include, for instance,
advanced maternal age
\citep{Attali2012, Lean2017, Berntsen2019},
certain causes of infertility and
comorbidities including obesity \citep{Pasquali2007, Maheshwari2007}.
In relation to obstetric interventions, one concern is that
the considerable emotional and financial investment implied by an
ART pregnancy might exacerbate defensive medical behaviour,
where physicians intervene more than necessary driven by the fear
of medical liability \citep[for evidence of its impact
    on c-section rates, see][]{Dubay99,Jena15,Betran18,Bertoli19,Fenizia19,Longo20}.

In this paper, we study the effect of ART treatment\footnote{} on obstetric and
perinatal outcomes.
We measure its impact on the probability of pre-term birth
and, more broadly, on the health of the baby. This includes
the impact of ART on the baby's gestational age at birth,
weight at birth and its APGAR 1 and 5 scores (a measure of infant health).
Finally, we study its impact on the rates of spontaneous labour
(rather than induced), preterm spontaneous labour and c-section,
which helps us understand whether ART pregnancies experience higher
rates of interventions, holding all else constant.

Our approach leverages the stochastic nature of conception, including
ART conception. Indeed, while factors such as age and reproductive health
influence the probability of conception, chance
(idiosyncratic factors) plays an important role.
Fertile couples in their 20s have a $25\%$ chance of conceiving spontaneously
in each menstrual cycle \citep{ACOG}, and a $41-47\%$
probability of conceiving via IVF per IVF cycle\footnote{
    Estimates computed on 7 February 2024 via the online
    government-funded IVF prediction tools:
    \url{YourIVFSuccess.com.au}, based on Australian data,
    \url{https://www.cdc.gov/art/ivf-success-estimator/index.html},
    based on US data, and
    \url{https://w3.abdn.ac.uk/clsm/opis/tool/ivf1}, based on UK data.
    The parameters set for the first prediction tool were:
    ``Age of person intending to carry the pregnancy: 22",
    ``Sperm provider's age: 22",
    ``Previous pregnancy: No",
    ``Main infertility diagnosis: I have not had any tests carried out,
    or I do not have medical infertility'',
    ``Previous IVF treatment: No".
    The outcome was "having a baby in the 1st complete egg retrieval cycle"
    For the second tool the parameters were:
    ``Age: 22 years'',
    ``Weight: 110 lbs'',
    ``Height: 5 feet'',
    ``Number of IVF cycles used: I've never used IVF'',
    ``Number of prior pregnancies: None '',
    ``Number of prior births: None'',
    ``Egg source: My own eggs'',
    ``Diagnosis: I don't know/no reason''.
    The outcome was ``cumulative chance of live birth after 1 retrieval and
    all transfers within 12 months.''
    For the third tool, the parameters were:
    ``What is your age? 22'',
    ``How many years have you been trying to conceive? 0'',
    ``Have you been pregnant before? No'',
    ``Do you have a problem with your tubes? No'',
    ``Do you have an ovulation problem? No'',
    ``Do you have a male factor fertility problem? No'',
    ``Do you have an unexplained fertility problem? No'',
    ``Do you plan to have IVF or ICSI? IVF''
    The outcome was ``chance of having your first baby after 1 complete
    cycle of treatment.''
}.
In either case, the residual variation in conception remains large
and is plausibly idiosyncratic.
The persistently low precision of clinical prediction models
for IVF pregnancies, even the most recent ones
\citep[C-indices between 0.6-0.7][]{McLernon2023},
is also consistent with some variation in IVF success
being idiosyncratic.

Our strategy to isolate this variation has two components.
First, we focus on births from women who have undergone ART treatment in
Australia and subsequently given birth.
We assign births whose conception depended on ART treatment
to the treatment group, and assign births that were
naturally conceived soon after ART treatment to the control group.
Specifically, our treatment group includes singleton births from mothers
who had a successful ART cycle, while our control group includes
singleton births from mothers who had an unsuccessful ART cycle and conceived
spontaneously between 3 and 12 months after the cycle
(see Figure \ref{fig:sample_construction}).
This removes the selection bias associated with comparing the outcomes of
fertile and subfertile women\textemdash populations that are different over
both observable and unobservable characteristics.

\begin{figure}
    \centering
    \includegraphics[width=0.8\textwidth]{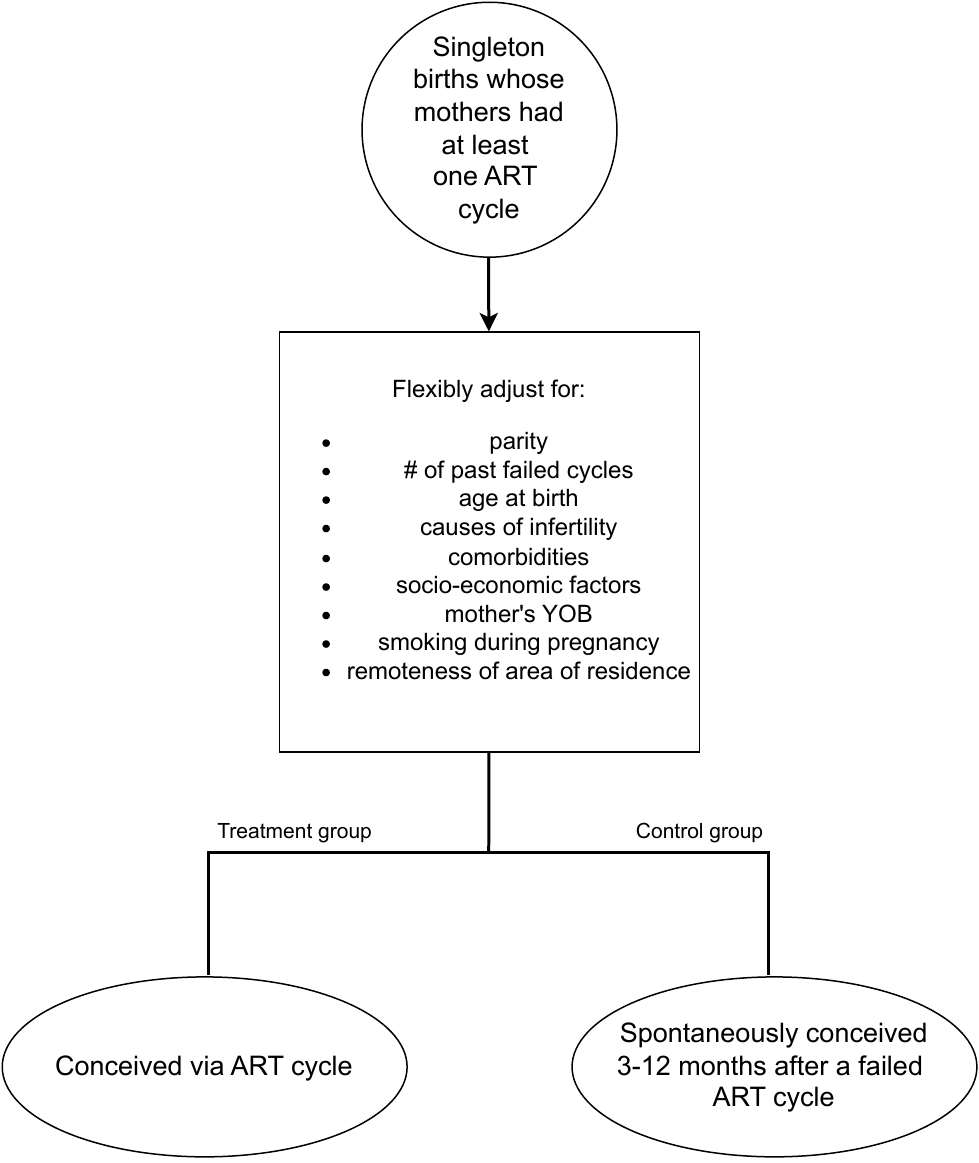}
    \caption{Diagram summarising the study}
    \label{fig:diagram}
\end{figure}

Second, we adjust for key confounding variables,
using both standard linear methods and a flexible approach\textemdash
double machine learning (DML) with random forests
(see Figure \ref{fig:diagram}).
Our linked administrative datasets allow us to observe both
cycle-level information and some clinical history of the mother,
which we use to control for key confounders
such as age, parity, number of prior ART cycles, socio-economic
characteristics and co-morbidities. Finally, we perform a number of
sensitivity analyses, studying how unobserved confounders might affect
our estimates \citep[via ][]{Cinelli2019}
and exploring how our results change using alternative
sample restrictions and model specifications.

We do not find evidence that ART increases the risk of preterm birth or
other adverse obstetric outcomes.
In fact, the effects of ART on gestational age at birth,
birth weight and APGAR 1 and 5 scores are not statistically different
from zero at conventional levels.
Our estimated null effects are precise, so that it is unlikely that
clinically significant effect sizes exist and have not been detected
\footnote{
    We can exclude the following (clinically small)
    effect sizes for the effect of ART on
    obstetric outcomes: effects between [-0.16, 0.10] weeks for gestational age
    at birth; between [-1.1, 1.9] p.p. for the risk of preterm birth;
    between [-0.07, 0.09] points for the APGAR1 score;
    between [-0.02, 0.1] points for the APGAR5 score;
    between [-14, 60] grams for the effect of ART on birth weight.
}.
Moreover, we estimate that ART has a (precise) null effect on the risk of preterm
spontaneous birth, an important adverse outcome that is not subject to medical
intervention and indicates poor infant health.
We also find that ART slightly \textit{decreases} the risk
of obstetric interventions, lowering the risk of a caesarean section (-5.1 p.p.)
and increasing the rate of spontaneous labour (+4 p.p.).

We contribute to the literature on ART by proposing
a new strategy that uses exogenous residual variation
in ART success to identify the causal effect of ART.
We focus on obstetric outcomes, but our strategy can be applied
to other child- and mother-related outcomes.
As \cite{Berntsen2019} point out, identifying the effect of ART on
perinatal outcomes as separate from that of infertility
remains a major challenge.
Studies that use all spontaneously conceived births\footnote{These are
    births from both fertile mothers and mothers with a history of infertility
    treatments.} to construct a comparison group for ART birth
\cite[e.g.][]{Farley2021}
can only identify this effect by relying heavily on their observed controls,
as the mothers of spontaneously conceived babies can be very different from ART
mothers in terms of both observable and unobservable characteristics.
Sibling studies such as \cite{Westvik2021}\textemdash where some siblings were
conceived spontaneously, while others via ART\textemdash help controlling
for maternal factors that do not change over time, but are confounded by
maternal age, birth order and all the time-varying maternal factors.
First, our approach implies conditioning on mothers with a history
of ART. This eliminates an important source of selection bias, which is
prevalent in studies that include births from fertile mothers.
Second, it allows us to compare\textemdash to ART births\textemdash
spontaneous births to similar mothers who also had recently undergone ART.
This helps us build a credible
counterfactual to conceiving via ART treatment while leveraging DML to
flexibly control for a number of key confounders, including age and parity.
Our strategy can be used in future research to study the effect of ART on
other outcomes of interest.

The rest of the paper is structured as follows. Section \ref{sec:lit_review}
reviews the literature on ART and obstetric outcomes. Section \ref{sec:data}
describes our sample and data sources, while Section \ref{sec:methods} details the
methods we use to estimate the effect of ART on obstetric outcomes.
In Section \ref{sec:desc_analysis}, we conduct a descriptive analysis.
In Section, \ref{sec:results} we report and discuss our findings.
our sample. Section \ref{sec:conclusion} closes with a review of the results.

\section{Related literature \label{sec:lit_review}}
In this section, we start by defining assisted reproduction technologies and
discussing related medical literature. We then discuss Health
Economics research on ART.

\subsection{Background and medical literature}
Assisted reproductive technology (ART) is a mainstream treatment for
infertility, with an estimated over two million cycles performed each year
globally \citep{Chambers2021}. There are over 500,000 ART-conceived births
each year globally, accounting for more than 5\% of total births in a number
of high-income countries \citep{EIM2022,ChoiVenetis22}.
ART treatment, otherwise known as IVF, involves a sequential procedure
where (i) the oocyte (egg) and sperm are joined together outside the body in a
specialised laboratory, (ii) the fertilised egg (embryo) is allowed to grow in
a protected environment for 4-6 days, and (iii) the embryo is transferred into
the woman's uterus with the hope of a pregnancy.
Perinatal outcomes are reported to be poorer in ART-conceived pregnancies
compared to pregnancies conceived after a spontaneous conception, and this
might be attributed either to the underlying infertility, to the ART treatment,
or both \citep{Berntsen2019}. Several meta-analyses and cohort studies have
compared perinatal outcomes for ART-conceived singletons with
spontaneously-conceived singletons and found that ART-conceived singletons
were at risk of poorer perinatal outcomes, including a higher risk preterm birth
\citep{Helmerhorst2004,jackson2004,Pandey2012,Marino2014,Qin17},
low birth weight
\citep{Helmerhorst2004,jackson2004,Pandey2012,Marino2014,Qin17},
and  perinatal mortality
\citep{Helmerhorst2004,jackson2004,Pandey2012,Marino2014,Qin17,Farley2021}.
Moreover, previous studies reported that compared to spontaneously-conceived
pregnancies, ART-conceived pregnancies are associated with increased obstetric
complications, such as caesarean section
\citep{Helmerhorst2004, Pandey2012, Buckett2007},
induction of labour \citep{Pandey2012}, gestational diabetes
\citep{Pandey2012}, and gestational hypertension
\citep{Pandey2012,Opdahl2015,Thomopoulos2013}.

However, given that these obstetric complications are relatively rare
the absolute risk due to ART remains quite small \citep{Pandey2012}.
A major challenge for research into the obstetric and perinatal outcomes
following ART is how to separate the contribution of underlying infertility
from the ART treatment per se. Currently, it remains unclear to what
extent these adverse obstetric and perinatal outcomes can be
attributed to the ART treatment, rather than the underlying infertility and
other risk factors correlated with ART use.
Our study addressed this issue.

\subsection{Health Economics literature}
There is growing interest in fertility treatments among economists,
as the extension of the fertility window they bring has the potential
to affect the childbirth and career decisions of women, among other things.
In particular, US states have progressively mandated that private insurers
offer coverage for fertility treatment.
Such mandates have informed a growing body of literature
about how the resulting increased access to
fertility treatments \citep{Hamilton2012}
has increased the birth rates for women over age 35
\citep{Schmidt2005,Schmidt2007}, increased multiple-birth rates
for all women \citep{Zaresani2023}, delayed the timing of
marriage \citep{Abramowitz2014} and, the extent to which  existing disparities
in treatment use have increased \citep{Bitler2006, Buckles2013},
with the largest utilization of in fertility treatments found in older,
more educated women \citep{Bitler2011}, and the extent to which access to
ART treatment have decreased the rate of child adoption

\cite{Hamilton2018} study these insurance mandates formally,
constructing and estimating a dynamic model for IVF patients'
choices. To reduce patient welfare while lowering medical costs,
they recommend policies that increase insurance coverage while
charging more for multiple births (twins, triplets, etc.),
as they have medical costs much higher than singleton births.
More broadly, \cite{Bhalotra2022} provide evidence that, compared to
singletons, twin births not only
decrease maternal and child health, they also lead to steeper decreases
in maternal earnings and lower subsequent fertility.
Their study leverages a Swedish policy mandating that IVF embryo transfers
must be single, rather than multiple, hence lowering the chances of a multiple
birth for IVF prospective mothers.

An international study of 30 countries that systematically quantified the
impact of consumer cost on assisted reproduction technology (ART)
utilization and number of embryos transferred found that the relative
cost that consumers pay for ART treatment (after taking account of insurance,
wages and disposable income) predicts the level of access to treatment as well
as the number of embryos transferred \citep{Chambers2014}.
After controlling for other explanatory
variables, affordability had a strong and robust association with utilization,
with a 10-percentage-point decrease in affordability predicted to, on average,
decrease ART utilization by 32\%. Affordability was also independently
associated with the number of embryos transferred with less affordability
treatment leading to more risky embryo transfer practices.

Affordability and access have not been an issue for Israeli
citizens since 1994, when the government made IVF free to all citizens
up to the birth of two children \cite{Gershoni2021, Gershoni2021AEJ}.
Comparing men's outcomes to those
of women, they find that this reform delayed the age at first marriage for
women and their age at first birth. It also led to an increase in the number
of women who have completed college and graduate education.
This latter finding is consistent with \cite{Kroeger2017}, who find US
evidence that increased coverage for fertility treatments
makes college-educated women more likely to be in professional occupations
and to complete a professional degree.

A different approach is taken by \cite{Lundborg2017},
who do not study the impact of IVF directly, and instead
use IVF success as an instrument for childbearing. They find that childbearing
(i.e. a successful IVF cycle) leads to a persistent decrease in women's
earnings. They work less when the children are young and move to lower-paid jobs
closer to home as the children get older\textemdash thus earning less.

Finally, our study contributes to the literature on
``defensive medicine'' and c-sections.
This literature finds robust cross-country evidence that part of the variation
in caesarean rates can be explained by the physician's fear of litigation
\citep{Dubay99,Jena15,Betran18,Bertoli19,Fenizia19,Longo20}.
Given the findings of this literature, one could expect this phenomenon
to be even more pronounced in IVF pregnancies, which not only tend to
be high risk, but might have also required a considerable financial
investment. Our study addresses our question, studying whether ART
leads to changes in caesarean rates.

\section{Data \label{sec:data}}
This study was approved by the NSW Population and Health Services Research
Ethics Committee (2017/HRE1202), the ACT Health Human Research Ethics
Committee (ETH.2.218.032), the Calvary Public Hospital Human Research and
Ethics Committee (3-2018), and the Australian Institute of Health and
Welfare Ethics Committee (AIHW) (EO2017/4/420),

Our cohort was extracted from a newly created medically assisted reproduction
(MAR) data linkage, described in detail elsewhere
\citep{ChambersChoi21,ChoiVenetis22}.
This linkage contains longitudinal demographic, treatment, and outcome data
for women who underwent ART treatment in New South Wales (NSW) and
Australian Capital Territory (ACT) between 2009 and 2017.
Combined, NSW/ACT represents Australia’s most populous
jurisdictions with more than 8.5 million residents and nearly
100,000 births per year (Australian Bureau of Statistics 2021).

Briefly, the MAR data linkage includes all births (mother and baby records)
recorded in NSW and ACT perinatal data collections (PDCs) that are
linked to the Australian and New Zealand Assisted Reproduction Database
(ANZARD), a clinical quality registry of all ART treatments and outcomes
performed across Australia and New Zealand. Given the licencing requirement
for fertility clinics to submit data to ANZARD, complete ascertainment can
be assumed. Over 96\% of ANZARD records were able to be linked to
jurisdictional datasets and over 94\% of births recorded in ANZARD also had a
birth recorded in the PDCs with very high concordance between birth outcomes
\citep{ChambersChoi21}. In addition, the linkage includes data from the
NSW and ACT Admitted Patient Data Collection (APDC) and the Registry of Births,
Deaths and Marriages (RBDM).

\subsection{Study cohort}
From the wider MAR data linkage, we make use of the ANZARD and PCD datasets,
which give us detailed information on ART treatment history of
mothers and on their babies' perinatal outcomes.

Our sample covers all the births that occurred between 2009-2017 in New South Wales.
We drop births in the Australian Capital Territory as they are only available
for public hospitals\textemdash contrary to the NSW data, which also
covers private-hospital births.
We also drop mothers with tubal disease, as they would not be able to
conceive without ART and thus would not be eligible for the control group.

\subsection{Treatment and control groups}
We construct the treatment group by selecting all ART-conceived births resulting
from ART treatment received in 2009-2017.
We construct the control group by selecting all births conceived
spontaneously (i.e. ART-independent births) within a 3-to-12-month window
after a failed ART cycle received in 2009-2017.
The 3-to-12 month windows was chosen to make the treatment and control groups as
similar as possible over unobserved characteristics and to  minimise the
risk of misclassification between ART and spontaneous births.
The upper bound of the window needs to strike a balance between bias and
variance. The smaller the window, the greater the similarity between
treatment and control groups, at the price of fewer observations.
This initial sample includes 21,102 births, of which 19,646 are in treatment
group and 1,456 are in the control group.

Then, we drop births with at least one missing observation,
7477 treatment births
(38.1\%) and 521 control births (35.8\%)
leaving us with 12,169 treatment births and 935 control births\footnote{
    These values change slightly across our outcomes at the estimation
    stage as, when we run the analysis for each outcome, we drop the births
    for which that outcome is missing.}.
These missing rates are mostly driven by the gestational diabetes
and diabetes mellitus variables, of which 32\% of the observations are
missing. These are important variables, either confounders or closing
a back-door path, and excluding them would likely bias our estimates.
Our other covariates have less than 0.5\% of missing observations
with the exception of ``smoking'', which has 3.6\%.

\begin{figure}
    \centering
    \includegraphics[width=0.8\textwidth]{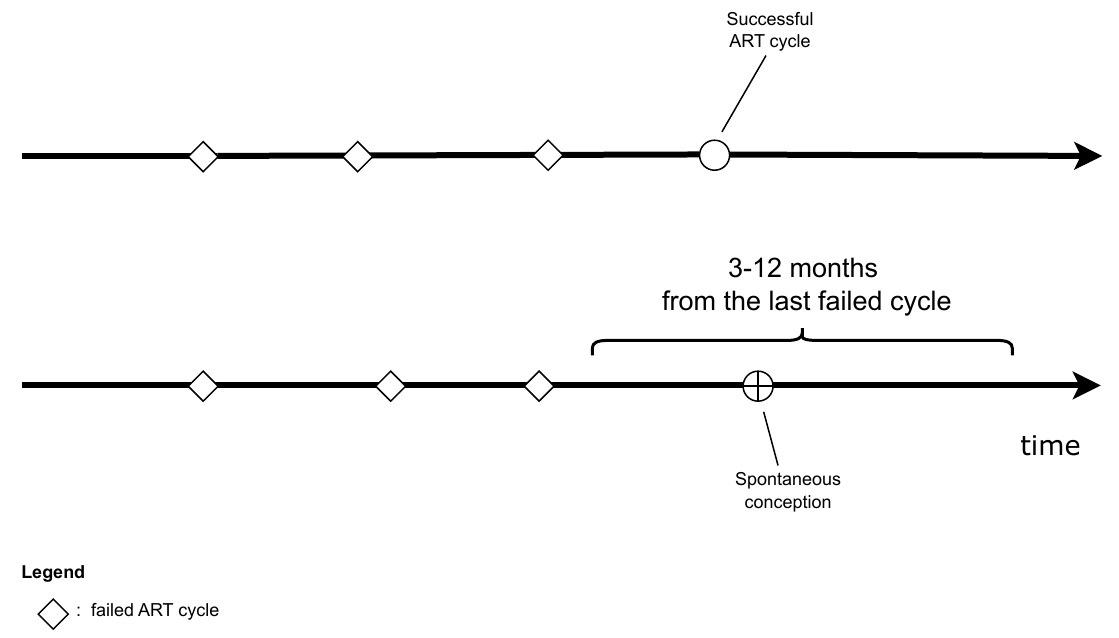}
    \caption{Diagram illustrating the control-group construction}
    \label{fig:sample_construction}
\end{figure}

The key issue informing the width of the conception window
is that prospective mothers experiencing an unsuccessful
ART cycle tend to schedule the next one at a close date, if they decide
to schedule one\footnote{
    In our sample of subfertile and ART-conceived singletons, the median
    time between unsuccessful cycles is 97 days. The 25th percentile is 56 days
    and the 75th percentile is 322 days.
}. Hence, if we compare a mother that has a successful ART cycle
with one that conceived spontaneously a long time after a failed cycle, we are
comparing a woman that might have undertaken several more cycles, had she been
unsuccessful, with a woman that did not do that.
These two women are likely different in important ways, and this would likely
be a source of selection bias in our analysis.
Figure \ref{fig:sample_construction} can help visualising how the control
group is constructed.

\subsection{Outcomes and covariates}
Our outcomes and covariates of interest are presented in Table \ref{tab:des}.
These covariates have been selected as they are key confounders\footnote{
    While pre-eclampsia, gestational hypertension and gestational diabetes are not
    confounders, it is important to include them as they might be blocking a
    backdoor path or mediate the effect of ART on obstetric outcomes},
the latter meaning that
they can both have an impact on our outcomes and on
the probability of a successful ART cycle\textemdash the treatment.

We complement the information in Table \ref{tab:des} with a few remarks.
First, APGAR scores are standard measures of newborn
health at birth and, as such, are strong predictors of neonatal death
and neurological disorders \citep[e.g.][]{Persson2018, Cnattingius2020},
among other adverse outcomes. Lower values imply lower newborn health.

Second, the confounder recording the number of previous failed ART cycles
was based on the number of autologous ART cycles\footnote{
    An autologous ART cycle is an ART cycle where the woman uses her oocyte
    or embryo, rather than one donated by a third party. In other words,
    we excluded donor cycles.}
, fresh or thaw\textemdash excluding freeze-all oocyte and/or embryo cycles
\footnote{
    These are cycles where a woman freezes her oocytes or embryos for
    future use.
}.
Women donating or freezing their oocytes and/or eggs were excluded
because they weren't at risk of pregnancy.

Third, the \textit{Socio-economic Indexes for Areas} \citep[SEIFA;][]{SEIFA2011}
variable used is the Index of Relative Socio-economic Disadvantage (IRSD).
It is constructed by the Australian Bureau of Statistics and calculated
using 2011 Census data, at the base level of Statistical Area Level 1 (SA1).
A number of variables are used to measure determinants of disadvantage at the
SA1 level, such as the proportion of unemployed or of single parents.
Then principal component analysis is used to determine the variable weights in
a data-driven way, resulting in the index. Finally, the index is divided into
quintiles, which we use in our analysis. We are able to assign this variable
thanks to SA1 information provided in the Perinatal Data Collection
dataset.

\begin{longtable}{llp{8cm}}
    \caption{Variable description and source \label{tab:des}}                                                                                                       \\
    \toprule
    Variable                              & Source & Description                                                                                                    \\
    \midrule
    \endfirsthead

    \endhead

    \multicolumn{3}{c}%
    {{\bfseries \tablename\ \thetable{} -- continued from previous page}}                                                                                           \\
    \toprule
    Variable                              & Source & Description                                                                                                    \\
    \midrule
    \endhead

    \hline \multicolumn{3}{r}{{Continued on next page}}                                                                                                             \\ \hline
    \endfoot

    \hline \hline
    \endlastfoot

    \textit{Outcomes}                     &        &                                                                                                                \\
    \hspace{5pt} Gestational age at birth & PDC    & No. of completed weeks of gestation at the time of birth                                                       \\
    \hspace{5pt} Preterm birth            & PDC    & Equal to one if the baby is born after $<37$ weeks of gestation and zero otherwise                             \\
    \hspace{5pt} Spontaneous labour       & PDC    & Equal to one if labour was not induced (spontaneous) and zero otherwise                                        \\
    \hspace{5pt} Preterm Spont. labour    & PDC    & Equal to one if the labour was spontaneous and the birth preterm, and zero otherwise                           \\
    \hspace{5pt} C-section                & PDC    & Equal to one if the baby was delivered via c-section and zero otherwise (i.e. vaginal delivery)                \\
    \hspace{5pt} APGAR 1 \& APGAR 5       & PDC    & Count variables with values between 0-10, measuring newborn health at 1 and 5 minute from birth, respectively. \\
    \hspace{5pt} Birth weight             & PDC    & Weight at birth, in grams    \vspace{5pt}                                                                      \\

    \textit{Covariates}                   &        &                                                                                                                \\
    \hspace{5pt} Age                      & PDC    & Mother's age at time of birth                                                                                  \\
    \hspace{5pt} Parity                   & PDC    & Mother's number of previous pregnancies greater than 20 weeks gestation                                        \\
    \hspace{5pt} No. of failed ART cycles & ANZARD & No. of past failed ART cycles                                                                                  \\
    \hspace{5pt} Maternal YOB             & PDC    & Maternal year of birth                                                                                         \\
    \hspace{5pt} SEIFA index              & ABS    & Quintiles of the 2011 Census Index of Relative Socio-economic Disadvantage \citep{SEIFA2011}                   \\
    \hspace{5pt} Remoteness of area       & PDC    & 4 levels of remoteness of mother's dwelling, each encoded into one binary variable.
    Derived from the Accessibility/Remoteness Index of Australia (ARIA+) \citep{abs2001australian}                                                                  \\
    \hspace{5pt} Born in Australia        & PDC    & Binary variable equal to one if the mother was born in Australia and zero otherwise                            \\
    \hspace{5pt} Diabetes mellitus        & PDC    & Binary variable for maternal diabetes mellitus                                                                 \\
    \hspace{5pt} Chronic hypertension     & PDC    & Binary variable for maternal chronic hypertension                                                              \\
    \hspace{5pt} Pre-eclampsia            & PDC    & Binary variable for maternal proteinuric gestational hypertension                                              \\
    \hspace{5pt} Gestational hypertension & PDC    & Binary variable for maternal non-proteinuric gestational hypertension                                          \\
    \hspace{5pt} Smoke                    & PDC    & Binary variable equal to one if the mother smoked at any time during the pregnancy                             \\
    \hspace{5pt} Endometriosis            & ANZARD & Binary variable equal to one if the treating clinician thinks
    subfertility is due endometriosis                                                                                                                               \\
    \hspace{5pt} Male infertility         & ANZARD & Binary variable equal to one if the treating clinician thinks
    subfertility is due to a male factor problem                                                                                                                    \\
    \hspace{5pt} Other infertility        & ANZARD & Binary variable equal to one if the treating clinician thinks
    subfertility is due to other female factors apart from tubal disease and endometriosis                                                                          \\
    \hspace{5pt} Unexplained infertility  & ANZARD & Binary variable equal to one if the treating clinician thinks
    there is clinical subfertility without apparent explanation                                                                                                     \\
\end{longtable}

\section{Methods} \label{sec:methods}
In this section, we define the ideal experiment that our observational
analysis approximates (Section \ref{sec:exp}), followed by a presentation
of the empirical methods we use (Section \ref{sec:emp_spec}).

\subsection{Ideal experiment} \label{sec:exp}
\begin{figure}
    \centering
    \includegraphics[width=1\textwidth]{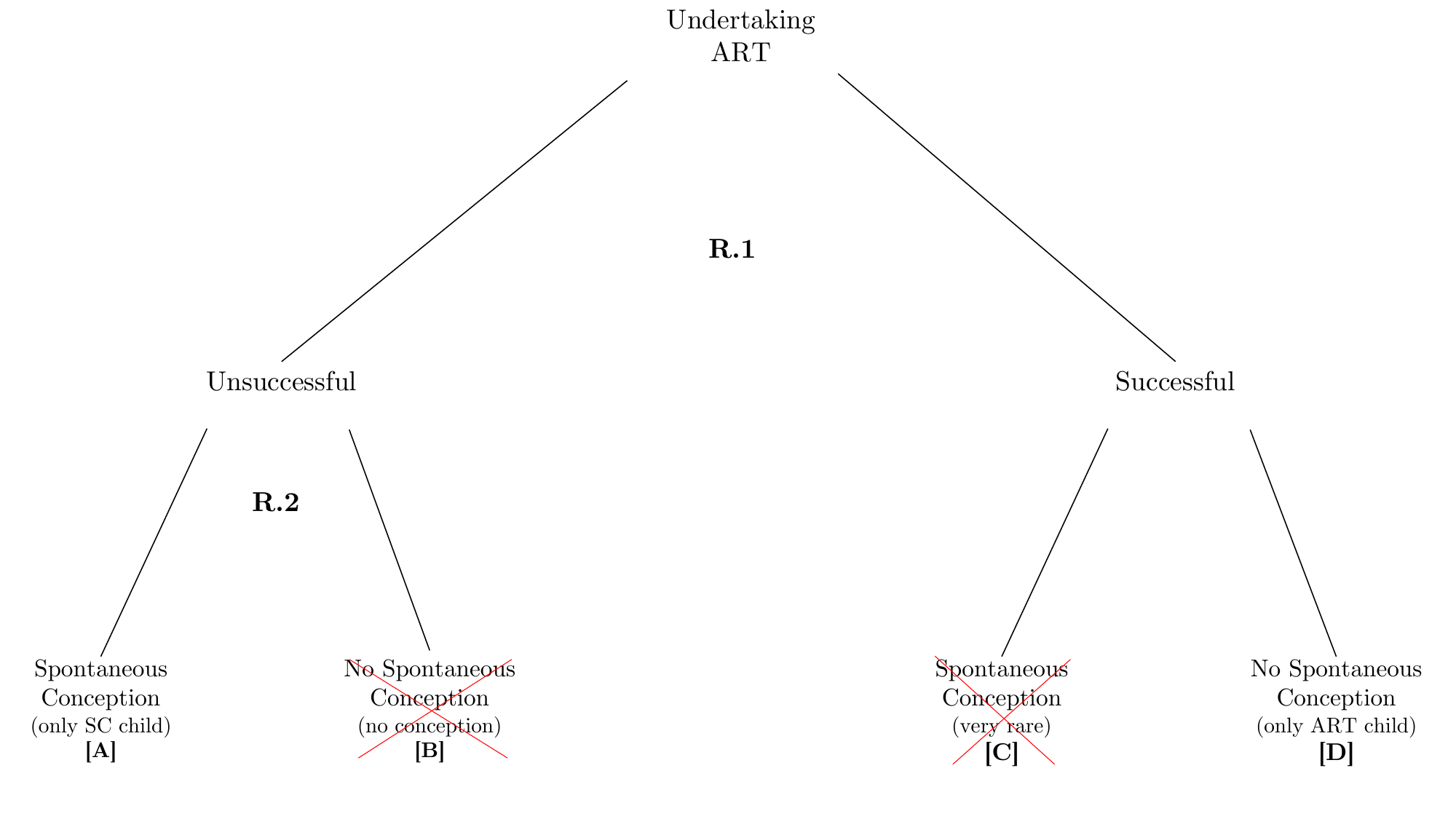}
    \caption{Diagram of the ideal experiment}
    \label{fig:random}
\end{figure}

Our study can be conceptualised as approximating a two-stage randomised
experiment. As shown in Figure \ref{fig:random}, in the
ideal experiment, we would take the whole population of prospective mothers
undergoing an ART cycle and determine the success of their cycle at random,
thus dividing them into two groups. Then, we would randomly ``assign'' a
spontaneous conception to each of these two groups, leaving us with four groups
(A, B, C, and D in Figure \ref{fig:random}). From these,
we exclude Group C as a spontaneous
conception shortly after a successful ART cycle
is almost impossible and we do not observe it. By this we mean that
within our 3-to-12-month window from a successful ART cycle\textemdash
hence resulting in a live birth 9 months after conception\textemdash
there are only 3 months for an ART mother to conceive again and
spontaneously. We also exclude Group B as it does not lead to a conception\textemdash
our obstetric outcomes are not defined for this group.
Instead, we focus on Groups A and D.
Notice that here there are
two ``treatments'': ART success and
spontaneous conception after a failed ART cycle.

If is this two-step randomisation model were the correct model of the world,
we should observe
excellent balance between our groups of interest\textemdash
Group A (spontaneous conception) and Group D (ART conception).
We do not impose this strong assumption though. Instead,
we make a conditional exogeneity/independence claim.
This means assuming
that the treatment assignments are independent of our outcomes
of interest after controlling for the relevant confounders\textemdash
variables causing changes in both treatment and outcome.

Given that our dataset includes key confounders and covariates,
we see this conditional independence assumption as credible.
Nonetheless, as no study can credibly claim to control for \textit{all}
confounders, we study how violations of the conditional independence
assumption affect our estimates in Section \ref{sec:ovb}.

\subsection{Empirical specifications} \label{sec:emp_spec}
Once we have defined our cohort, we take two approaches to estimate the effect
of ART on obstetric outcomes.
First, we estimate the following model via OLS:

\begin{equation}
    Y_{i} = \tau {D}_{i} + \mathbf{X}_{i}^{\prime} \boldsymbol{\beta} + \varepsilon_{i},
    \label{eq:ols}
\end{equation}

where, $Y_i$ is the outcome variable, $D_i$ is the treatment variable,
and matrix
$$
    \mathbf{X}_{i} = \left(X_{1i}, \ldots, X_{pi}\right)
$$
consists of the covariates listed in Table \ref{tab:des}\textemdash
which enter the equation linearly\textemdash and $\varepsilon_{i}$ is
the error term.
Assuming that D is conditionally exogenous/independent, i.e.
$$
    \mathbb{E}\left[\varepsilon_i \mid D_i, \mathbf{X}_i \right]=0,
$$
and that $\tau$ is homogeneous across the treatment and control groups,
then $\tau$ measures the average treatment effect (ATE) of
treatment $D_i$ on outcome $Y_i$.

Second, we take a double machine learning approach and estimate the
treatment effects using an Interactive Regression Model \citep{chern18}.
Given a binary treatment variable, the Interactive Regression Model estimates
fully heterogeneous average treatment effects. Here, ``fully heterogeneous''
means that each of the two response curves
\textemdash the potential outcomes as a function of $\mathbf{X}$\textemdash
is allowed to be a different nonparametric function.
Vectors $\mathbf{Y_i}$ and $\mathbf{D_i}$ are modelled such that
\begin{eqnarray}
    {Y_i} = g_0({D_i}, \mathbf{X_i}) + {U_i},
    & \mathbb{E}({U_i} \mid \mathbf{X_i}, {D_i}) = {0},
    \label{eq:dml_y}\\
    {D_i} = m_0(\mathbf{X_i}) + {V_i},
    & \mathbb{E}({V_i} \mid \mathbf{X_i}) = {0}
    \label{eq:dml_d}.
\end{eqnarray}

We still target and estimate the ATE, here written as:
$$
    \tau=\mathbb{E}\left[g_0({1}, \mathbf{X}_i)-g_0({0}, \mathbf{X}_i)\right].
$$

As mentioned above, in this framework we do not need to assume homogeneous
treatment effects.
Variables $\mathbf{X}_i$ are allowed to affect both the outcome $\mathbf{Y}$ and
the treatment $\mathbf{D}$, each via a different function\textemdash
$m_0(\mathbf{X})$ and $g_0(\mathbf{X})$, respectively.
In particular, $m_0(\mathbf{X})$ is the propensity score and as such, its
estimate can be inspected and used to test assumptions such as common
support.
This can help dissipate the black-box-type concerns typical of
machine-learning applications.

Moreover, the Interactive Regression Model identifies the ATE
under the \textit{unconfoundedness} (or
\textit{conditional independence}) assumption \citep{rosenbaum1983}.
This assumption has three parts.
First, it assumes that the probability of a birth being conceived via ART
(vs. spontaneously) does not depend of the probability of other babies being
conceived via ART (no network/spillover effects).
Second, it assumes that  the probability of a birth being conceived via ART
does not depend on their potential outcome (no selection into treatment).
This means that a birth's probability of being conceived via ART cannot
depend, for instance, on how likely the birth is born pre-term.
Third, it assumes that no birth is conceived via ART with certainty
(common support assumption).

Selection into treatment is generally a major concern in studies on ART,
as women are not randomly allocated to ART treatment, but rather
choose to undertake ART treatment based on their fertility status.
We deal with this issue by only selecting mothers
that have undergone at least one
ART cycle. Hence, this assumption is not of concern in our study.
The same argument applies to point the no network/spillover effects assumption.
Finally, the partially stochastic nature of ART success (i.e. it is not
a deterministic process), combined with an inspection of Figure
\ref{fig:ps_balance_anz12}, should convince that the common support
assumption holds.

Our final step involves a formal discussion of omitted variable bias.
Using the tools developed in \cite{Cinelli2019}, we examine how sensitive
our OLS estimates are to the omission of confounding variables\textemdash
typically because they are not observed by the analyst.

\section{Descriptive analysis} \label{sec:desc_analysis}
We start from a descriptive analysis of our sample.
Table \ref{tab:tab:cov_balance_anzard_12}
reports difference in means across treatment groups.
We find a clinically important difference in maternal age, which is higher
by 1.9 years (36.6-34.7 years) in mothers that spontaneously conceived
after a failed ART \textemdash
at an age where maternal fertility is in significant decline.
This is also reflected in a lower maternal year of birth.
In our sample, mothers who spontaneously conceived after ART
are less likely to be at their first cycle
($0.25 - 0.36 = - 0.10$ p.p.) and have had more unsuccessful cycles before their
pregnancy ($3.1 - 1.6 = 1.5$ cycles).
A last noticeable difference is in the prevalence of male infertility,
which is 7 p.p. higher in mothers of ART-conceived babies.
The two groups are clinically similar in other respects,
such as number of past pregnancies, socio-economic disadvantage,
co-morbidities and cause of infertility.

\newgeometry{top=5mm, bottom=10mm, left=5mm, right=5mm}
\thispagestyle{empty}

\begin{table}
    \centering
    \footnotesize
    \caption{\label{tab:tab:cov_balance_anzard_12}Covariate balance table}
    \begin{threeparttable}
        \begin{tabular}[t]{lllllll}
            \toprule
            \multicolumn{1}{c}{ }              & \multicolumn{2}{c}{Control} & \multicolumn{2}{c}{Treatment} & \multicolumn{2}{c}{Difference}                             \\
            \cmidrule(l{3pt}r{3pt}){2-3} \cmidrule(l{3pt}r{3pt}){4-5} \cmidrule(l{3pt}r{3pt}){6-7}
            Variable                           & Mean                        & Obs.                          & Mean                           & Obs.  & Value   & p-value \\
            \midrule
            Age                                & 36.78                       & 935                           & 34.75                          & 12169 & 2.024   & 0.0000  \\
                                               & (5.13)                      &                               & (4.48)                         &       & (0.163) &         \\
            First cycle                        & 0.26                        & 935                           & 0.36                           & 12169 & -0.099  & 0.0000  \\
                                               & (0.44)                      &                               & (0.48)                         &       & (0.015) &         \\
            No. of past unsuccessful cycles    & 3.16                        & 935                           & 1.66                           & 12169 & 1.501   & 0.0000  \\
                                               & (2.53)                      &                               & (2.18)                         &       & (0.073) &         \\
            Maternal year of birth             & 1975.81                     & 935                           & 1977.98                        & 12169 & -2.166  & 0.0000  \\
                                               & (5.21)                      &                               & (4.66)                         &       & (0.167) &         \\
            SEIFA                              & 3.49                        & 935                           & 3.49                           & 12169 & 0.001   & 0.9776  \\
                                               & (1.46)                      &                               & (1.41)                         &       & (0.048) &         \\
            Remoteness: major city             & 0.9                         & 935                           & 0.89                           & 12169 & 0.011   & 0.2843  \\
                                               & (0.3)                       &                               & (0.31)                         &       & (0.011) &         \\
            Remoteness: inner regional         & 0.07                        & 935                           & 0.09                           & 12169 & -0.014  & 0.1058  \\
                                               & (0.26)                      &                               & (0.28)                         &       & (0.01)  &         \\
            Remoteness: outer regional         & 0.02                        & 935                           & 0.02                           & 12169 & 0.005   & 0.3292  \\
                                               & (0.15)                      &                               & (0.13)                         &       & (0.005) &         \\
            Remoteness: remote and very remote & 0                           & 935                           & 0                              & 12169 & -0.001  & 0.2863  \\
                                               & (0.03)                      &                               & (0.05)                         &       & (0.001) &         \\
            Mother born in Australia           & 0.61                        & 935                           & 0.66                           & 12169 & -0.053  & 0.0013  \\
                                               & (0.49)                      &                               & (0.47)                         &       & (0.016) &         \\
            Had past pregnancies               & 0.4                         & 935                           & 0.32                           & 12169 & 0.086   & 0.0000  \\
                                               & (0.49)                      &                               & (0.47)                         &       & (0.017) &         \\
            No. of past pregnancies            & 0.53                        & 935                           & 0.37                           & 12169 & 0.154   & 0.0000  \\
                                               & (0.77)                      &                               & (0.62)                         &       & (0.027) &         \\
            Diabetes mellitus                  & 0.01                        & 935                           & 0.01                           & 12169 & 0.002   & 0.5584  \\
                                               & (0.12)                      &                               & (0.11)                         &       & (0.004) &         \\
            Gestational diabetes               & 0.12                        & 935                           & 0.09                           & 12169 & 0.033   & 0.0028  \\
                                               & (0.33)                      &                               & (0.29)                         &       & (0.011) &         \\
            Chronic hypertension               & 0.01                        & 935                           & 0.01                           & 12169 & 0.002   & 0.5394  \\
                                               & (0.11)                      &                               & (0.1)                          &       & (0.003) &         \\
            Pre-eclampsia                      & 0.02                        & 935                           & 0.02                           & 12169 & -0.001  & 0.8918  \\
                                               & (0.14)                      &                               & (0.14)                         &       & (0.004) &         \\
            Gestational hypertension           & 0.04                        & 935                           & 0.05                           & 12169 & -0.008  & 0.2416  \\
                                               & (0.2)                       &                               & (0.22)                         &       & (0.007) &         \\
            Smoked during pregnancy            & 0.02                        & 935                           & 0.01                           & 12169 & 0.008   & 0.0996  \\
                                               & (0.14)                      &                               & (0.11)                         &       & (0.004) &         \\
            Endometriosis                      & 0.11                        & 935                           & 0.11                           & 12169 & -0.008  & 0.4604  \\
                                               & (0.31)                      &                               & (0.32)                         &       & (0.011) &         \\
            Other cause of infertility         & 0.38                        & 935                           & 0.37                           & 12169 & 0.011   & 0.4964  \\
                                               & (0.49)                      &                               & (0.48)                         &       & (0.016) &         \\
            Male infertility                   & 0.35                        & 935                           & 0.43                           & 12169 & -0.079  & 0.0000  \\
                                               & (0.48)                      &                               & (0.49)                         &       & (0.016) &         \\
            Unexplained infertility            & 0.34                        & 935                           & 0.29                           & 12169 & 0.055   & 0.0006  \\
                                               & (0.48)                      &                               & (0.45)                         &       & (0.016) &         \\
            \bottomrule
        \end{tabular}
        \begin{tablenotes}
            \item \textit{Notes: } This table compares the covariate means of our treatment group\textemdash all ART births\textemdash and control group\textemdash spontaneous births conceived 3-to-12 months after a failed ART cycle. Each mean's standard errors is reported below it between parentheses. Taking the difference between the reported control and treated means typically returns a value inconsistent with the reported difference. This is due to rounding.
        \end{tablenotes}
    \end{threeparttable}
\end{table}

\restoregeometry

\begin{figure}
    \centering
    \includegraphics[width=1\textwidth]{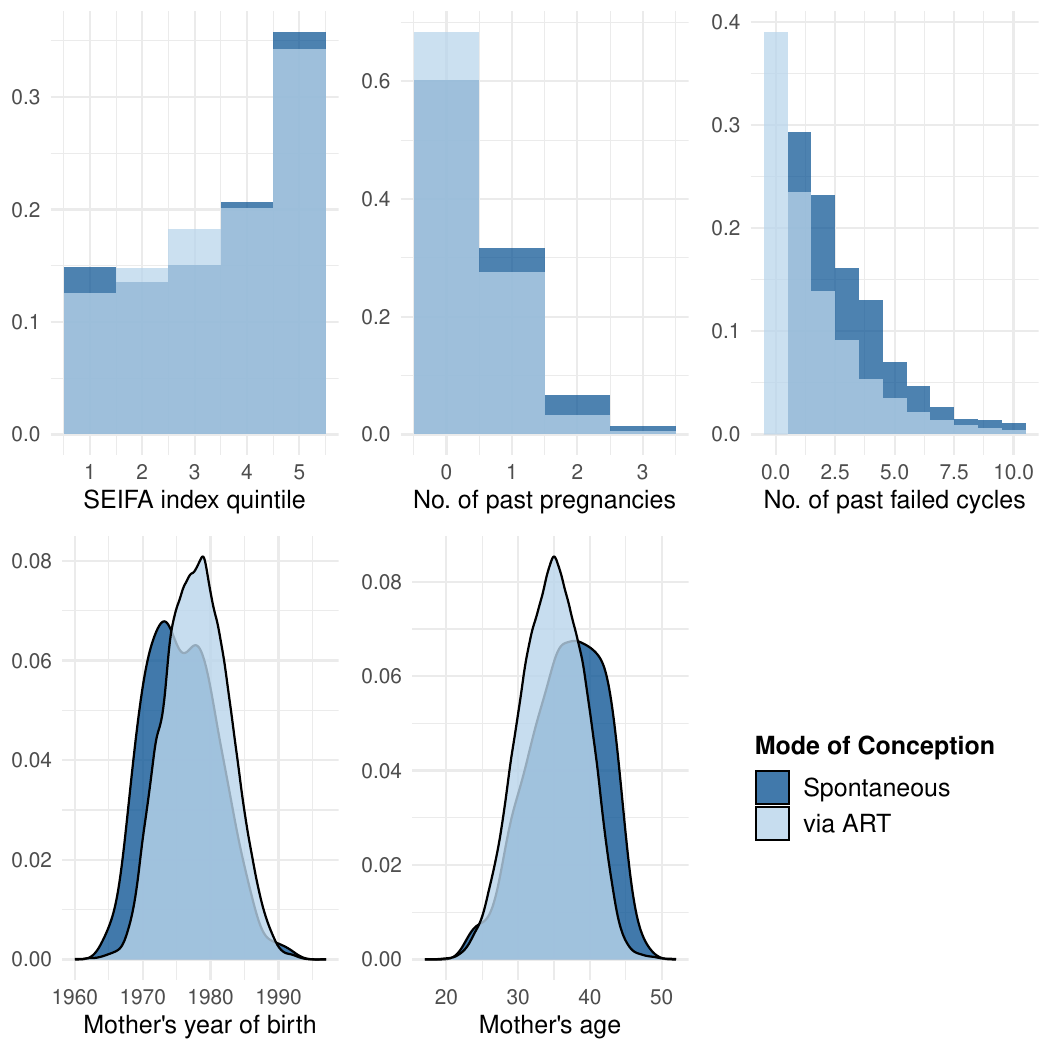}
    \caption{Densities for non-binary covariates}
    \label{fig:densities_nat_anz}
\end{figure}

Figure \ref{fig:densities_nat_anz}
plots the densities of a selection of covariates. This is to reassure us
that the distribution of the covariates are similar across the
two treatment groups,
so that the observed differences in means are simply
capturing a distribution shift.

\section{Results} \label{sec:results}
In this section, we take our models to the data and estimate the effect of
ART on obstetric outcomes. In Section \ref{sec:ols}, we estimate a linear regression
model via OLS. This is best approach under the assumption that the
model is approximately linear, and sometimes even beyond linearity
\citep{hansen2022}.
Because we do not know if the true data-generating process is approximately
linear or not, in Section \ref{sec:dml} we estimate our model flexibly
using double machine learning.
Finally, Section \ref{sec:ovb} studies how omitted variable
bias (and misspecification)
in our linear regression model could affect our results.

Average treatment effects for binary outcomes (see Table \ref{tab:des})
should be interpreted as percentage point changes. For instance,
a coefficient of 0.046 for \textit{spontaneous labour} means that
we estimate that ART treatment decreases the probability of a spontaneous
labour by 4.6 percentage points (p.p.). The coefficients for
gestational age, APGAR 1 and 5, and birth weight should be interpreted as
shifts in levels, i.e. a 0.05 coefficient means that ART treatment
increases gestational age/APGAR1/APGAR5/birth weight by 0.05 weeks of
gestation/APGAR1 points/APGAR5 points/grams.
The above applies to both OLS and DML estimates.

\subsection{Linear regression} \label{sec:ols}
Table \ref{tab:tab:ols}
reports the OLS estimates of Equation \ref{eq:ols} under our full
sample (column 1), and under further sample restrictions (columns 2-6)
\footnote{When restricting the sample, we adjust the model accordingly, so
    that, for instance, we drop the \textit{Number of previous pregnancies}
    variable if we are conditioning on having no previous pregnancies.}.
In column 2, we only include births whose mother
had no previous pregnancies longer than 20 weeks of gestation
(nulliparous) \textit{nor}
previous unsuccessful ART cycles.
In column 3, we only include births whose mother
had no previous pregnancies longer than 20 weeks of gestation.
In column 4, we only include, in the control group, births whose mother
had a single previous unsuccessful ART cycles, and in the treatment group,
ART births conceived at the first ART cycle.
Finally, in column 5 we drop all our covariates and instead stratify by
our most important covariates.
More precisely, we estimate a difference in means across treatment
groups in a sample restricted to nulliparous mothers at their first ART cycle,
aged less than 40 and without comorbidities (selected primips).

\begin{table}

    \caption{\label{tab:tab:ols}OLS: Effect of IVF conception on perinatal outcomes}
    \centering
    \begin{threeparttable}
        \begin{tabular}[t]{lllllll}
            \toprule
                                            & (1)       & (2)       & (3)       & (4)       & (5)       & (6)       \\
            \midrule
            \textit{Spontaneous labour}     & 0.042     & -0.017    & 0.028     & 0.008     & 0.028     & -0.017    \\
                                            & (0.016)   & (0.041)   & (0.02)    & (0.031)   & (0.025)   & (0.054)   \\
                                            & {}[0.006] & {}[0.668] & {}[0.157] & {}[0.786] & {}[0.263] & {}[0.756] \\
            \textit{Preterm spont. labour } & 0.009     & 0.01      & 0.014     & 0.006     & 0.014     & 0.006     \\
                                            & (0.006)   & (0.014)   & (0.008)   & (0.01)    & (0.01)    & (0.019)   \\
                                            & {}[0.156] & {}[0.481] & {}[0.075] & {}[0.587] & {}[0.157] & {}[0.751] \\
            \textit{C-section}              & -0.035    & 0.046     & -0.016    & 0.014     & 0.001     & -0.006    \\
                                            & (0.017)   & (0.04)    & (0.021)   & (0.031)   & (0.026)   & (0.053)   \\
                                            & {}[0.039] & {}[0.252] & {}[0.437] & {}[0.657] & {}[0.964] & {}[0.915] \\
            \textit{Gestational age}        & -0.077    & -0.154    & -0.145    & -0.07     & -0.222    & -0.086    \\
                                            & (0.086)   & (0.193)   & (0.12)    & (0.151)   & (0.14)    & (0.285)   \\
                                            & {}[0.375] & {}[0.423] & {}[0.225] & {}[0.642] & {}[0.113] & {}[0.764] \\
            \textit{Pre-term birth}         & 0.01      & 0.027     & 0.017     & 0.017     & 0.022     & 0.015     \\
                                            & (0.01)    & (0.02)    & (0.013)   & (0.016)   & (0.015)   & (0.027)   \\
                                            & {}[0.286] & {}[0.179] & {}[0.192] & {}[0.294] & {}[0.152] & {}[0.58]  \\
            \textit{APGAR1}                 & -0.001    & 0.202     & 0.02      & 0.134     & 0.002     & 0.364     \\
                                            & (0.057)   & (0.154)   & (0.078)   & (0.109)   & (0.091)   & (0.211)   \\
                                            & {}[0.983] & {}[0.192] & {}[0.795] & {}[0.218] & {}[0.981] & {}[0.085] \\
            \textit{APGAR5}                 & 0.012     & 0.149     & -0.017    & 0.126     & -0.031    & 0.306     \\
                                            & (0.041)   & (0.108)   & (0.055)   & (0.081)   & (0.065)   & (0.168)   \\
                                            & {}[0.764] & {}[0.169] & {}[0.756] & {}[0.117] & {}[0.634] & {}[0.068] \\
            \textit{Birth weight}           & 24.057    & -25.465   & 10.402    & 13.195    & -8.251    & 15.619    \\
                                            & (20.749)  & (47.036)  & (27.358)  & (36.889)  & (32.334)  & (67.219)  \\
                                            & {}[0.246] & {}[0.588] & {}[0.704] & {}[0.721] & {}[0.799] & {}[0.816] \\
                                            &           &           &           &           &           &           \\
            \textit{Obs. by treatment}      & 935/12169 & 153/3493  & 558/8311  & 268/4717  & 376/7199  & 91/2661   \\
            \bottomrule
        \end{tabular}
        \begin{tablenotes}
            \item Notes: This table reports average treatment effect
            estimates for the effect of an IVF conception on perinatal outcomes,
            estimated via OLS.
            Each column reports estimates for a different model specification.
            Each row reports effects for a different outcome.
            Column (1) reports the preferred estimates, which use all sample and
            confounders. Column (2): baseline if preterm equal 1.
            Column (3): first pregnancy and cycle. Column (4): first
            pregnancy. Column (5) first cycle.
            Column (6) restricts the sample to women at their first
            pregnancy and of less than 40 years of age.
            Standard errors are reported between parentheses, while p-values
            between square brackets.
        \end{tablenotes}
    \end{threeparttable}
\end{table}

The estimated treatment effects are, in general, clinically small.
Focusing on our baseline estimates (Column 1),
we find that ART has no significant effect on the rate of preterm births,
on the rate of spontaneous preterm births,
and on the average gestational age at birth\textemdash
neither at the statistical nor at the clinical level.
Taking preterm birth as an example,
the 0.008 coefficient, with a 0.01 standard error, indicates that
our best estimate for the effect of ART treatment of preterm birth
is 0.8 p.p., but we cannot exclude that the true estimate is equal to zero
at the 95\% confidence level. The implied confidence interval is
    [$-0.0116, 0.0276$], so that we cannot exclude effects ranging from
-1.16 p.p. to +2.76 p.p\footnote{Confidence intervals, while omitted in
    Table \ref{tab:tab:ols} due to space constraints, are reported for the
    double machine learning estimates, which are very similar to
    the OLS estimates and are reported
    Tables \ref{tab:tab:dml_forest_ate_anzard_12}
    and \ref{tab:tab:dml_forest_ate_anzard_12_trim_15}}.

The same applies to birth weight and to the APGAR 1 and 5 scores,
whose estimated effects are not statistically significant and are clinically
small.
To give an idea of the magnitudes of these statistically insignificant
estimates, the effect of ART on APGAR1 is estimated to be $-0.013$ points,
on a 1-to-5 scale, while the effect on gestational age at birth is
estimated to be $-0.057$ weeks, which less than 10 hours.
As shown by the p-values in the square brackets, these estimates
are not statistically significant at the 1\% level or below, despite
small standard errors.
Considering that gestational age is measured in weeks, the APGAR scores are
integers between zero and five and that birth weight is measured grams,
we can notice how imposing additional restrictions on the sample
does not shift the estimates in a clinically significant way for the most part.
Understandably, our estimates' standard errors and statistical significance
varies as the sample gets smaller.

Moreover,
ART increases the chances of a spontaneous labour by 4.6 p.p., implying that
ART mothers are less likely overall to be
subject to a labour induction than control
mothers, who conceived spontaneously after a failed ART cycle.
This reduction in interventions in also found at the mode-of-delivery level,
where ART reduces the overall chances of a C-section by 5 p.p.
Similarly to the estimated impacts on gestational age, APGAR scores and birth
weight, imposing further sample restrictions increases our standard errors
but does not have a clinically significant impact on our estimated
treatment effects.

The estimates from restricted samples used in Columns 2-5 are largely
consistent with our baseline estimates in Column 1.
The coefficients for c-section and
spontaneous labour, which are statistically significant in Column 1,
remain consistent in their sign (with one exception).
They do lose statistical significance though, as the sample size
drops\textemdash which is an issue particularly for the control group.
The coefficients for statistically
The coefficients for the other outcomes, not significant in Column 1,
remain mostly small, while becoming more noisy due to reduced sample
sizes.

\subsection{Double machine learning} \label{sec:dml}
\begin{figure}
    \centering
    \includegraphics[width=1\textwidth]{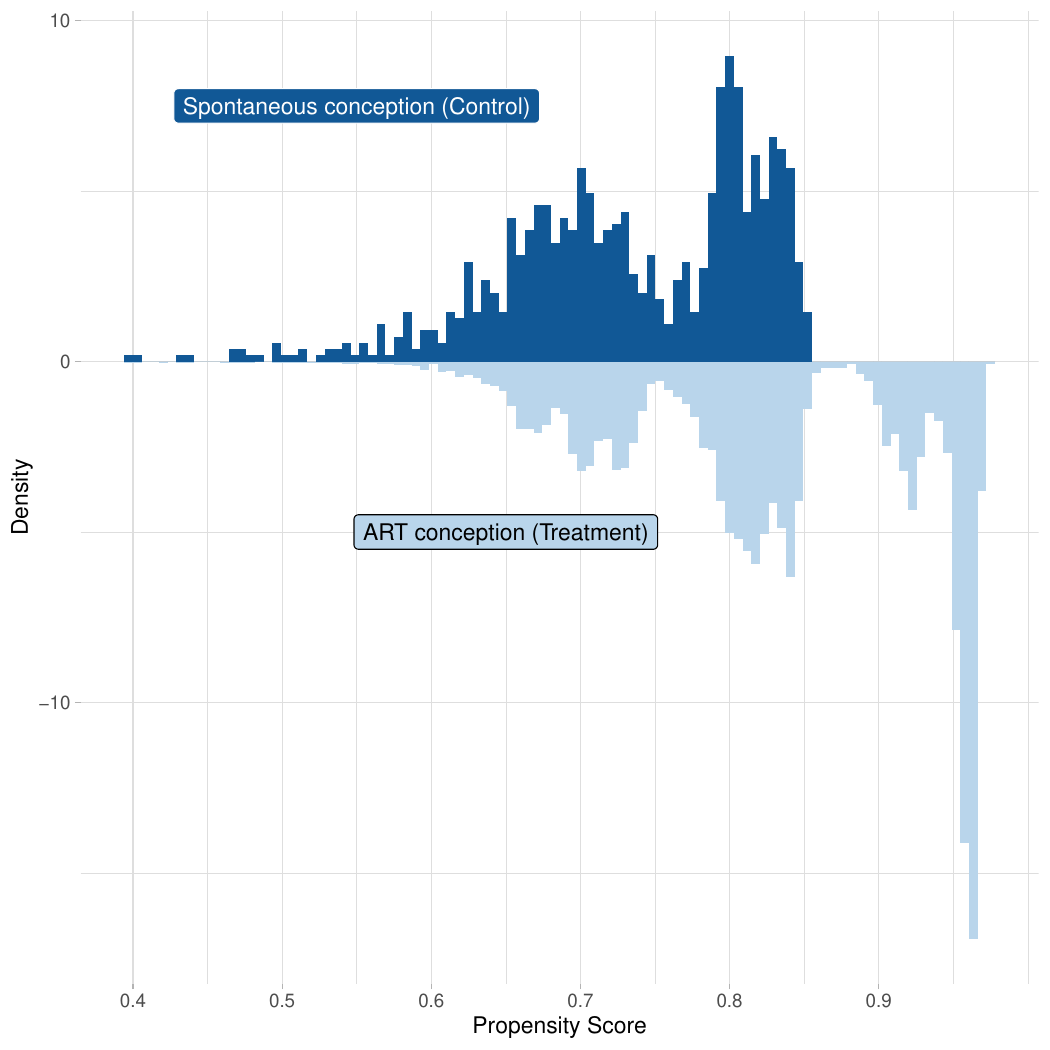}
    \caption{Propensity score density by treatment group}
    \label{fig:ps_balance_anz12}
\end{figure}

We start the analysis of our double machine learning model by inspecting
its propensity score estimates,
which the model uses to weight each observation and ultimately
estimate the treatment effects.
Figure \ref{fig:ps_balance_anz12}
shows the estimated propensity score by treatment group.
This is useful for two reasons.

First, we can check whether the common support assumption is met.
This implies that no unit should have a score of one or zero, which we observe.
Units close to zero or one would be of concern too as they would be assigned a
high weight, and we have some observations between a cluster of observation
around 0.95, with a maximum of 0.976. We take this into account in our analysis
by trimming all observations outside the $[0.15, 0.85]$ propensity score
interval.

Second, Figure \ref{fig:ps_balance_anz12}
allows us to inspect the degree of balance between the two treatment groups.
We can see that the two distributions overlap over most of the support,
with the notable exception of the $[0.85, 0.976]$ interval.
Again, we take this into account
by trimming all observations outside the $[0.15, 0.85]$ propensity score
interval.

\begin{figure}
    \centering
    \includegraphics[width=0.8\textwidth]{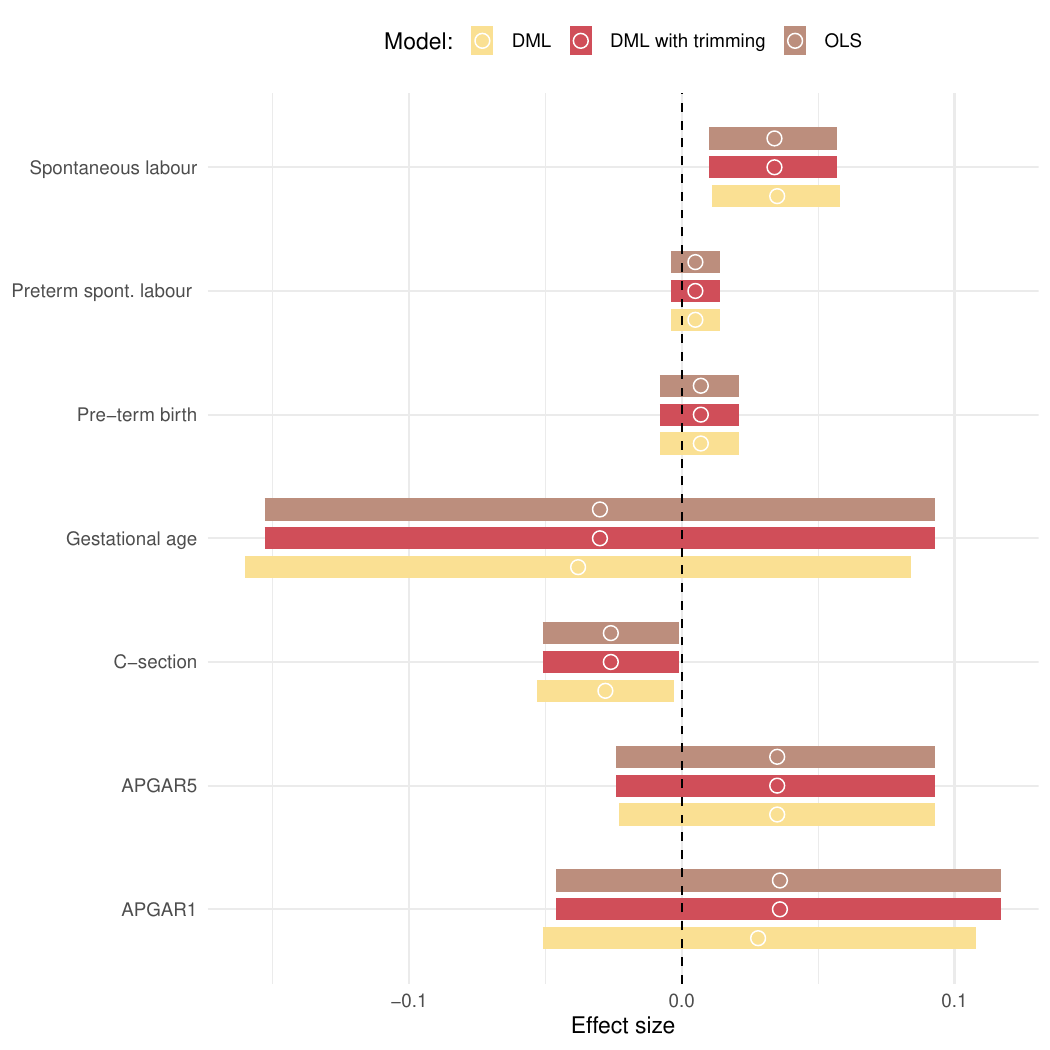}
    \caption{ATE estimates across outcomes and models}
    \label{fig:forest_nobweight}
\end{figure}

\begin{figure}
    \centering
    \includegraphics[width=0.8\textwidth]{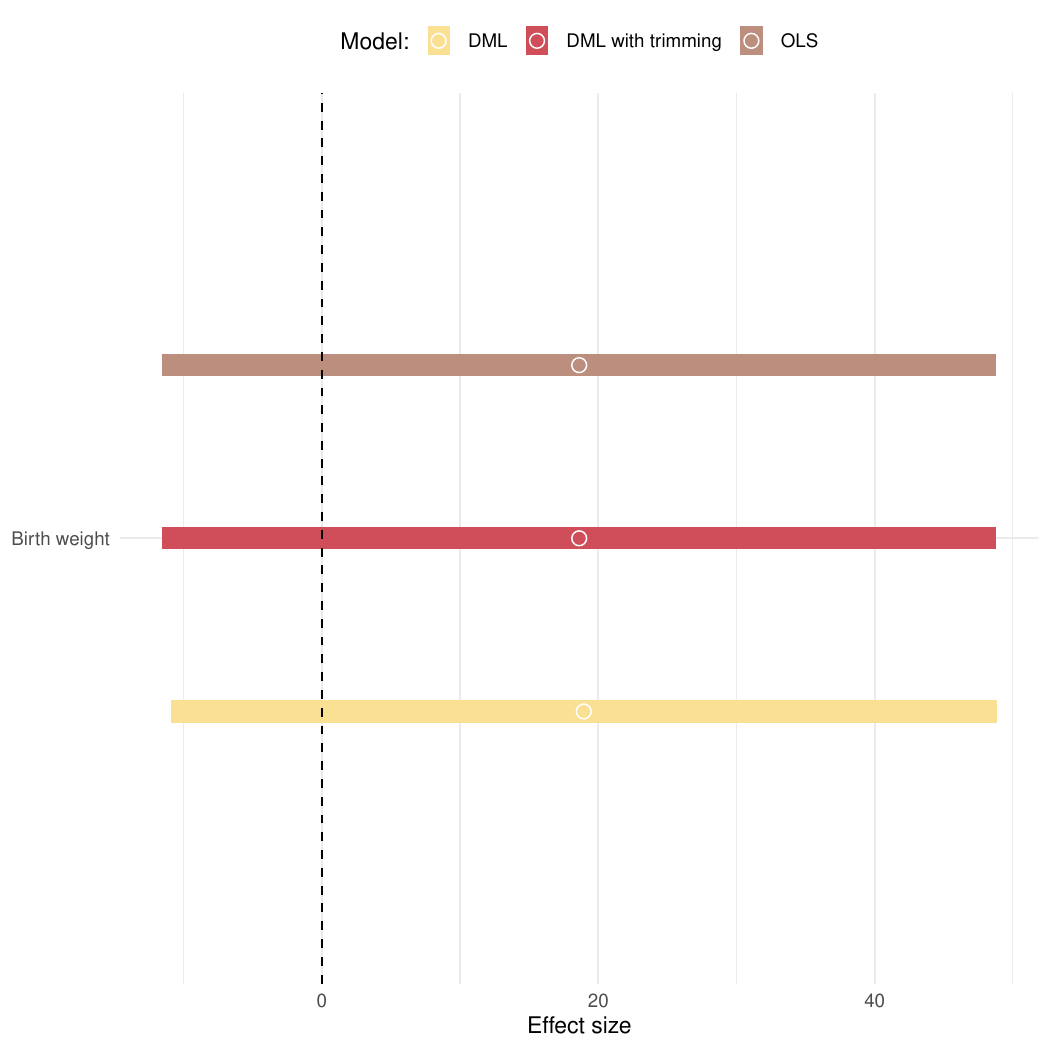}
    \caption{Birth weight ATE estimates across models}
    \label{fig:forest_bweight}
\end{figure}

Table \ref{tab:tab:dml_forest_ate_anzard_12}
reports the baseline DML treatment effect estimates, while
Table \ref{tab:tab:dml_forest_ate_anzard_12_trim_15}
reports the DML estimates after trimming the propensity score at 0.15
and 0.85.
Not only the baseline DML estimates are robust to trimming,
they are also virtually identical to the OLS estimates
(Table \ref{tab:tab:ols}
, Column 1), as can be visualised in Figures \ref{fig:forest_bweight}
and \ref{fig:forest_nobweight}.

\begin{table}

    \caption{\label{tab:tab:dml_forest_ate_anzard_12}DML: ATE estimates of ART conception on obstetric outcomes}
    \centering
    \begin{threeparttable}
        \begin{tabular}[t]{lrrrrrr}
            \toprule
                                  & Coef.  & s.e.   & l-CI    & r-CI   & p-value & Obs.  \\
            \midrule
            Spontaneous labour    & 0.035  & 0.012  & 0.011   & 0.058  & 0.0034  & 13101 \\
            Preterm spont. labour & 0.005  & 0.005  & -0.004  & 0.014  & 0.2453  & 13101 \\
            C-section             & -0.028 & 0.013  & -0.053  & -0.003 & 0.0258  & 13104 \\
            Gestational age       & -0.038 & 0.062  & -0.160  & 0.084  & 0.5413  & 13104 \\
            Pre-term birth        & 0.007  & 0.007  & -0.008  & 0.021  & 0.3608  & 13104 \\
            APGAR1                & 0.028  & 0.041  & -0.051  & 0.108  & 0.4865  & 13085 \\
            APGAR5                & 0.035  & 0.030  & -0.023  & 0.093  & 0.2404  & 13087 \\
            Birth weight          & 18.953 & 15.248 & -10.932 & 48.838 & 0.2139  & 13089 \\
            \bottomrule
        \end{tabular}
        \begin{tablenotes}
            \item \textit{Notes}: This tables reports our double machine learning estimates for the average
            treatment effect (ATE) of ART on obstetric outcomes. For each outcome (Column 1), we report the point estimate of the ATE (Column 2), its standard error (Column 3), its left and right 95\% confidence interval extrema (Columns
            4 and 5), its p-value (Column 6) and the number of observations (Column 7).
        \end{tablenotes}
    \end{threeparttable}
\end{table}

\begin{table}

    \caption{\label{tab:tab:dml_forest_ate_anzard_12_trim_15}DML: ATE estimates of ART conception on obstetric outcomes with trimming at 0.15}
    \centering
    \begin{threeparttable}
        \begin{tabular}[t]{lrrrrrr}
            \toprule
                                  & Coef.  & s.e.   & l-CI    & r-CI   & p-value & Obs.  \\
            \midrule
            Spontaneous labour    & 0.034  & 0.012  & 0.010   & 0.057  & 0.0050  & 13101 \\
            Preterm spont. labour & 0.005  & 0.005  & -0.004  & 0.014  & 0.2983  & 13101 \\
            C-section             & -0.026 & 0.013  & -0.051  & -0.001 & 0.0418  & 13104 \\
            Gestational age       & -0.030 & 0.063  & -0.153  & 0.093  & 0.6294  & 13104 \\
            Pre-term birth        & 0.007  & 0.007  & -0.008  & 0.021  & 0.3622  & 13104 \\
            APGAR1                & 0.036  & 0.041  & -0.046  & 0.117  & 0.3899  & 13085 \\
            APGAR5                & 0.035  & 0.030  & -0.024  & 0.093  & 0.2488  & 13087 \\
            Birth weight          & 18.621 & 15.402 & -11.567 & 48.809 & 0.2267  & 13089 \\
            \bottomrule
        \end{tabular}
        \begin{tablenotes}
            \item \textit{Notes}: This tables reports our double machine learning estimates for the average
            treatment effect (ATE) of ART on obstetric outcomes, after trimming births with propensity scores
            under 0.15 and above 0.85.
            For each outcome (Column 1), we report the point estimate of the ATE (Column 2), its standard error (Column 3), its left and right 95\% confidence interval extrema (Columns
            4 and 5), its p-value (Column 6) and the number of observations (Column 7).
        \end{tablenotes}
    \end{threeparttable}
\end{table}


\subsection{Omitted variable bias analysis} \label{sec:ovb}
This similarity between the estimates from the linear regression and the
flexible DML models is evidence that the data-generating process is not
highly non-linear and, conversely, that it is approximately linear.
Hence, the linear regression estimates should be taken seriously.
One last concern remains, that some important confounders have not been
included in our model and would drastically change our results if
included.

\begin{figure}[htb]
    \centering
    \thispagestyle{empty}
    \begin{tikzpicture}
        \node (img1)  {\includegraphics[scale=0.45]{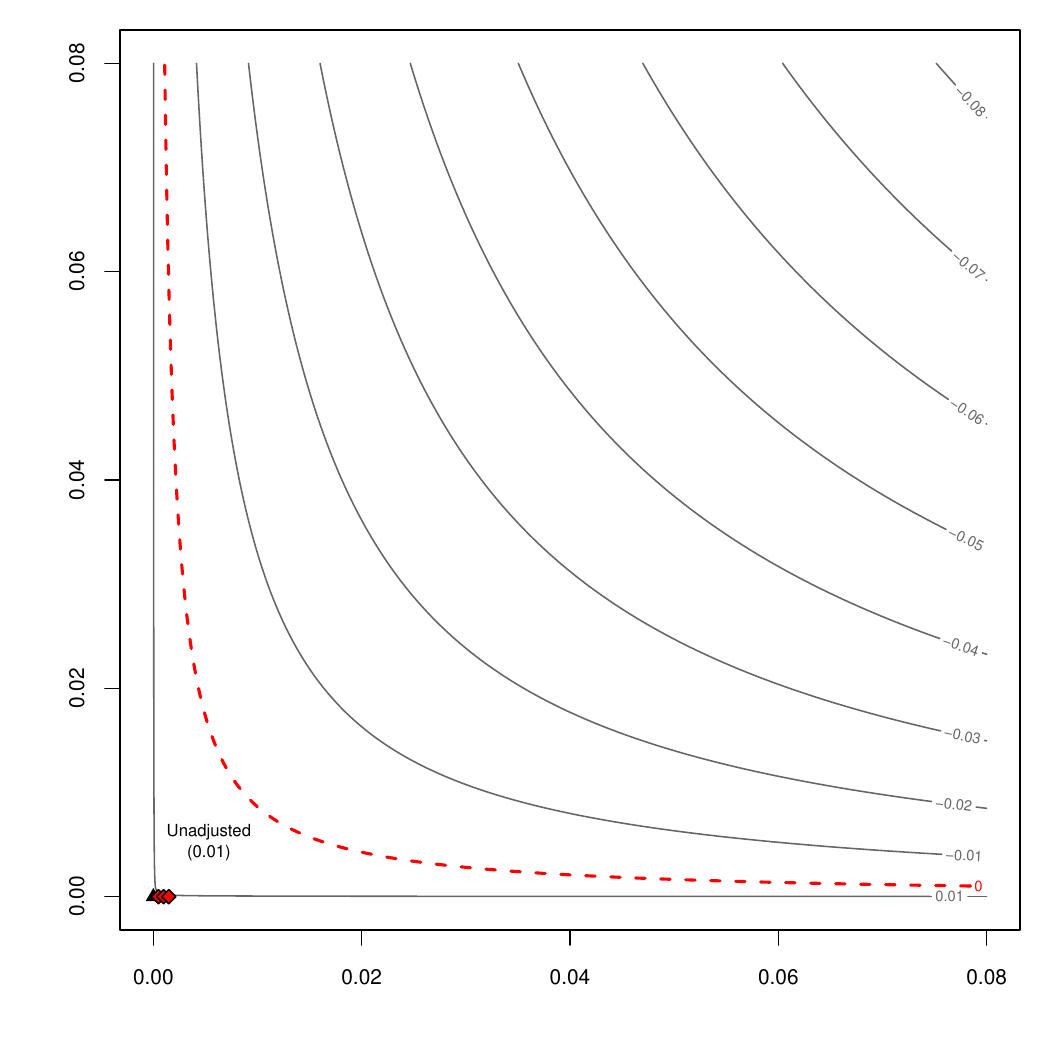}};
        \node[above=of img1, node distance=0cm, yshift=-1cm,font=\color{black}] {coefficient};
        \node[left=of img1, node distance=0cm, rotate=90, anchor=center,yshift=-1cm,font=\color{black}] {\tiny Partial $R^2$ of confounder(s) with the outcome};
        \node[left=of img1, node distance=0cm, rotate=90, anchor=center,yshift=0cm,font=\color{black}] {Maternal age at birth};
        \node[right=of img1,yshift=0cm, xshift=-1cm] (img2)  {\includegraphics[scale=0.45]{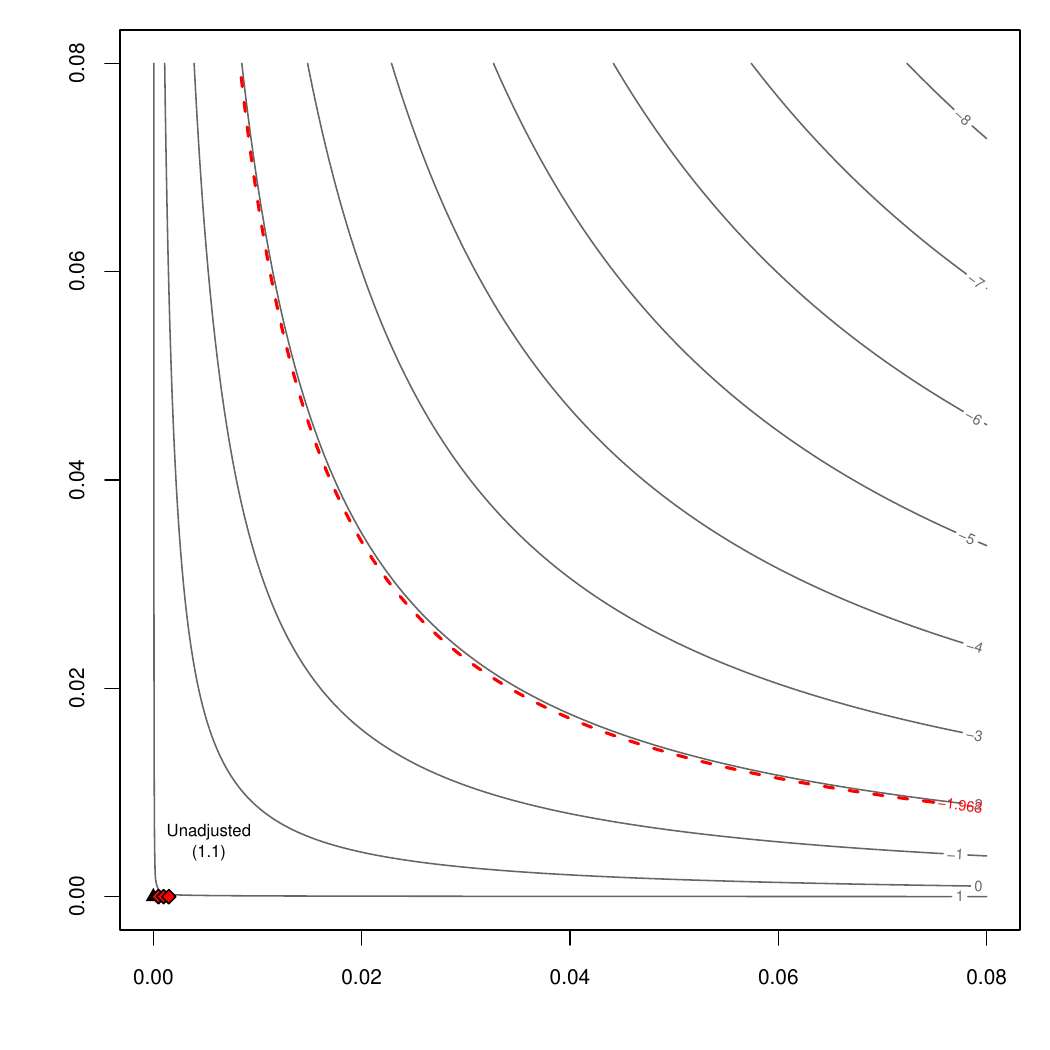}};
        \node[above=of img2, node distance=0cm, yshift=-1cm,font=\color{black}] {t-value};
        \node[below=of img1,yshift=1.5cm] (img3)  {\includegraphics[scale=0.45]{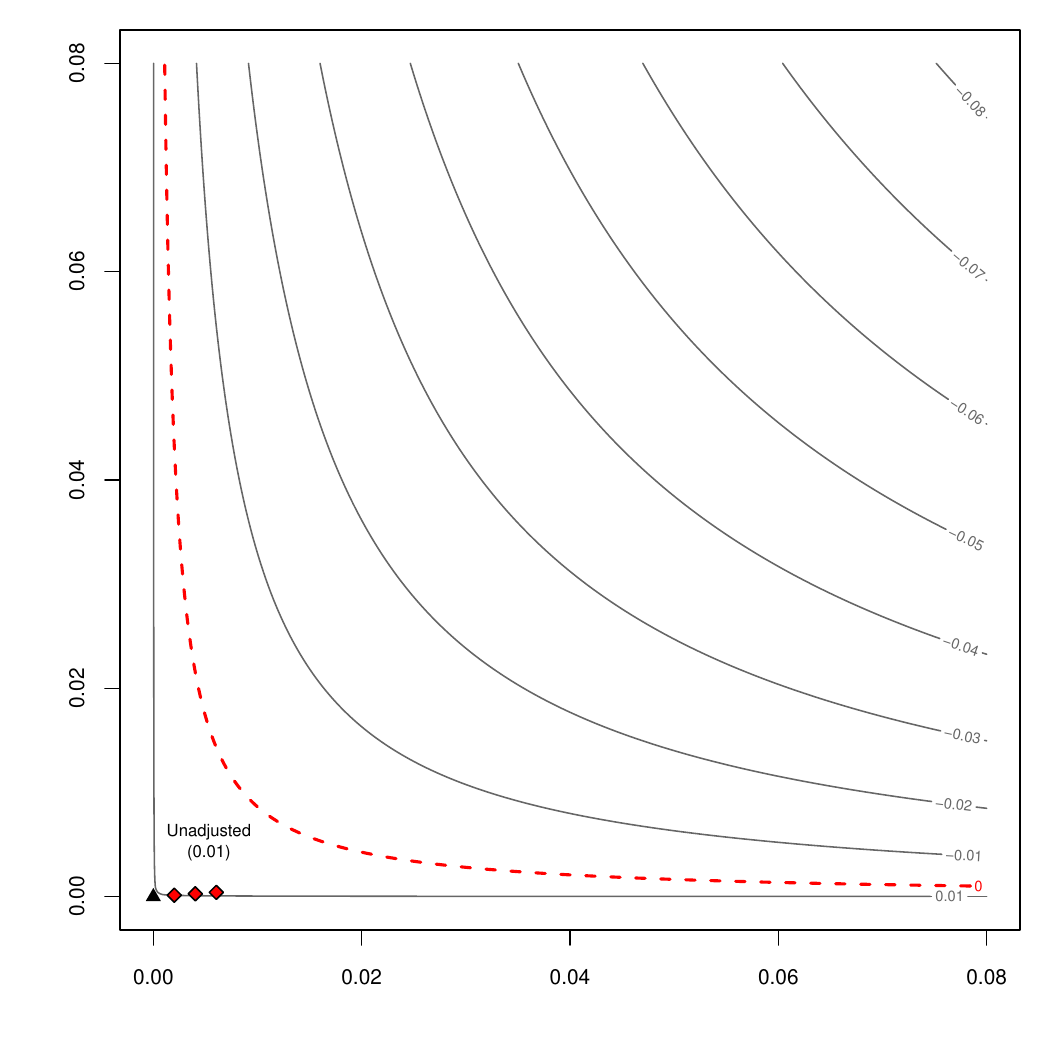}};
        \node[left=of img3, node distance=0cm, rotate=90, anchor=center,yshift=-1cm,font=\color{black}] {\tiny Partial $R^2$ of confounder(s) with the outcome};
        \node[left=of img3, node distance=0cm, rotate=90, anchor=center,yshift=0cm,font=\color{black}] {No. of past pregnancies};
        \node[below=of img2,yshift=1.5cm] (img4)  {\includegraphics[scale=0.45]{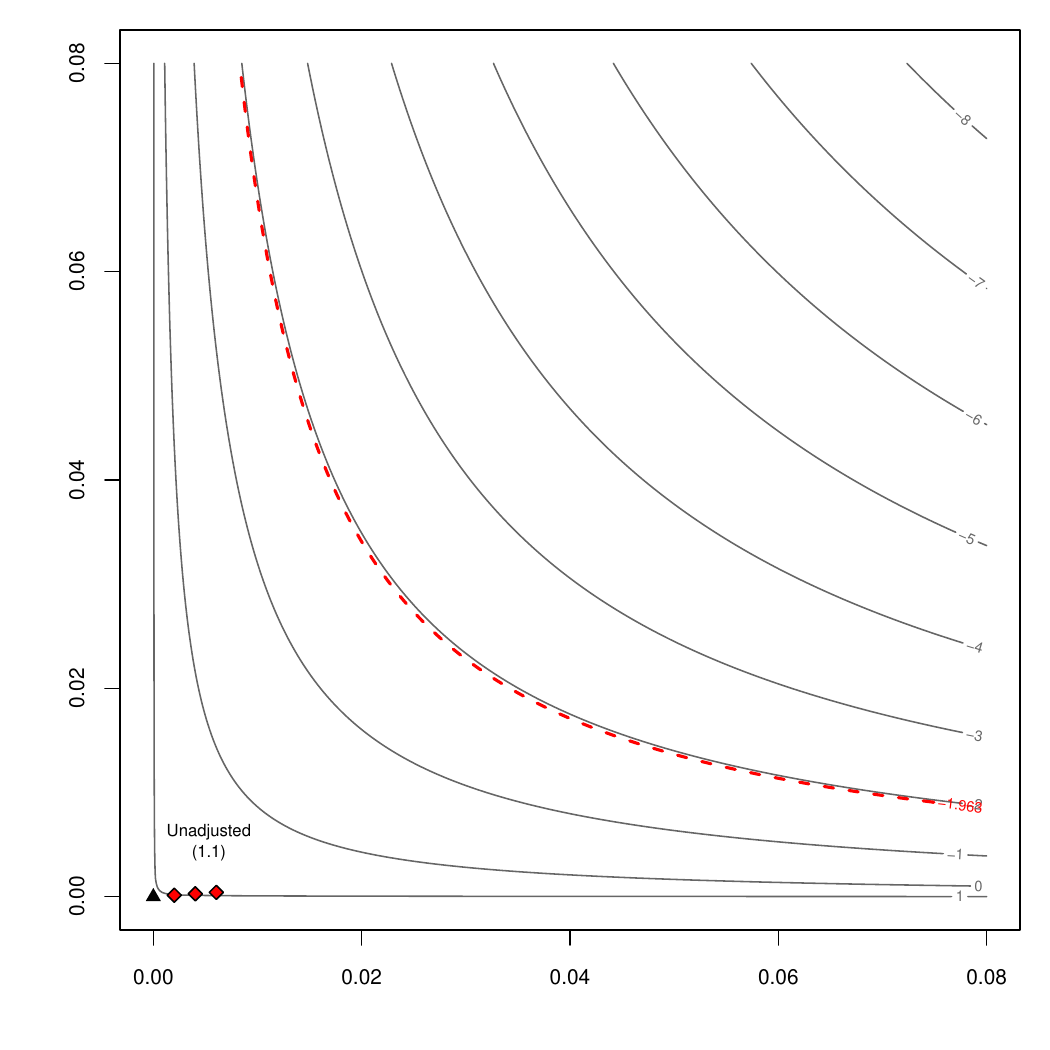}};
        \node[below=of img3,yshift=1.5cm] (img5)  {\includegraphics[scale=0.45]{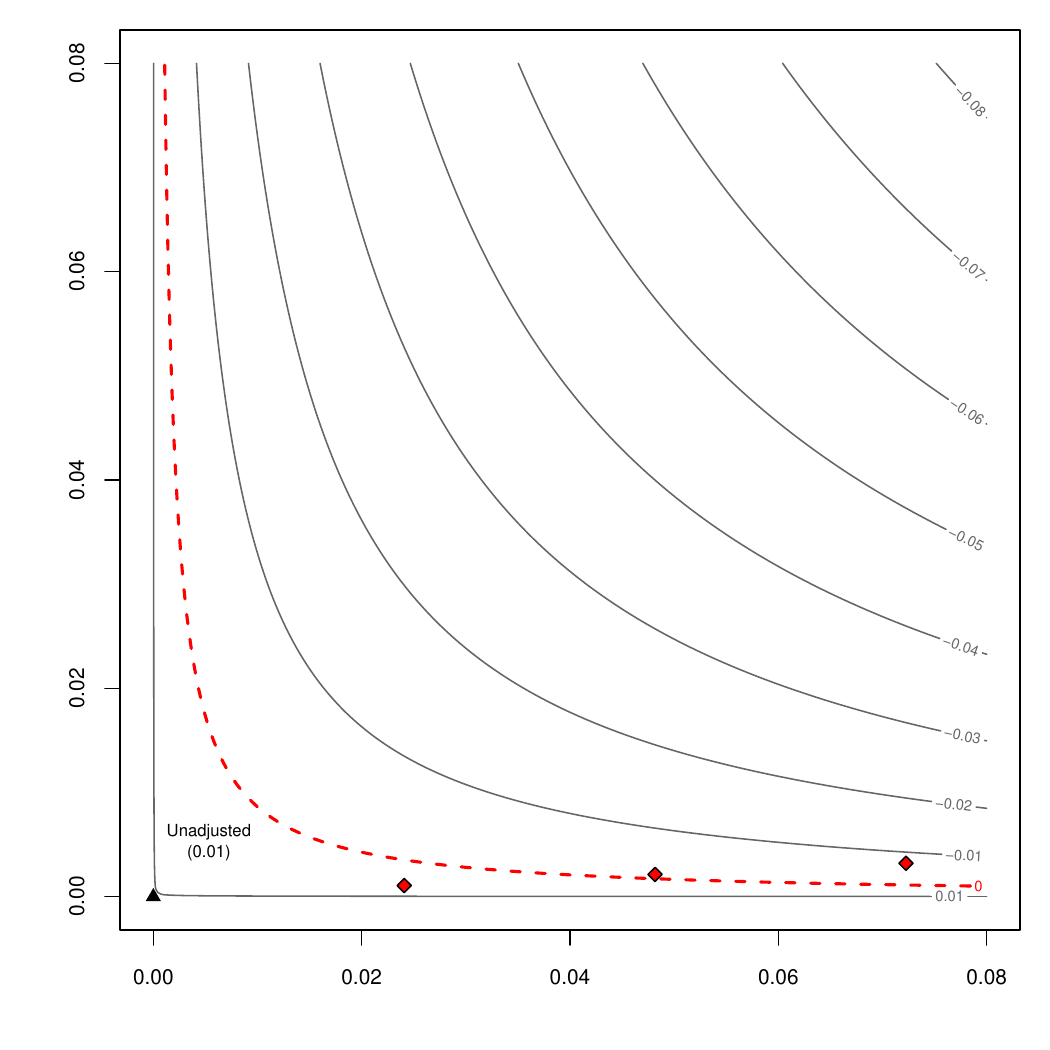}};
        \node[below=of img5, node distance=0cm, yshift=1.3cm,font=\color{black}] {\tiny Partial $R^2$ of confounder(s) with the treatment};
        \node[left=of img5, node distance=0cm, rotate=90, anchor=center,yshift=-1cm,font=\color{black}] {\tiny Partial $R^2$ of confounder(s) with the outcome};
        \node[left=of img5, node distance=0cm, rotate=90, anchor=center,yshift=0cm,font=\color{black}] {No. of past failed cycles};
        \node[below=of img4,yshift=1.5cm] (img6)  {\includegraphics[scale=0.45]{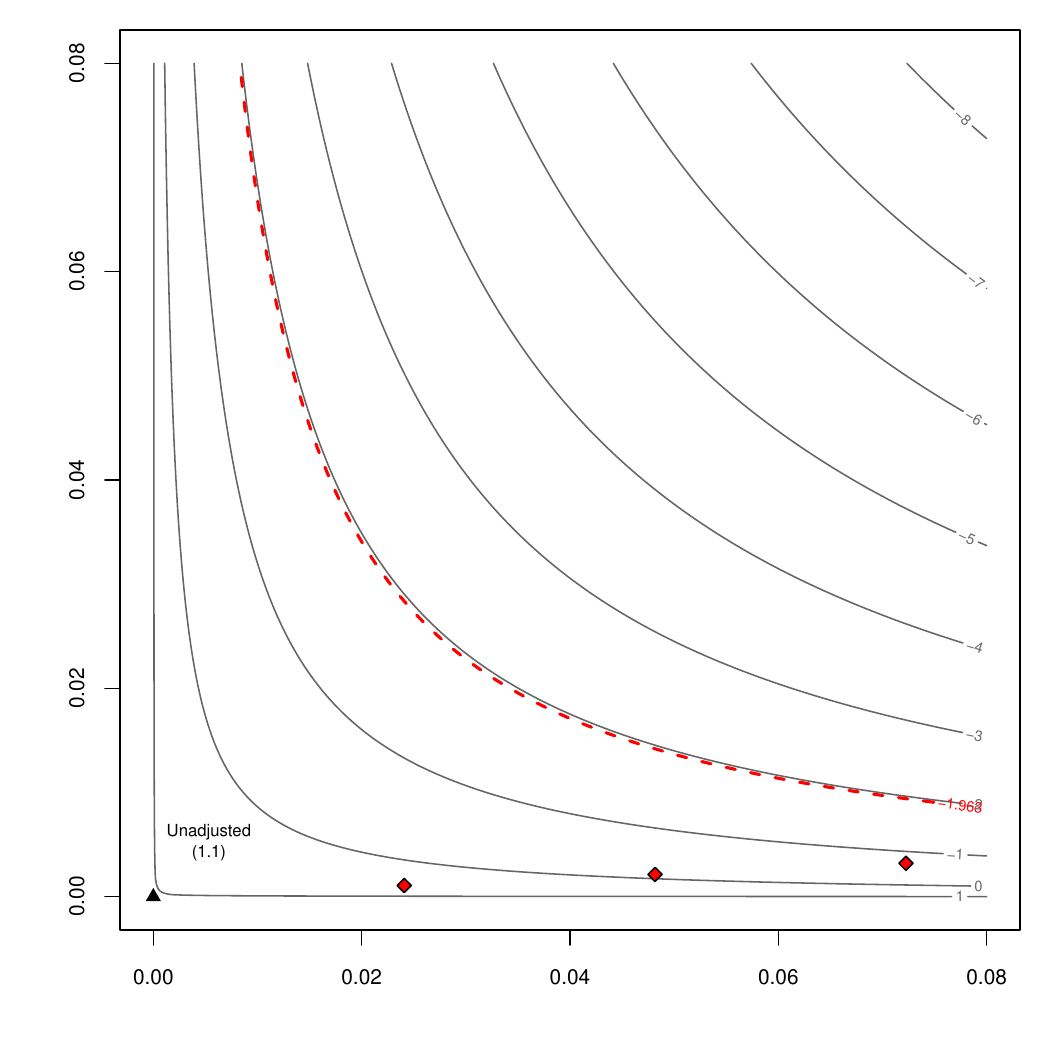}};
        \node[below=of img6, node distance=0cm, yshift=1.3cm,font=\color{black}] {\tiny Partial $R^2$ of confounder(s) with the treatment};
    \end{tikzpicture}
    \caption{Sensitivity of pre-term-birth ATE estimate to unobserved confounding}
    \label{fig:sensemakr_plot_gestage_pre}
\end{figure}

We study this possibility graphically using the tools developed in
\cite{Cinelli2019}.
In Figure \ref{fig:sensemakr_plot_gestage_pre}, we show what would happen in
to our baseline linear regression estimate (left column)
and t-static (right column) if we were not controlling for
one or more confounders once/twice/thrice as strong as maternal age (top row),
number of past pregnancies (middle row), or number of past cycles (bottom).
While the estimated coefficient (t-value) is depicted as a black triangle,
the hypothetical coefficients (t-values) under the x1, x2 and x3 scenarios
are represented by red rhombi, progressively apart from the triangle.

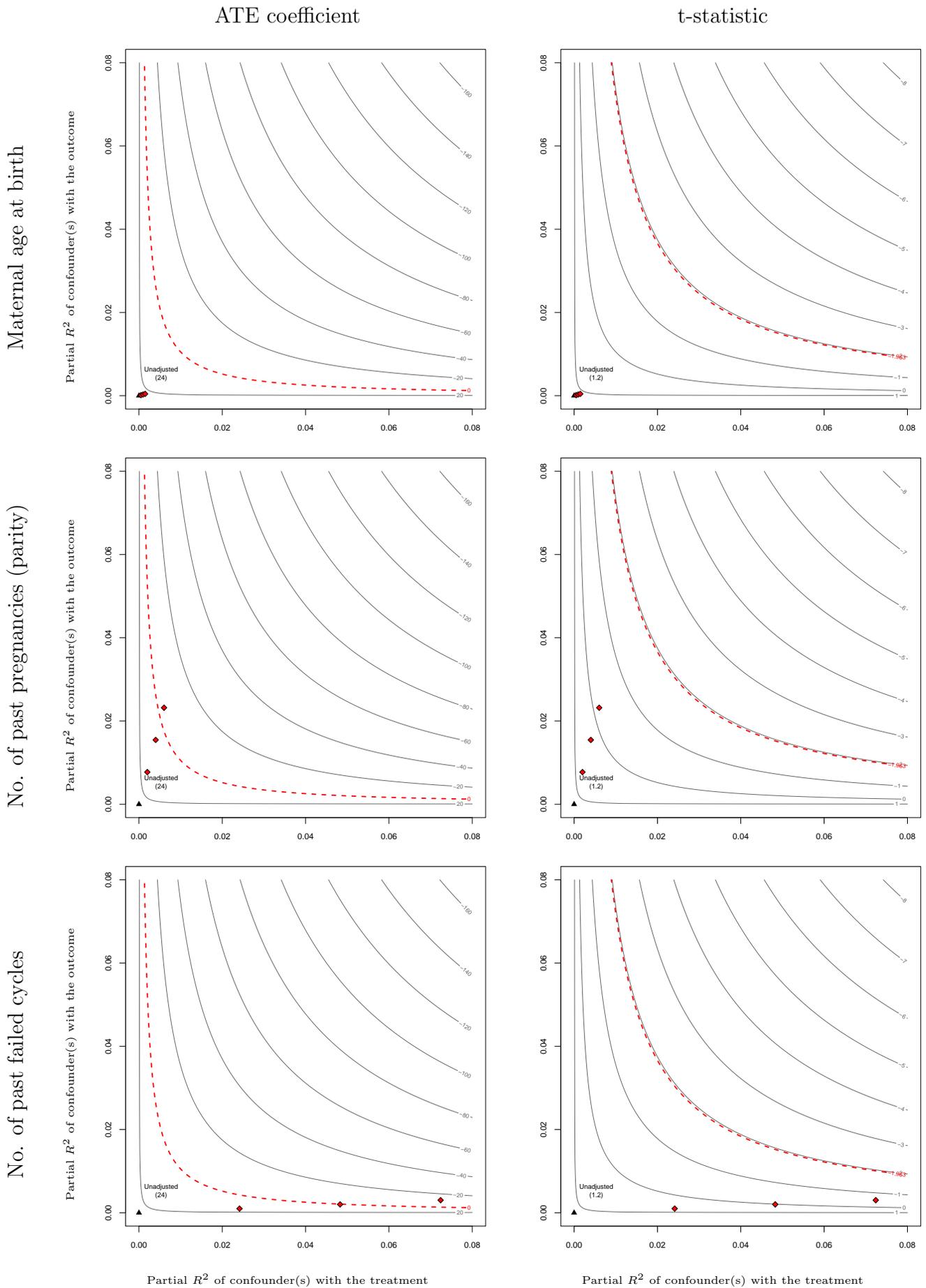
\begin{figure}[htb]
    \centering
    \thispagestyle{empty}
    \begin{tikzpicture}
        \node (img1)  {\includegraphics[scale=0.45]{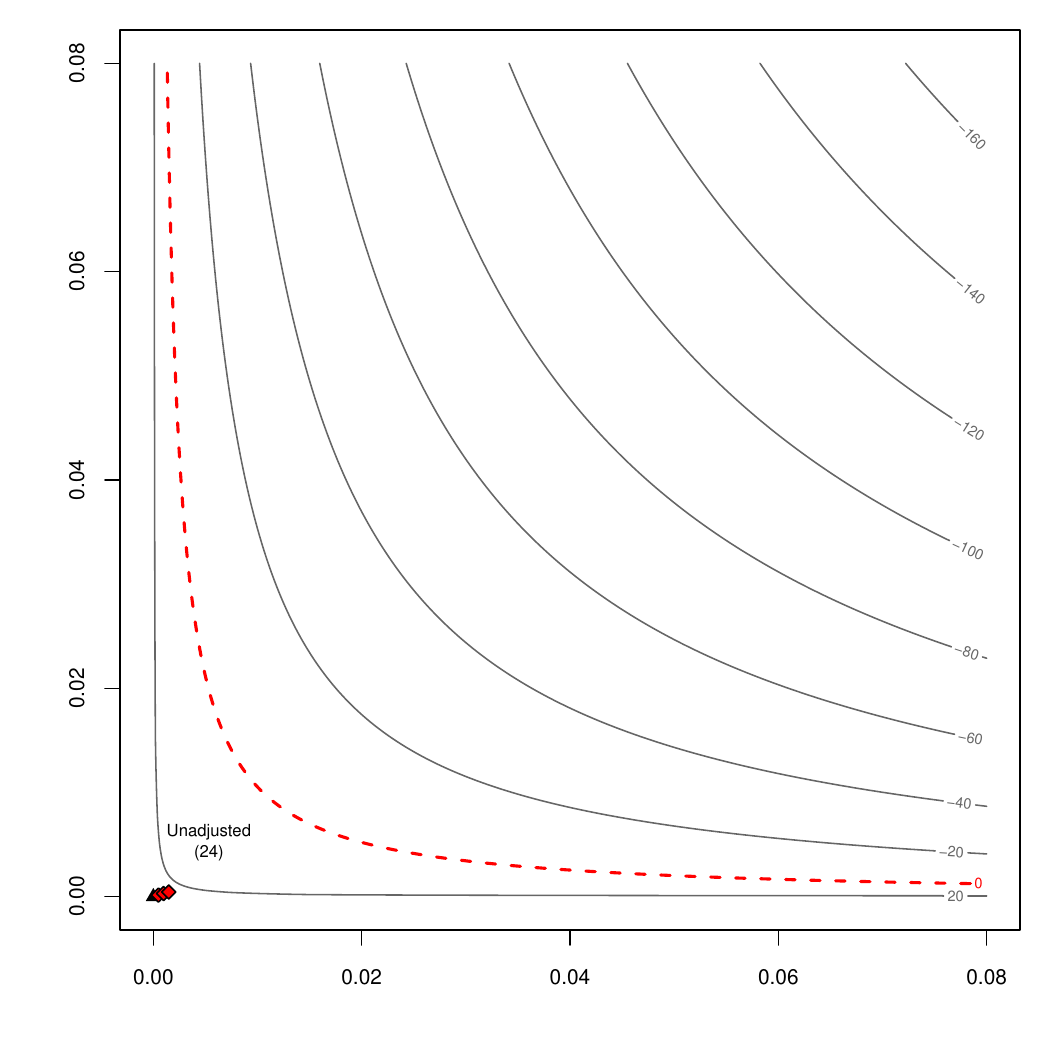}};
        \node[above=of img1, node distance=0cm, yshift=-1cm,font=\color{black}] {ATE coefficient};
        \node[left=of img1, node distance=0cm, rotate=90, anchor=center,yshift=-1cm,font=\color{black}] {\tiny Partial $R^2$ of confounder(s) with the outcome};
        \node[left=of img1, node distance=0cm, rotate=90, anchor=center,yshift=0cm,font=\color{black}] {Maternal age at birth};
        \node[right=of img1,yshift=0cm, xshift=-1cm] (img2)  {\includegraphics[scale=0.45]{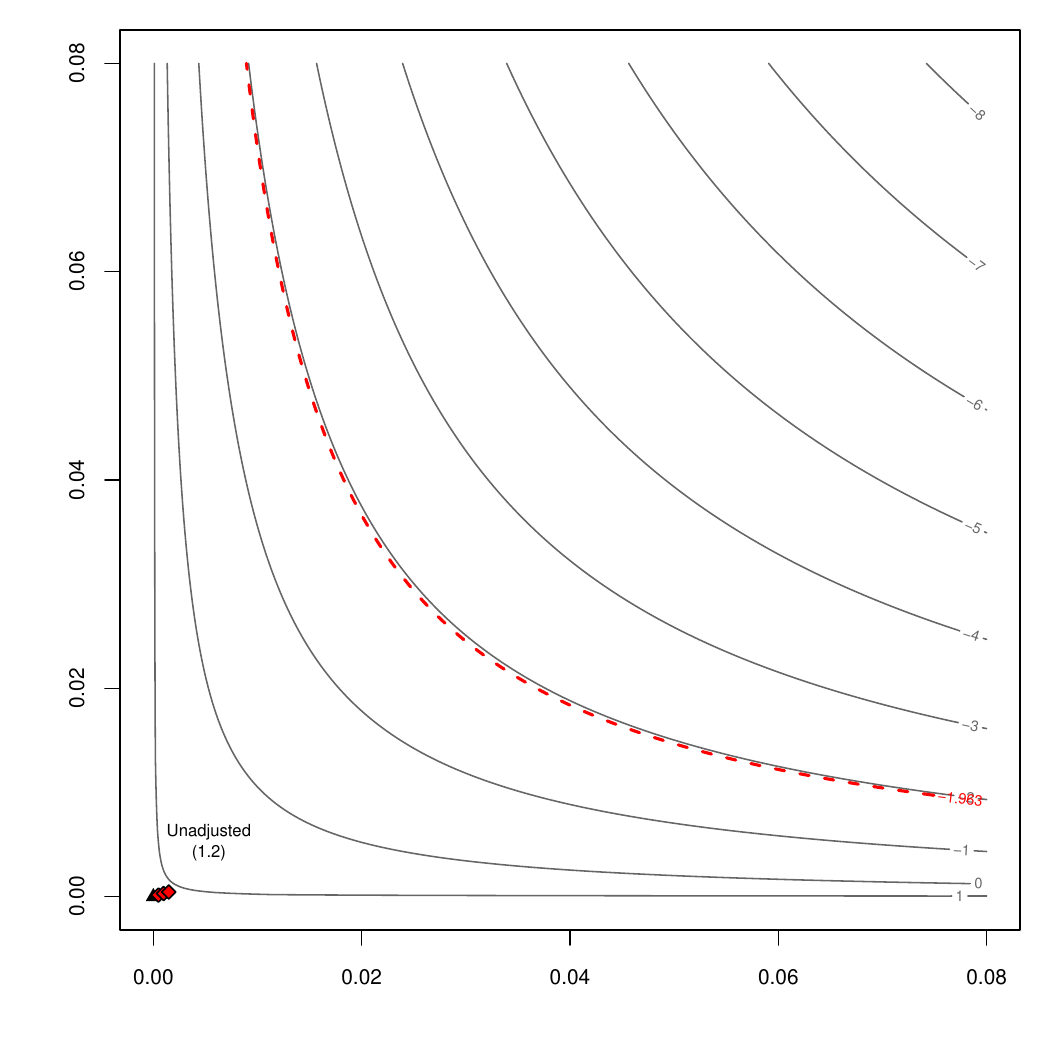}};
        \node[above=of img2, node distance=0cm, yshift=-1cm,font=\color{black}] {t-statistic};
        \node[below=of img1,yshift=1.5cm] (img3)  {\includegraphics[scale=0.45]{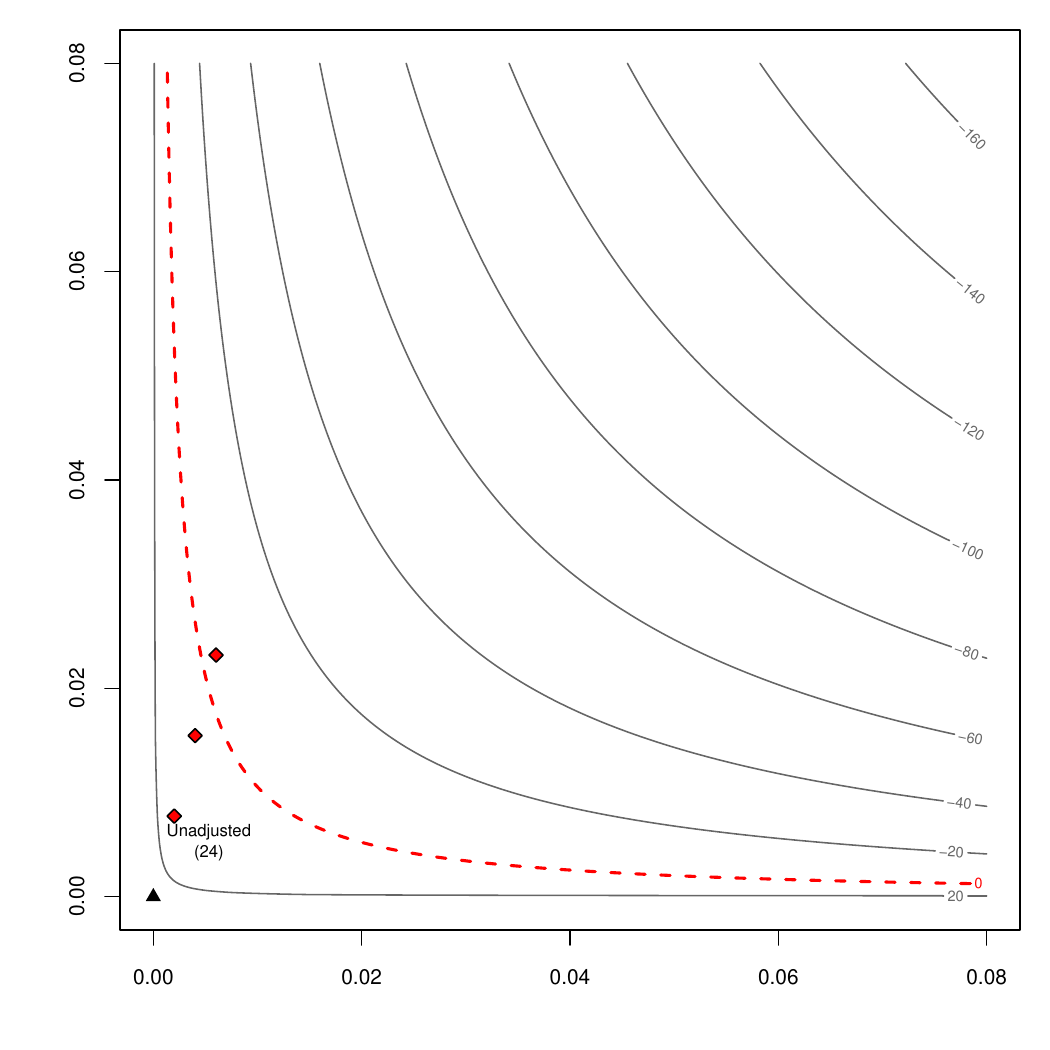}};
        \node[left=of img3, node distance=0cm, rotate=90, anchor=center,yshift=-1cm,font=\color{black}] {\tiny Partial $R^2$ of confounder(s) with the outcome};
        \node[left=of img3, node distance=0cm, rotate=90, anchor=center,yshift=0cm,font=\color{black}] {No. of past pregnancies (parity)};
        \node[below=of img2,yshift=1.5cm] (img4)  {\includegraphics[scale=0.45]{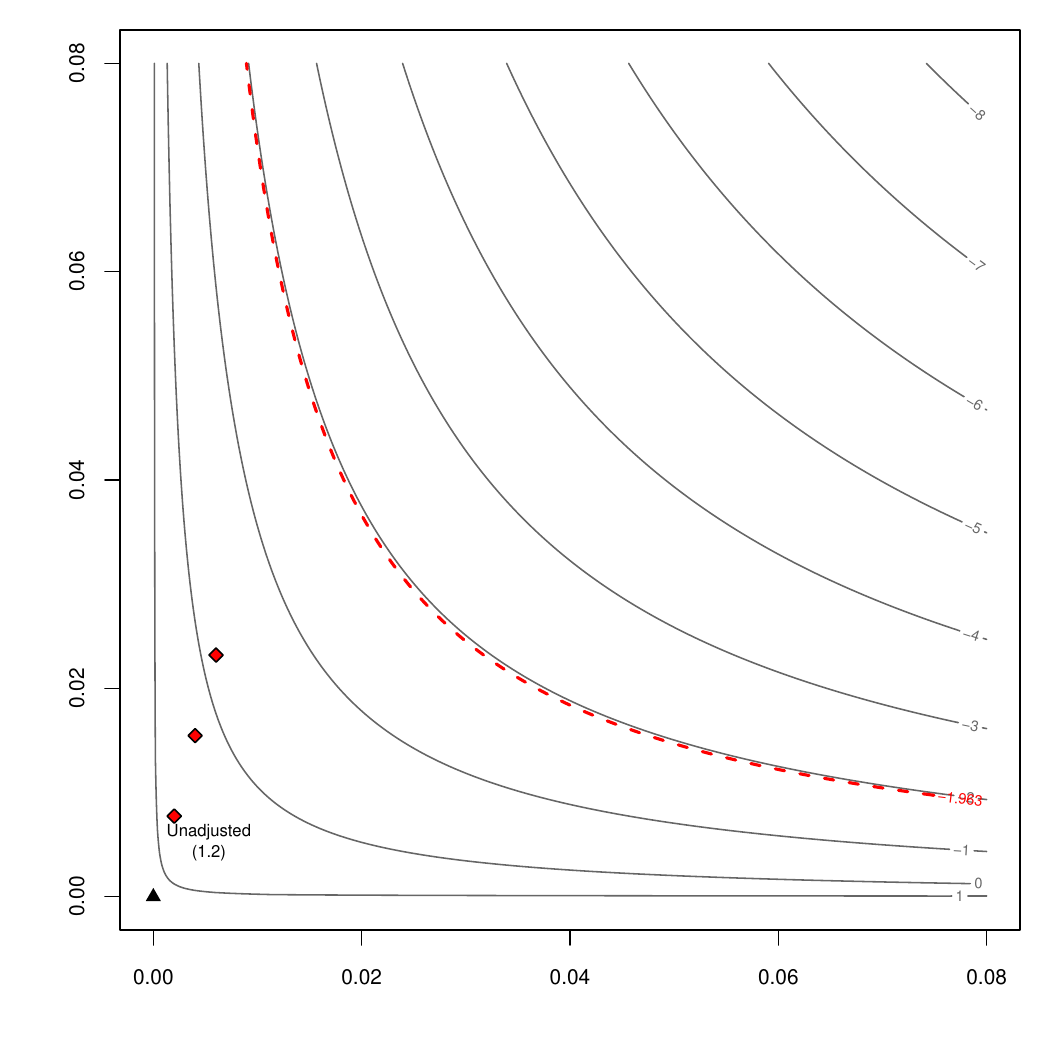}};
        \node[below=of img3,yshift=1.5cm] (img5)  {\includegraphics[scale=0.45]{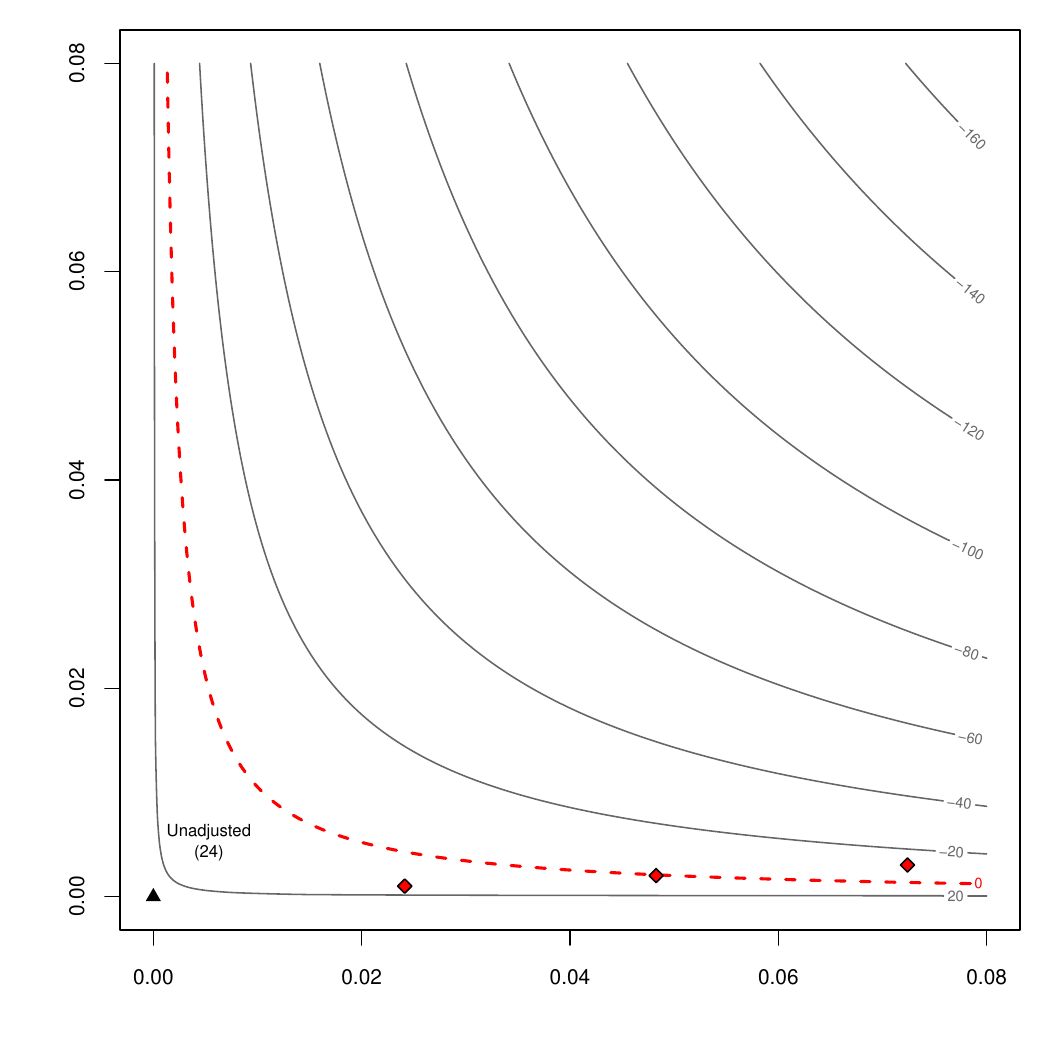}};
        \node[below=of img5, node distance=0cm, yshift=1.3cm,font=\color{black}] {\tiny Partial $R^2$ of confounder(s) with the treatment};
        \node[left=of img5, node distance=0cm, rotate=90, anchor=center,yshift=-1cm,font=\color{black}] {\tiny Partial $R^2$ of confounder(s) with the outcome};
        \node[left=of img5, node distance=0cm, rotate=90, anchor=center,yshift=0cm,font=\color{black}] {No. of past failed cycles};
        \node[below=of img4,yshift=1.5cm] (img6)  {\includegraphics[scale=0.45]{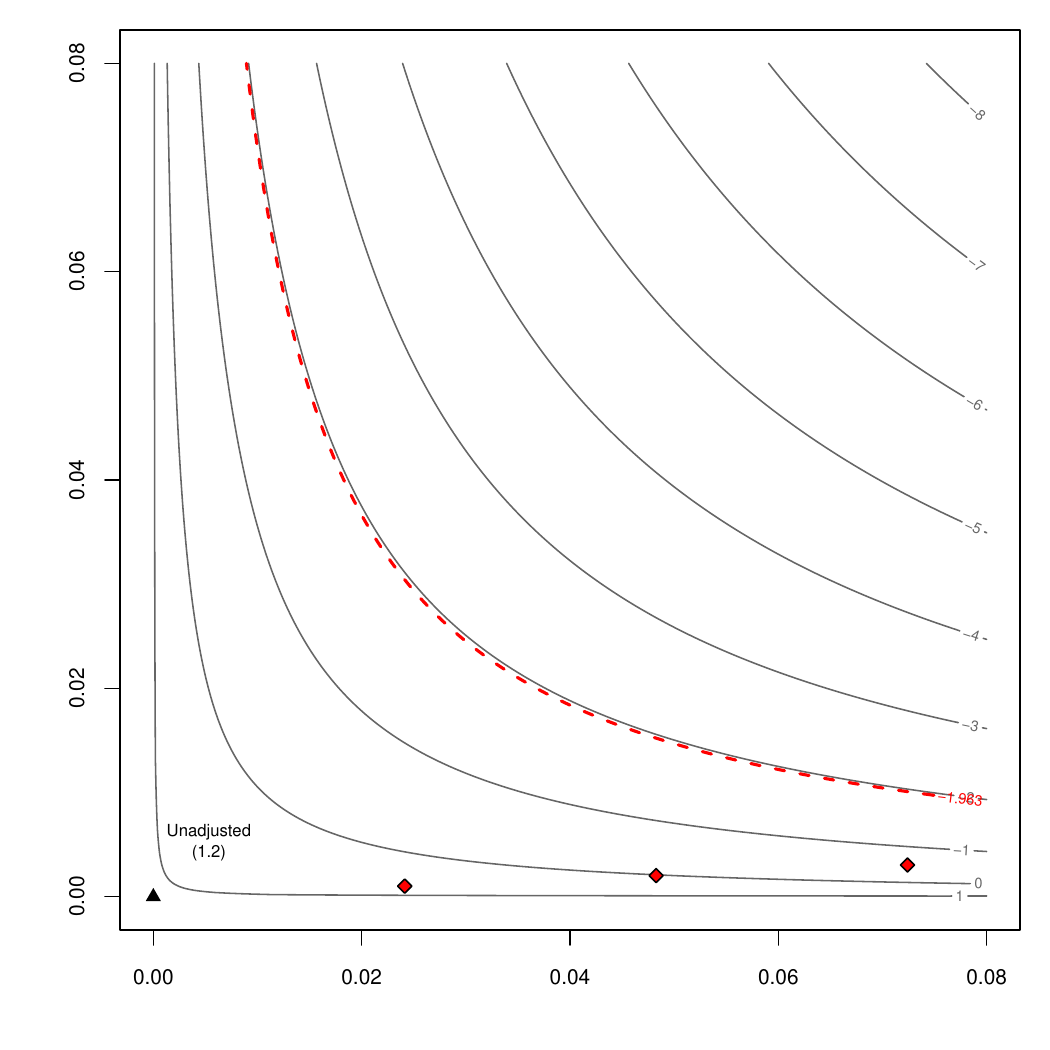}};
        \node[below=of img6, node distance=0cm, yshift=1.3cm,font=\color{black}] {\tiny Partial $R^2$ of confounder(s) with the treatment};
    \end{tikzpicture}
    \caption{Sensitivity of birth-weight ATE estimate to unobserved confounding}
    \label{fig:sensemakr_plot_bweight}
\end{figure}

In a coefficient plot, a rhombus crossing the red line (at zero) means that
the coefficient would change sign under that scenario.
We see this for birth weight (Figure \ref{fig:sensemakr_plot_bweight}),
in presence of an unobserved confounder three times as strong
as the \textit{parity} variable. In this case though, the rhombus is
almost on the zero line and hence treatment effect would be
virtually equal to zero.

In a t-statistic plot, a rhombus crossing the red line (at 1.96) means that
either that (i) a statistically insignificant coefficient (at the 5\% level)
would become significant in that scenario or (ii), vice versa, that
a statistically insignificant coefficient would become significant.
We do not see this case, as no t-statistic associated with a statistically
insignificant baseline coefficient crosses the 1.96 line.
We see case (ii)for spontaneous labour in presence of an unobserved
confounder three times as strong as \textit{Maternal age}, which would make
the ATE coefficient insignificant while moving it closer to zero, at 3 p.p.
The plots for the remaining outcomes are included in the appendix.
The plots for the C-section outcome resemble those for \textit{Spontaneous
    labour}, while the plots for the APGAR scores and for
\textit{Gestational age at birth} resemble those for \textit{Birth weight}.

Overall, these plots do not change the interpretation of our results.
Even after using the three strongest confounders as a benchmark for unobserved
confounders, with still find evidence that the effect of ART on
obstetric outcomes is either small (\textit{Spontaneous labour},
\textit{C-section}) or precisely zero (gestational age outcomes, APGAR score
outcomes and \textit{Birth weight}).
These results, taken as a whole, are evidence that there no clinically
significant differences between ART babies and spontaneously-conceived babies,
at least in subfertile mothers.

\newgeometry{top=5mm, bottom=10mm, left=5mm, right=5mm}     

\begin{figure}[htb]
    \centering
    \thispagestyle{empty}
    \begin{tikzpicture}
        \node (img1)  {\includegraphics[scale=0.45]{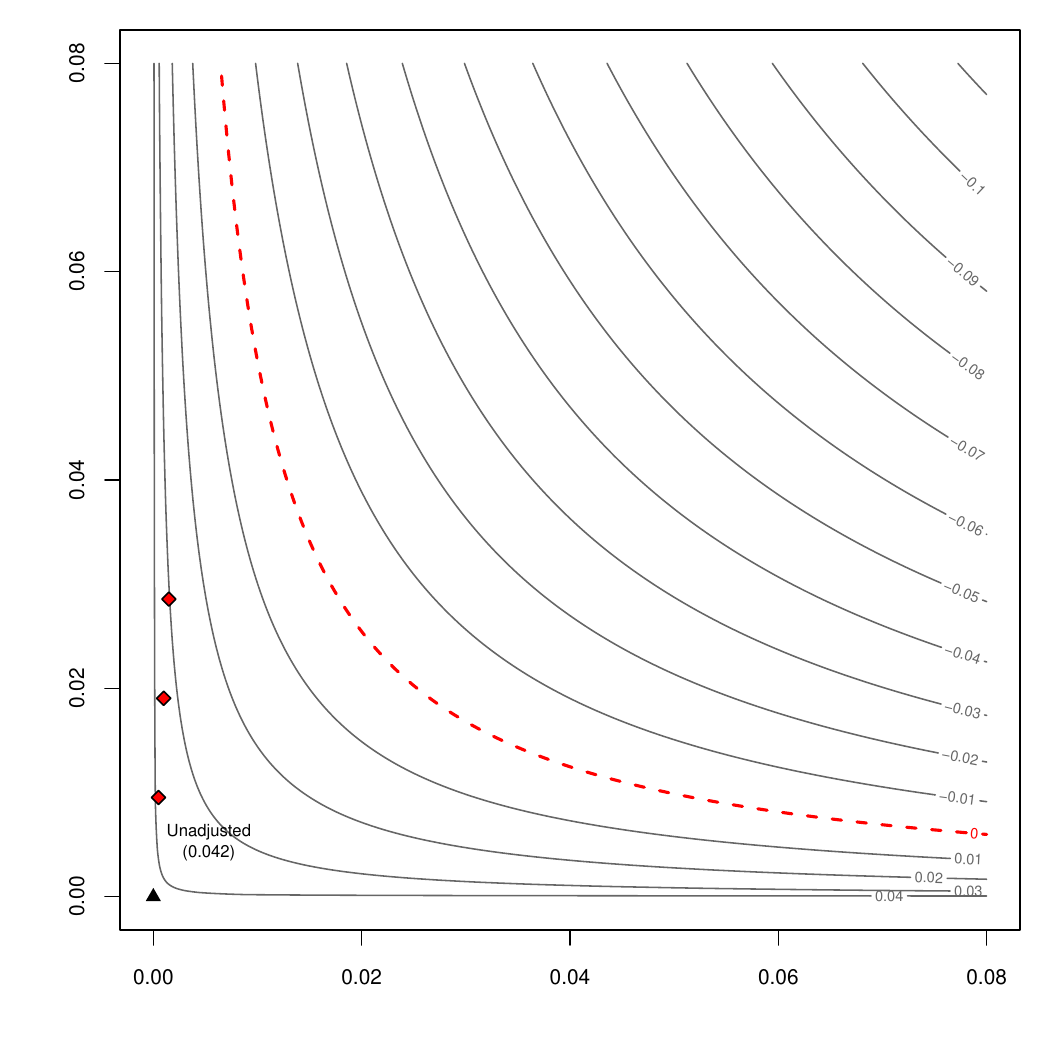}};
        \node[above=of img1, node distance=0cm, yshift=-1cm,font=\color{black}] {coefficient};
        \node[left=of img1, node distance=0cm, rotate=90, anchor=center,yshift=-1cm,font=\color{black}] {\tiny Partial $R^2$ of confounder(s) with the outcome};
        \node[left=of img1, node distance=0cm, rotate=90, anchor=center,yshift=0cm,font=\color{black}] {Maternal age at birth};
        \node[right=of img1,yshift=0cm, xshift=-1cm] (img2)  {\includegraphics[scale=0.45]{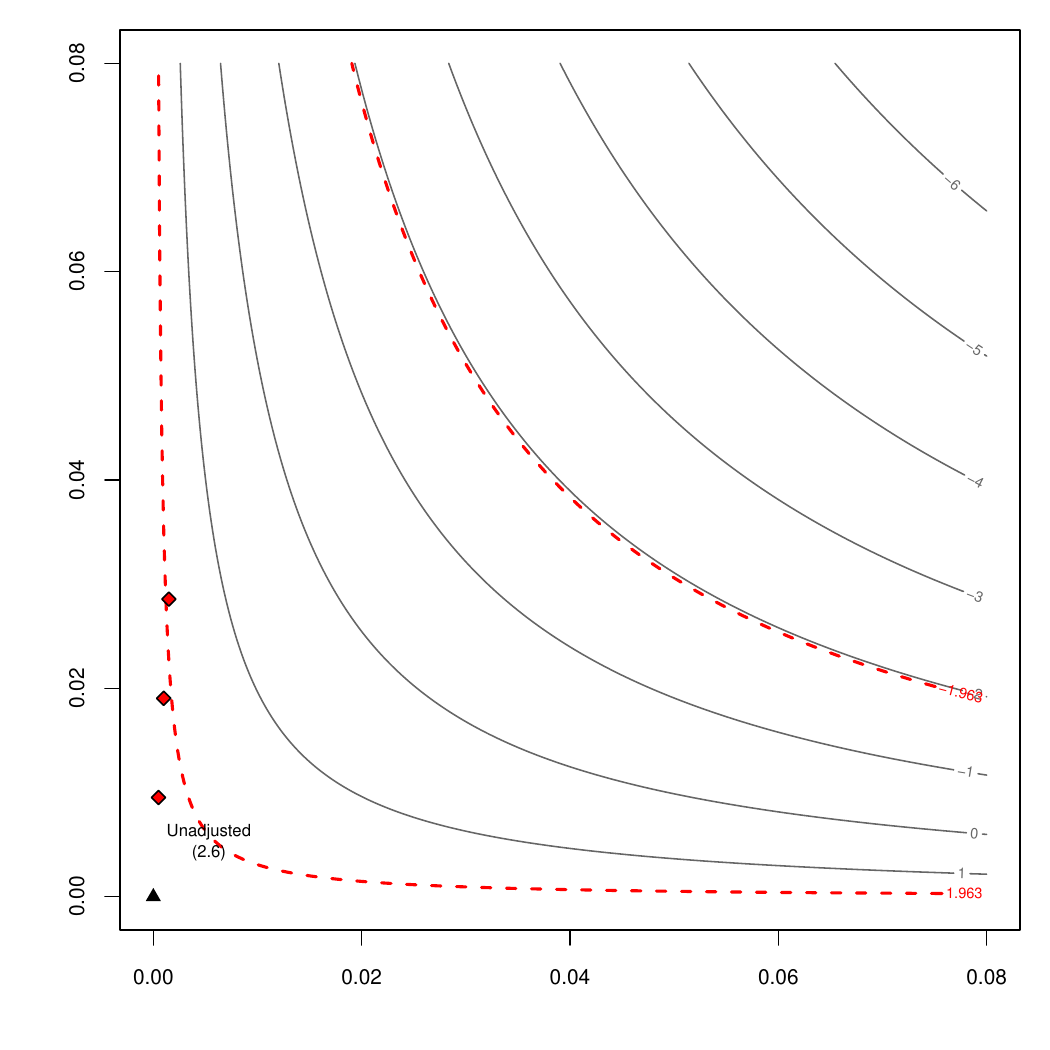}};
        \node[above=of img2, node distance=0cm, yshift=-1cm,font=\color{black}] {t-value};
        \node[below=of img1,yshift=1.5cm] (img3)  {\includegraphics[scale=0.45]{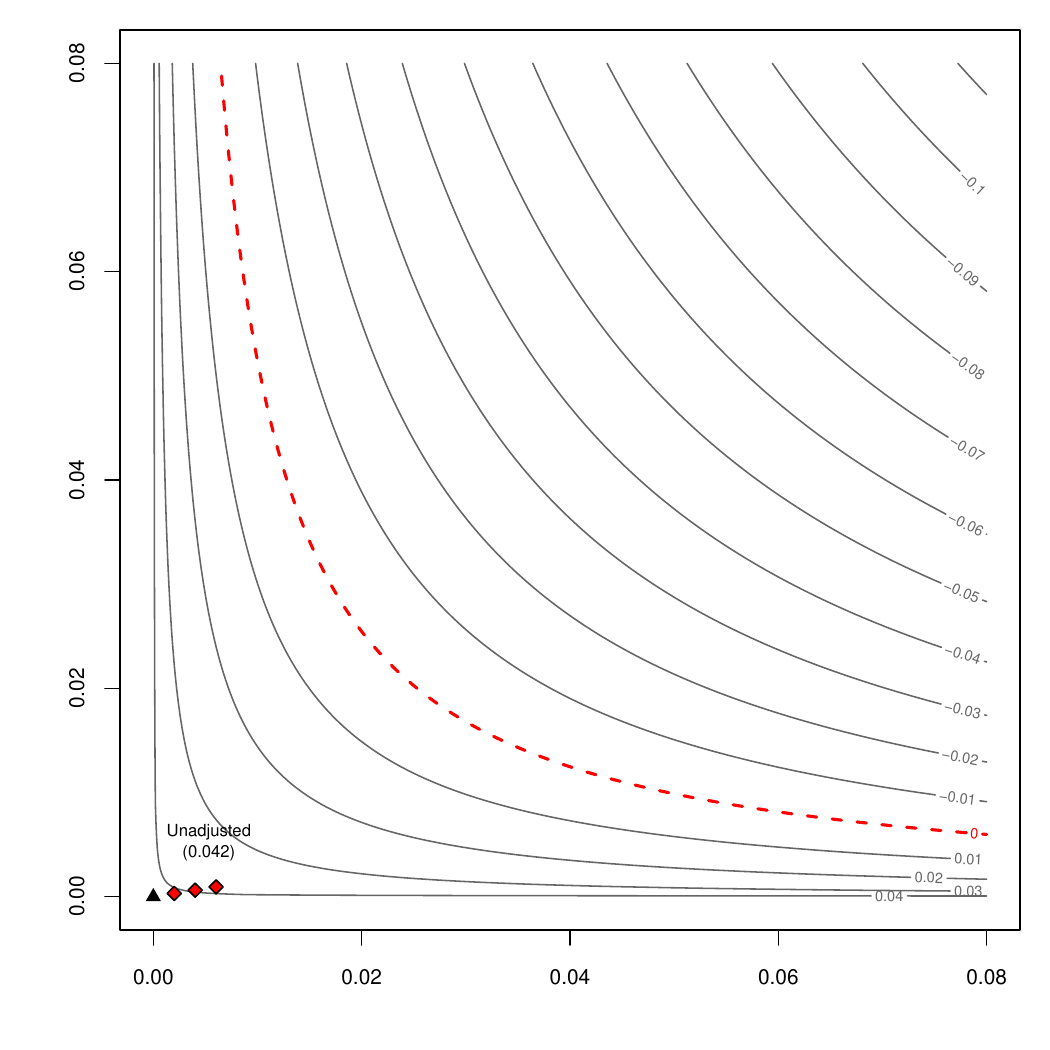}};
        \node[left=of img3, node distance=0cm, rotate=90, anchor=center,yshift=-1cm,font=\color{black}] {\tiny Partial $R^2$ of confounder(s) with the outcome};
        \node[left=of img3, node distance=0cm, rotate=90, anchor=center,yshift=0cm,font=\color{black}] {No. of past pregnancies};
        \node[below=of img2,yshift=1.5cm] (img4)  {\includegraphics[scale=0.45]{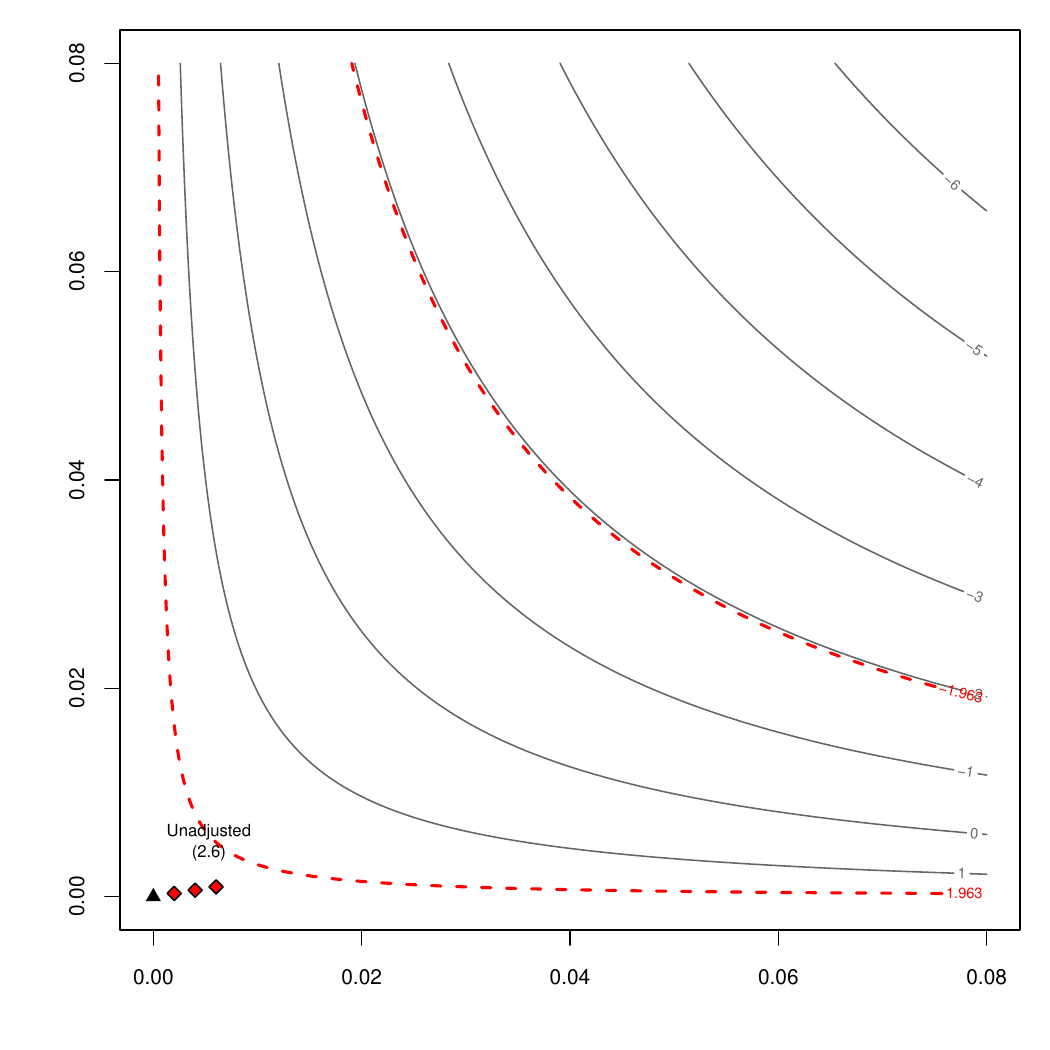}};
        \node[below=of img3,yshift=1.5cm] (img5)  {\includegraphics[scale=0.45]{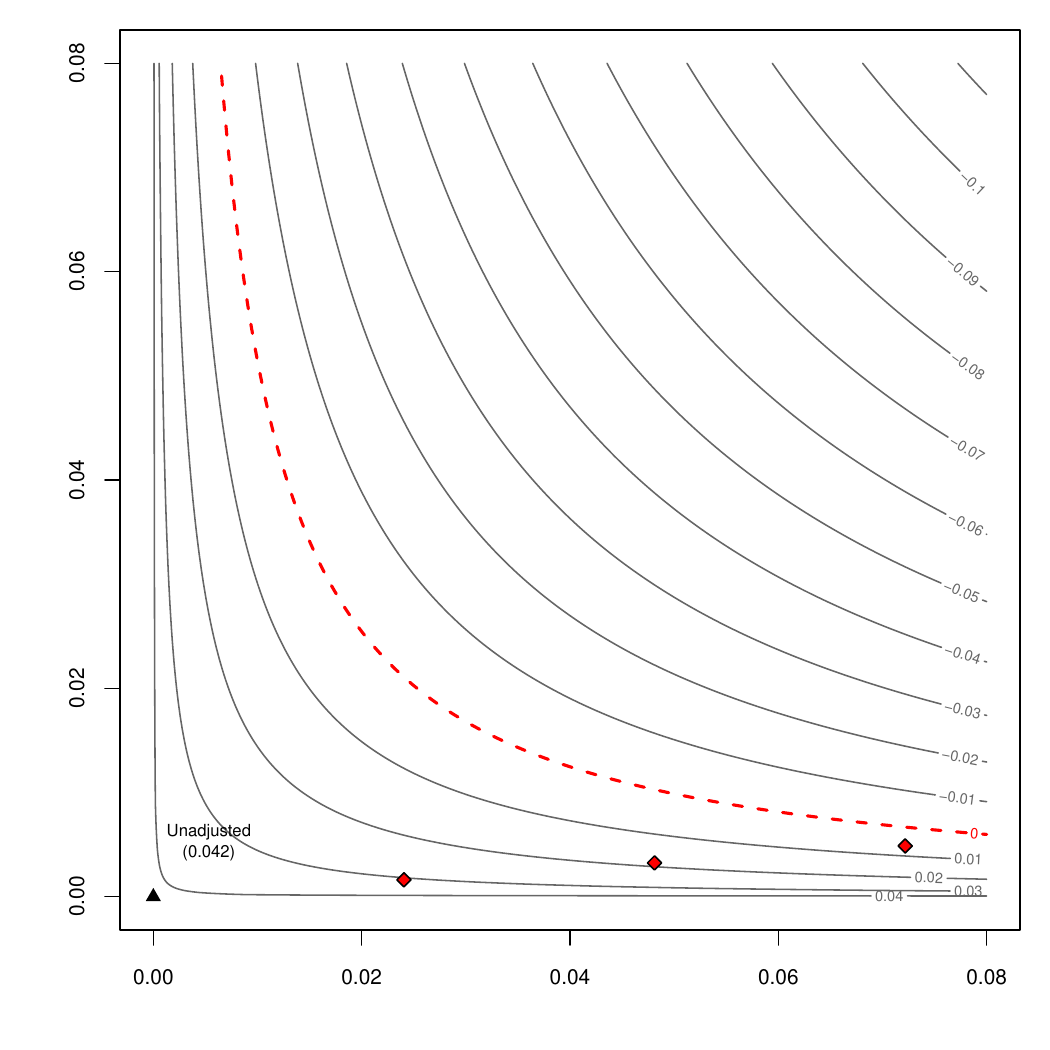}};
        \node[below=of img5, node distance=0cm, yshift=1.3cm,font=\color{black}] {\tiny Partial $R^2$ of confounder(s) with the treatment};
        \node[left=of img5, node distance=0cm, rotate=90, anchor=center,yshift=-1cm,font=\color{black}] {\tiny Partial $R^2$ of confounder(s) with the outcome};
        \node[left=of img5, node distance=0cm, rotate=90, anchor=center,yshift=0cm,font=\color{black}] {No. of past failed cycles};
        \node[below=of img4,yshift=1.5cm] (img6)  {\includegraphics[scale=0.45]{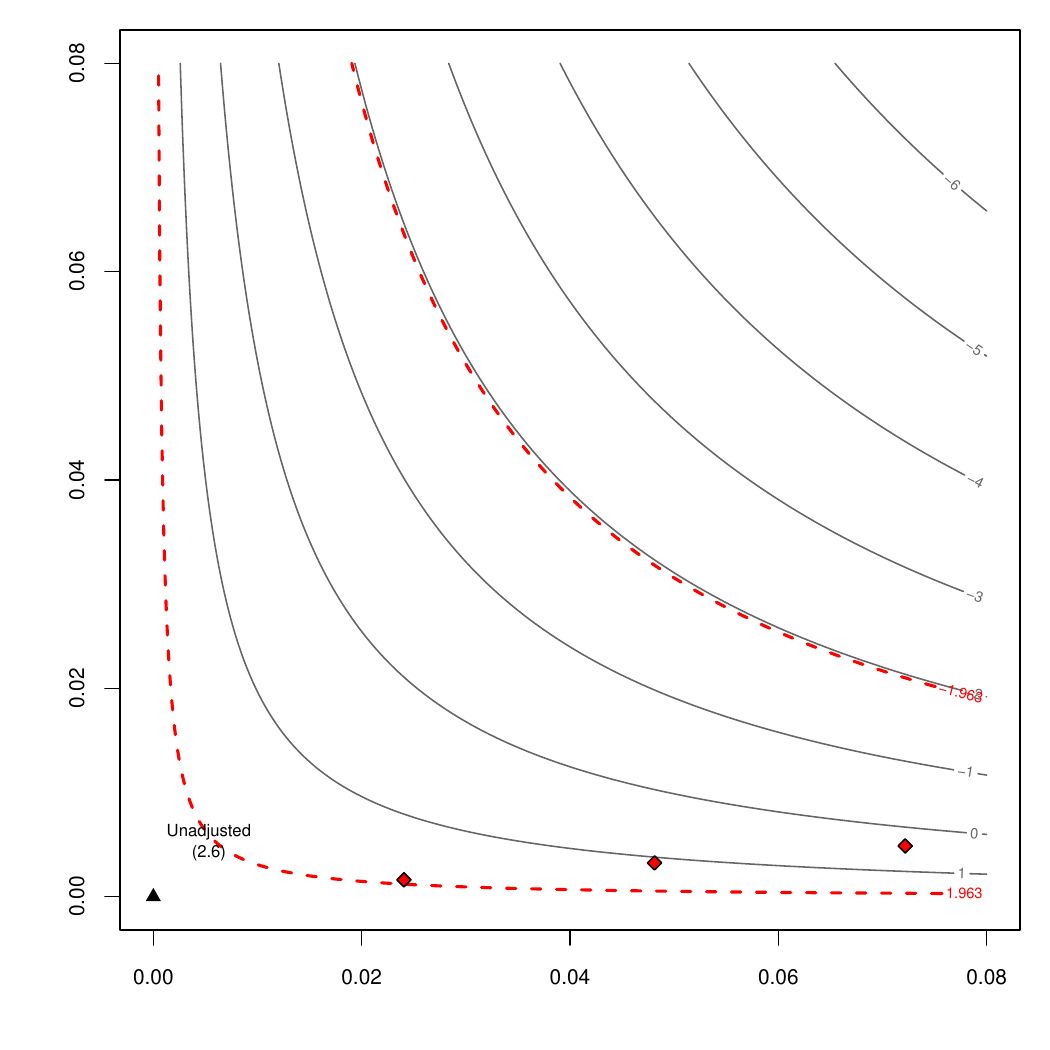}};
        \node[below=of img6, node distance=0cm, yshift=1.3cm,font=\color{black}] {\tiny Partial $R^2$ of confounder(s) with the treatment};
    \end{tikzpicture}
    \caption{Sensitivity of spontaneous-labour ATE estimate to unobserved confounding}
    \label{fig:sensemakr_plot_spontaneous_labour}
\end{figure}
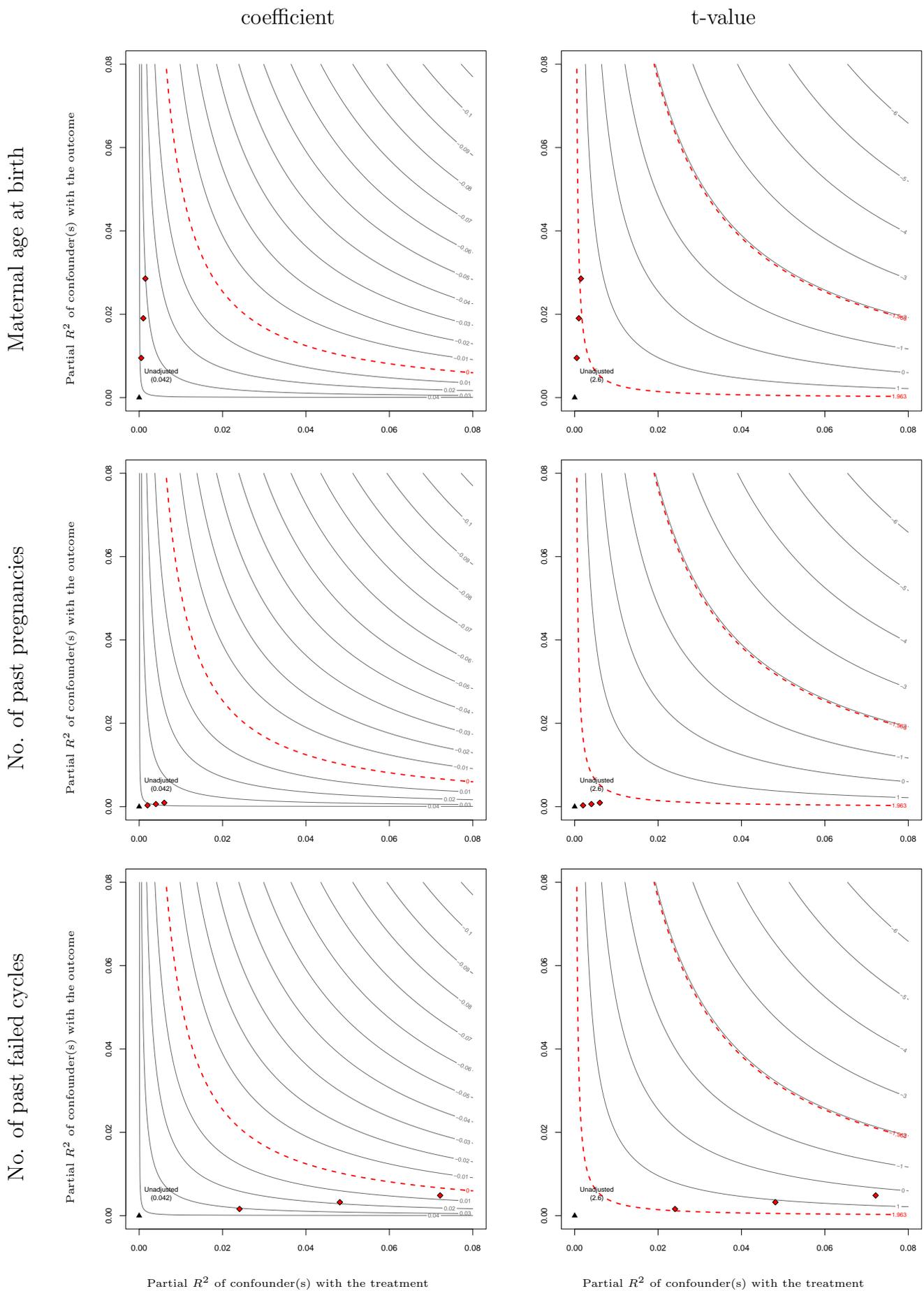
\restoregeometry

\section{Conclusion} \label{sec:conclusion}
We study the effect of ART on obstetric outcomes. We use the outcomes of babies
conceived spontaneously shortly after a failed ART cycle to build the
counterfactual outcome,
i.e. the outcome if that same baby was instead conceived
via ART.
To do this, (i) we select babies born 3-12 months after a failed ART cycle
as the control group and ART-conceived babies as the treatment group;
(ii) we estimate the average treatment effect of ART on preterm birth,
spontaneous preterm birth,
gestational age at birth, APGAR score and birth weight, using a linear model
and a more flexible double machine learning model; (iii) we study omitted
variable bias formally, using a selection of strong confounders as a benchmark
for unobserved confounders.
To our knowledge this is the first study to evaluated obstetric outcomes by
constructing cohorts from women who have undergone ART,
but who have conceived spontaneous or as a result of treatment,
thus removing much of the bias and confounding associated with previous studies.

We find that conceiving a baby via ART does not affect the risk of
preterm birth, preterm spontaneous labour,
the gestational age at birth, the birth weight of the baby nor
their APGAR scores, in subfertile mothers.
Our estimated null effects are precise, leaving us unable to rule out
only ART effects that are clinically small
\footnote{
    We can exclude the following effect sizes for the effect of ART on
    obstretric outcomes: effects between [-0.16, 0.10] weeks for gestational age
    at birth; between [-1.1, 1.9] p.p. for the risk of preterm birth;
    between [-0.07, 0.09] points for the APGAR1 score;
    between [-0.02, 0.1] points for the APGAR5 score;
    between [-14, 60] grams for the effect of ART on birth weight.
}.
Moreover, we find that ART has a small negative impact ($~-5 p.p.$)
on the chance of a c-section and of an induction of labour. While
the sign of these effect is robust to large levels of unobserved confounding,
such large levels of confounding effects would further reduce the magnitude
of these estimates and make them statistically insignificant.

Overall, these results support the hypothesis that ART does not independently
affect obstetric risk in subfertile women. This study indicates that while
ART treatment is an indicator of poorer obstetric and perinatal outcomes in
singleton pregnancies
(i.e. because of coexisting risk factors such as advanced maternal age and
conditions associated with infertility), ART treatment per se does not
appear to be an independent risk factor.
These findings should be reassuring for women undergoing ART and
fertility clinicians and hopefully provide context around
ART-conceived pregnancies.

\bibliographystyle{apa}
\bibliography{ivf_nat_biblio}

\appendix

\newgeometry{top=5mm, bottom=10mm, left=5mm, right=5mm}     
\begin{figure}[htb]
    \centering
    \thispagestyle{empty}
    \begin{tikzpicture}
        \node (img1)  {\includegraphics[scale=0.45]{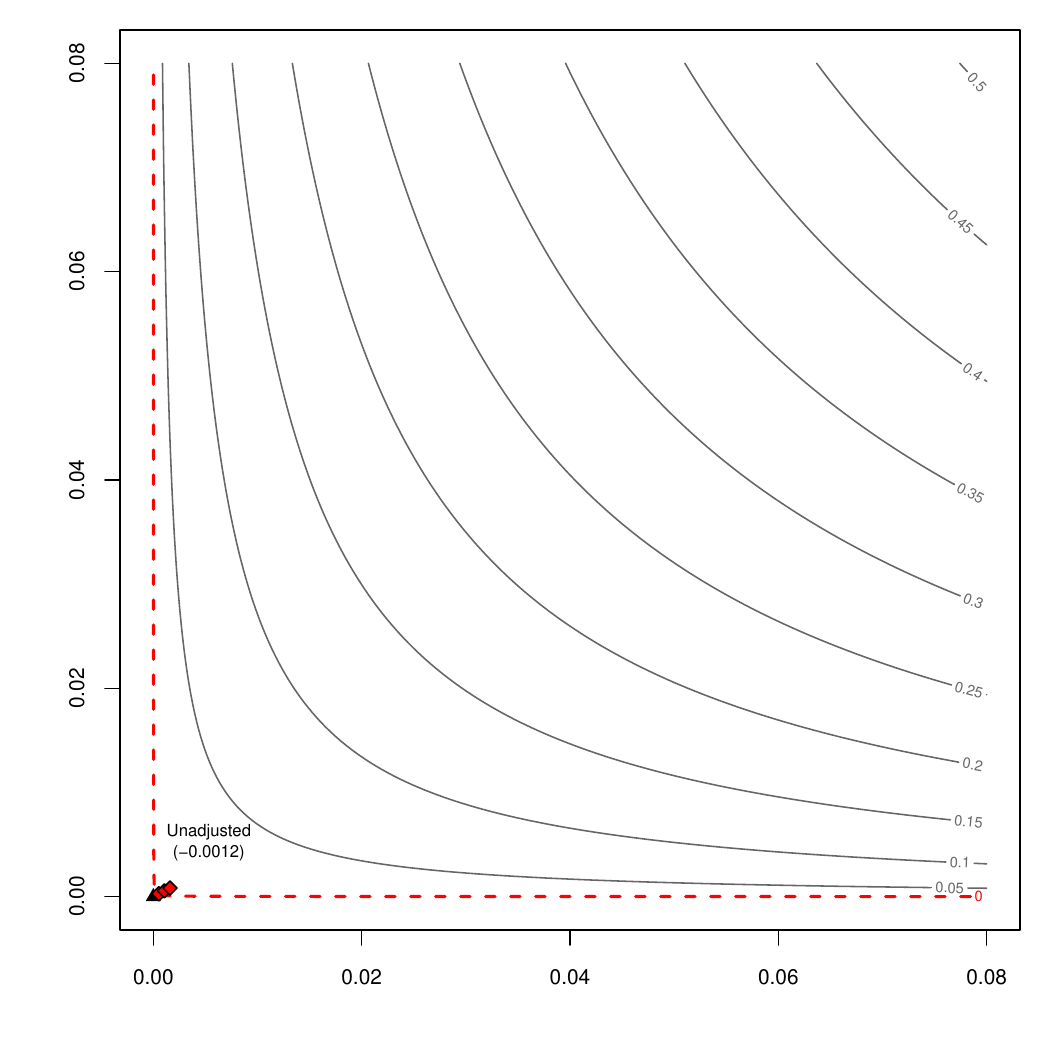}};
        \node[above=of img1, node distance=0cm, yshift=-1cm,font=\color{black}] {ATE coefficient};
        \node[left=of img1, node distance=0cm, rotate=90, anchor=center,yshift=-1cm,font=\color{black}] {\tiny Partial $R^2$ of confounder(s) with the outcome};
        \node[left=of img1, node distance=0cm, rotate=90, anchor=center,yshift=0cm,font=\color{black}] {Maternal age at birth};
        \node[right=of img1,yshift=0cm, xshift=-1cm] (img2)  {\includegraphics[scale=0.45]{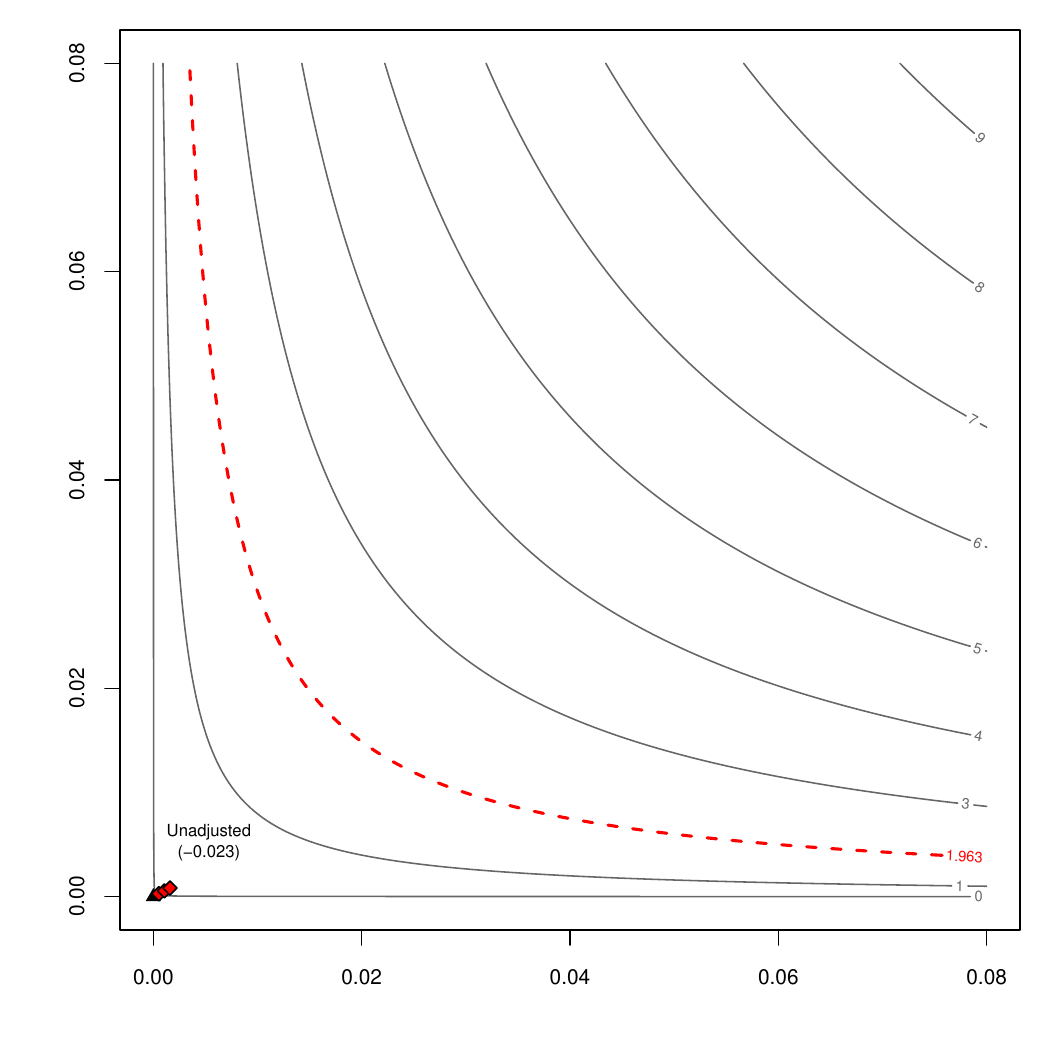}};
        \node[above=of img2, node distance=0cm, yshift=-1cm,font=\color{black}] {t-statistic};
        \node[below=of img1,yshift=1.5cm] (img3)  {\includegraphics[scale=0.45]{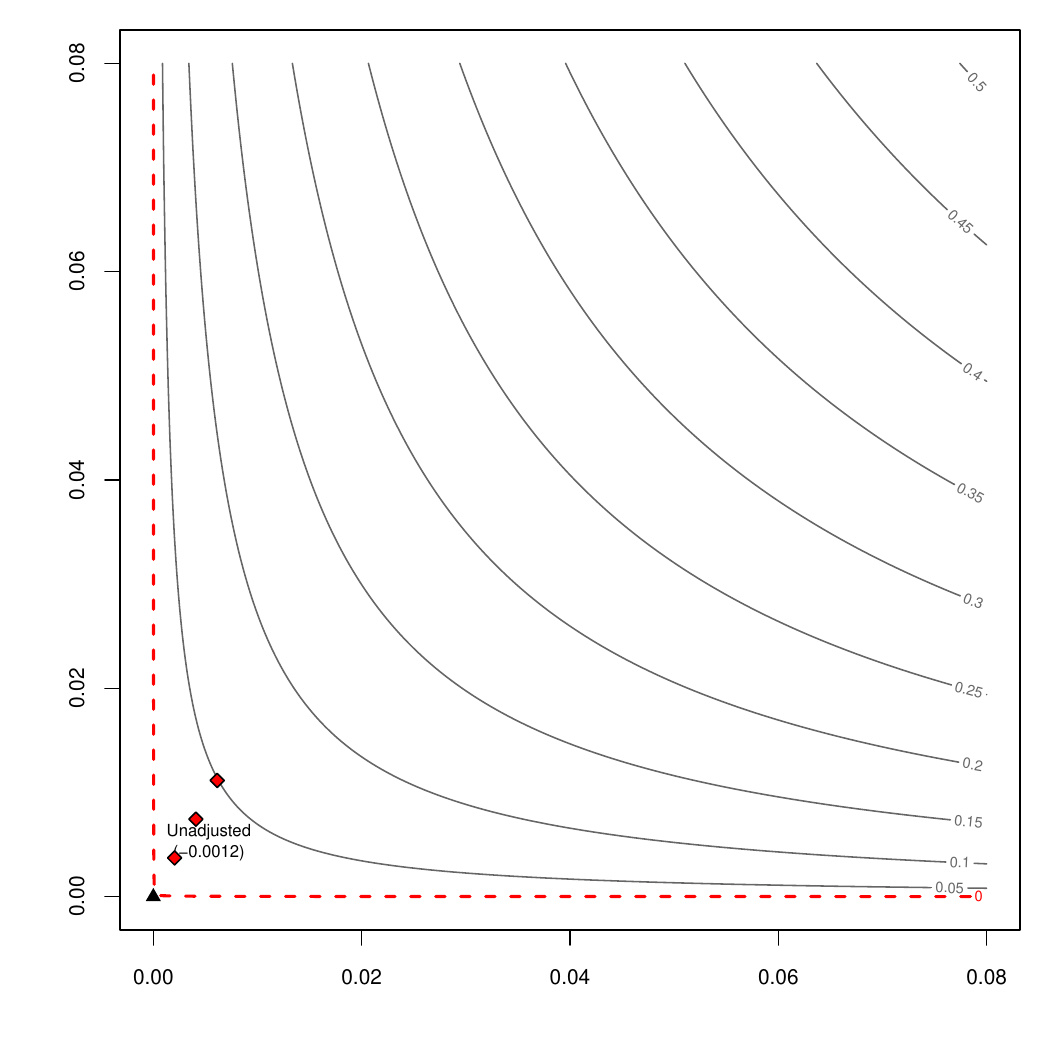}};
        \node[left=of img3, node distance=0cm, rotate=90, anchor=center,yshift=-1cm,font=\color{black}] {\tiny Partial $R^2$ of confounder(s) with the outcome};
        \node[left=of img3, node distance=0cm, rotate=90, anchor=center,yshift=0cm,font=\color{black}] {No. of past pregnancies (parity)};
        \node[below=of img2,yshift=1.5cm] (img4)  {\includegraphics[scale=0.45]{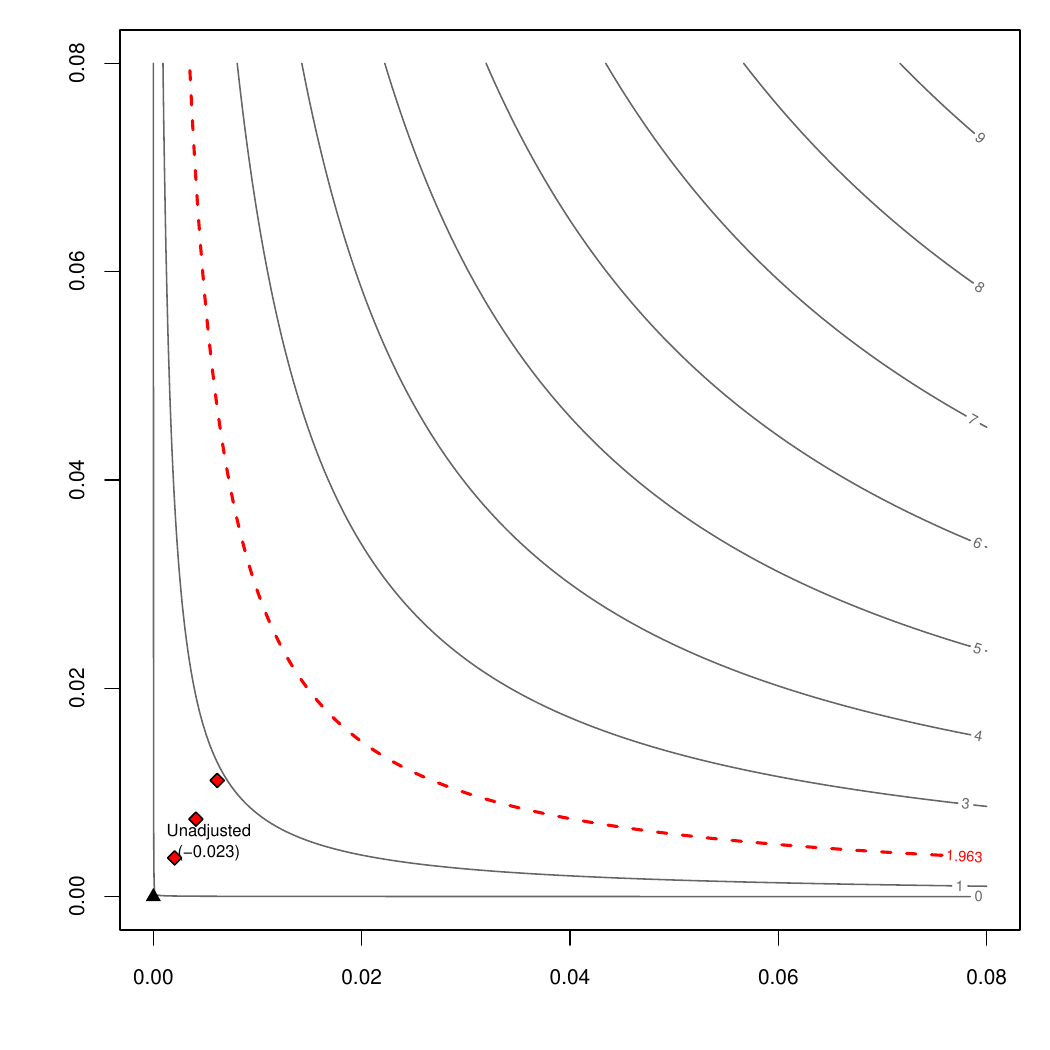}};
        \node[below=of img3,yshift=1.5cm] (img5)  {\includegraphics[scale=0.45]{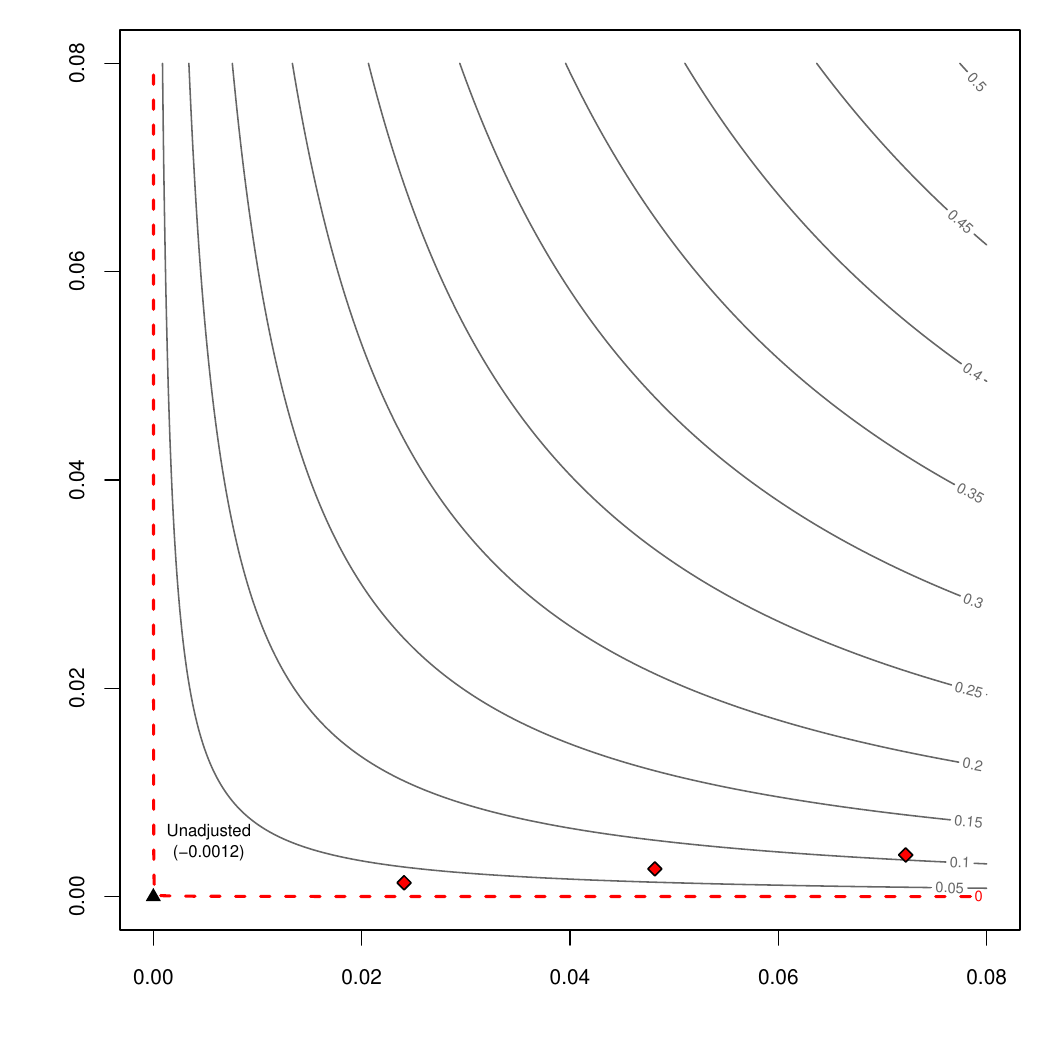}};
        \node[below=of img5, node distance=0cm, yshift=1.3cm,font=\color{black}] {\tiny Partial $R^2$ of confounder(s) with the treatment};
        \node[left=of img5, node distance=0cm, rotate=90, anchor=center,yshift=-1cm,font=\color{black}] {\tiny Partial $R^2$ of confounder(s) with the outcome};
        \node[left=of img5, node distance=0cm, rotate=90, anchor=center,yshift=0cm,font=\color{black}] {No. of past failed cycles};
        \node[below=of img4,yshift=1.5cm] (img6)  {\includegraphics[scale=0.45]{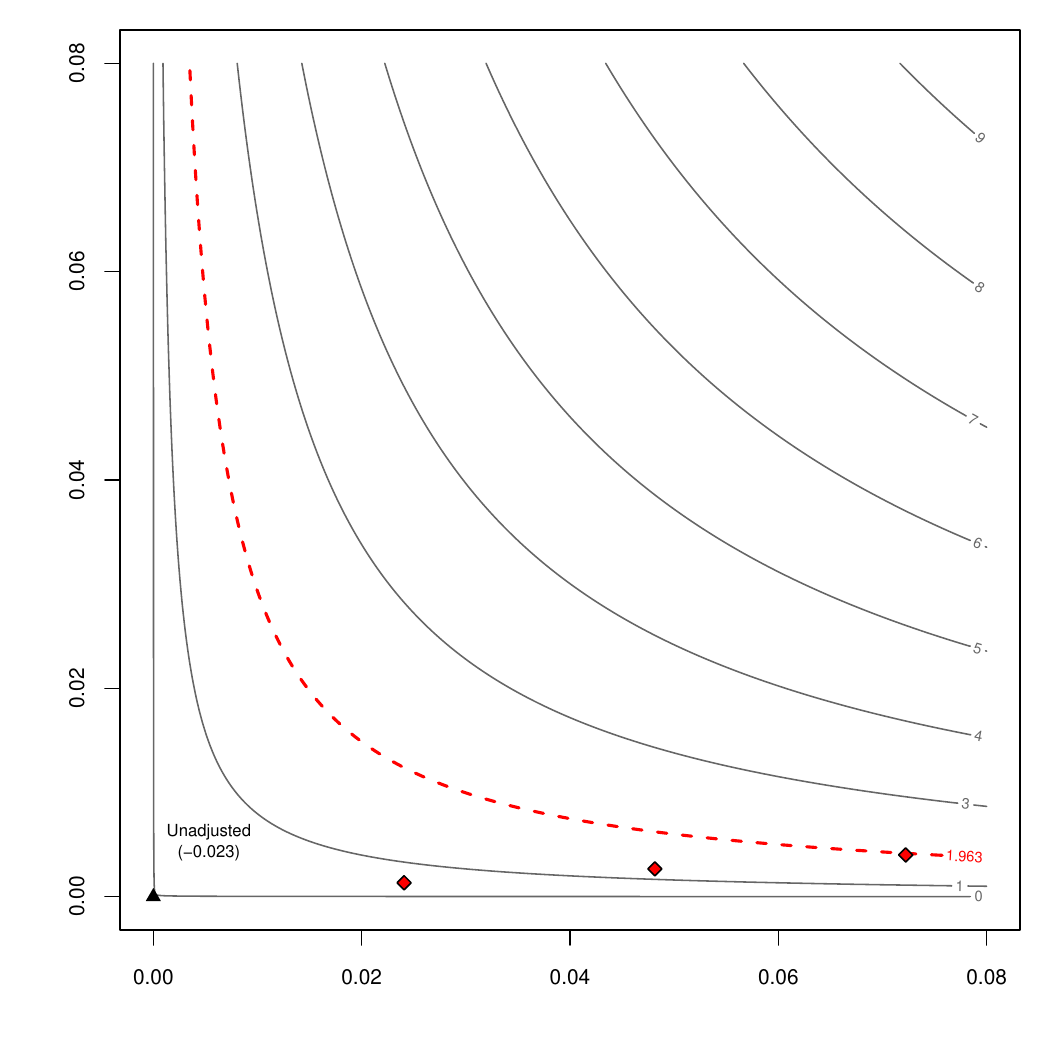}};
        \node[below=of img6, node distance=0cm, yshift=1.3cm,font=\color{black}] {\tiny Partial $R^2$ of confounder(s) with the treatment};
    \end{tikzpicture}
    \caption{Sensitivity of APGAR1 ATE estimate to unobserved confounding}
    \label{fig:sensemakr_plot_APGAR1}
\end{figure}

\begin{figure}[htb]
    \centering
    \thispagestyle{empty}
    \begin{tikzpicture}
        \node (img1)  {\includegraphics[scale=0.45]{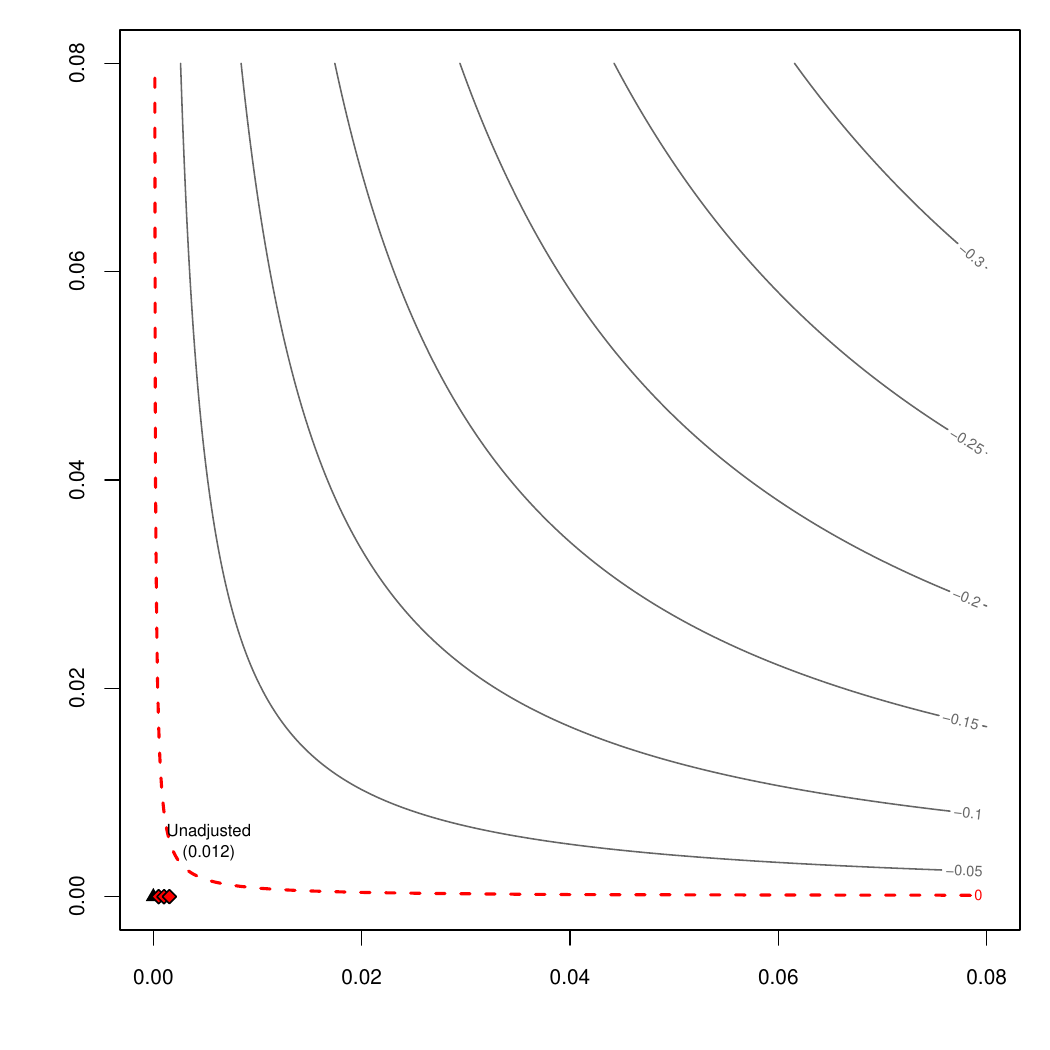}};
        \node[above=of img1, node distance=0cm, yshift=-1cm,font=\color{black}] {ATE coefficient};
        \node[left=of img1, node distance=0cm, rotate=90, anchor=center,yshift=-1cm,font=\color{black}] {\tiny Partial $R^2$ of confounder(s) with the outcome};
        \node[left=of img1, node distance=0cm, rotate=90, anchor=center,yshift=0cm,font=\color{black}] {Maternal age at birth};
        \node[right=of img1,yshift=0cm, xshift=-1cm] (img2)  {\includegraphics[scale=0.45]{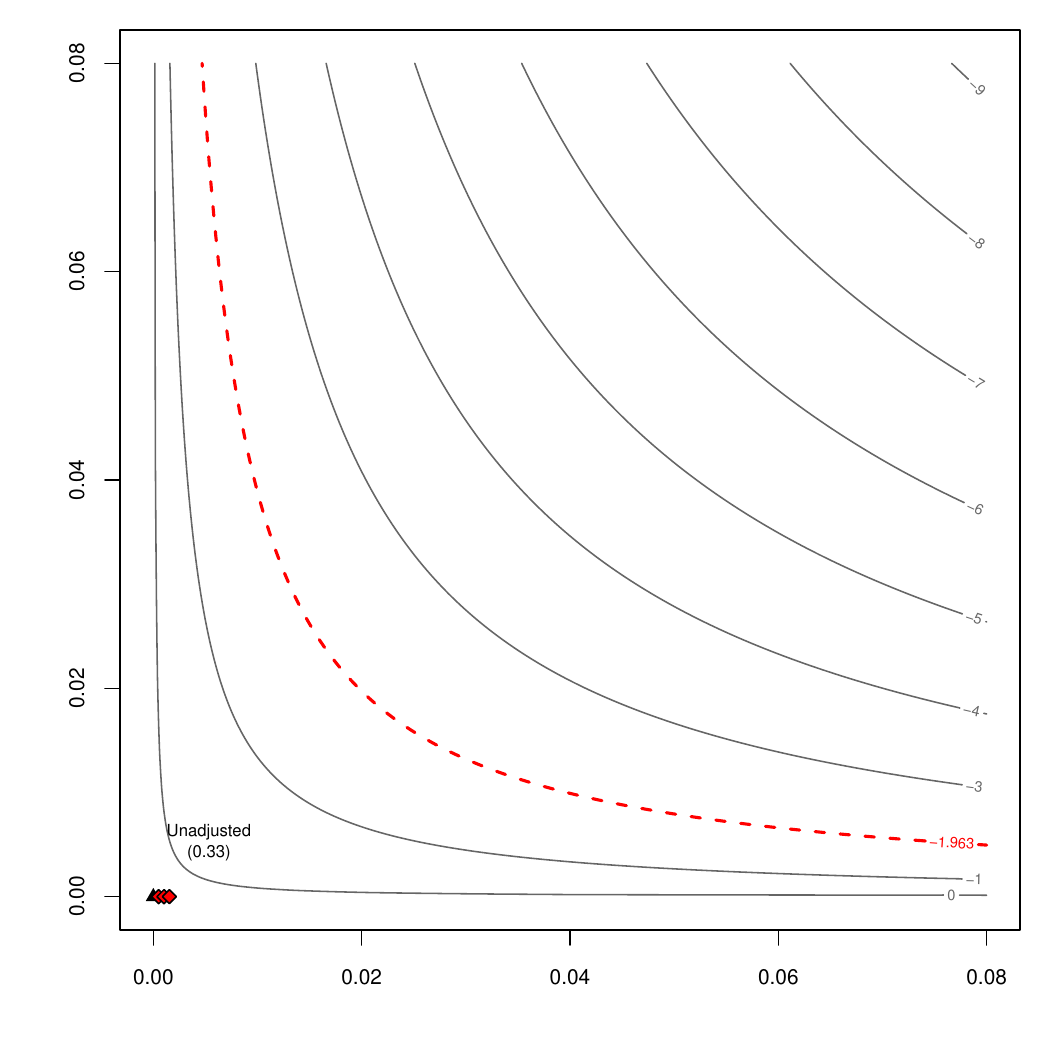}};
        \node[above=of img2, node distance=0cm, yshift=-1cm,font=\color{black}] {t-statistic};
        \node[below=of img1,yshift=1.5cm] (img3)  {\includegraphics[scale=0.45]{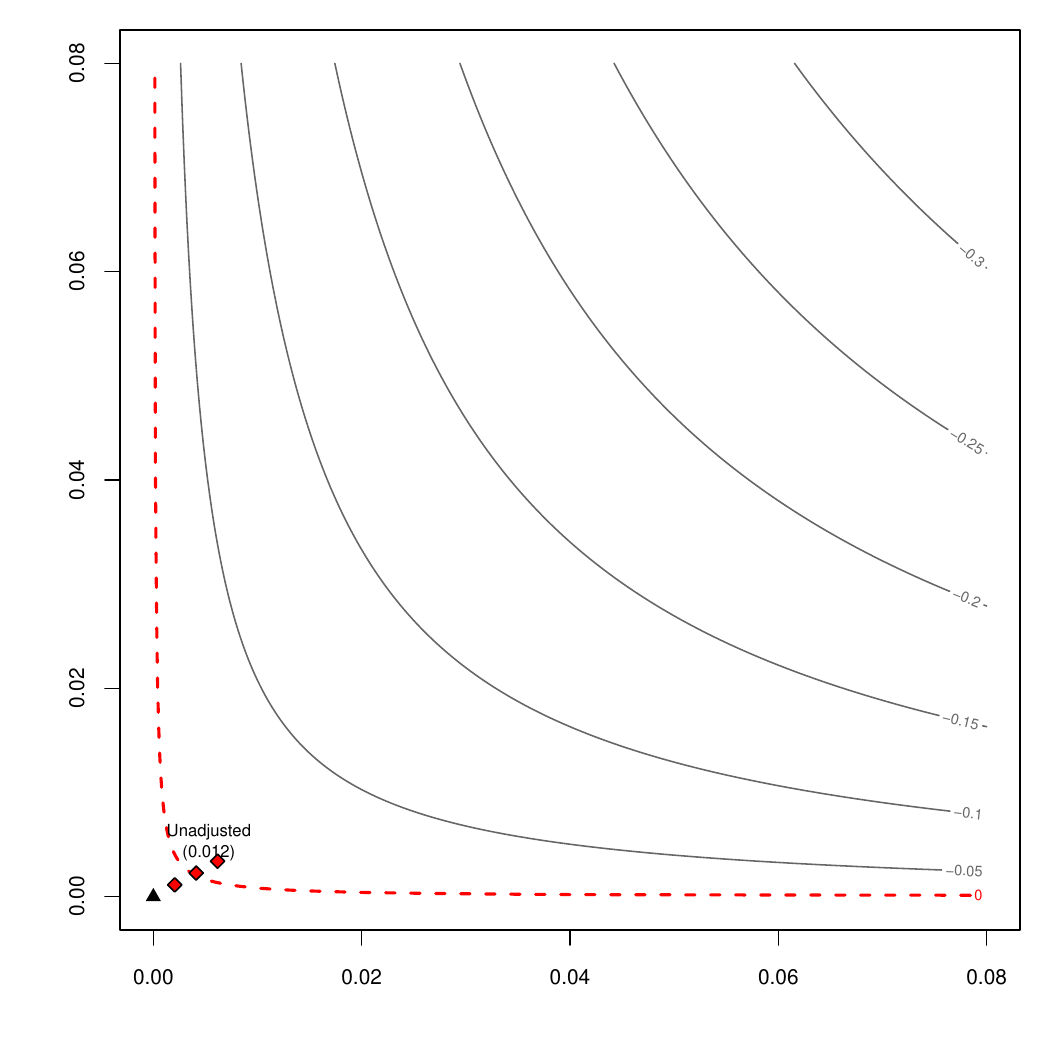}};
        \node[left=of img3, node distance=0cm, rotate=90, anchor=center,yshift=-1cm,font=\color{black}] {\tiny Partial $R^2$ of confounder(s) with the outcome};
        \node[left=of img3, node distance=0cm, rotate=90, anchor=center,yshift=0cm,font=\color{black}] {No. of past pregnancies (parity)};
        \node[below=of img2,yshift=1.5cm] (img4)  {\includegraphics[scale=0.45]{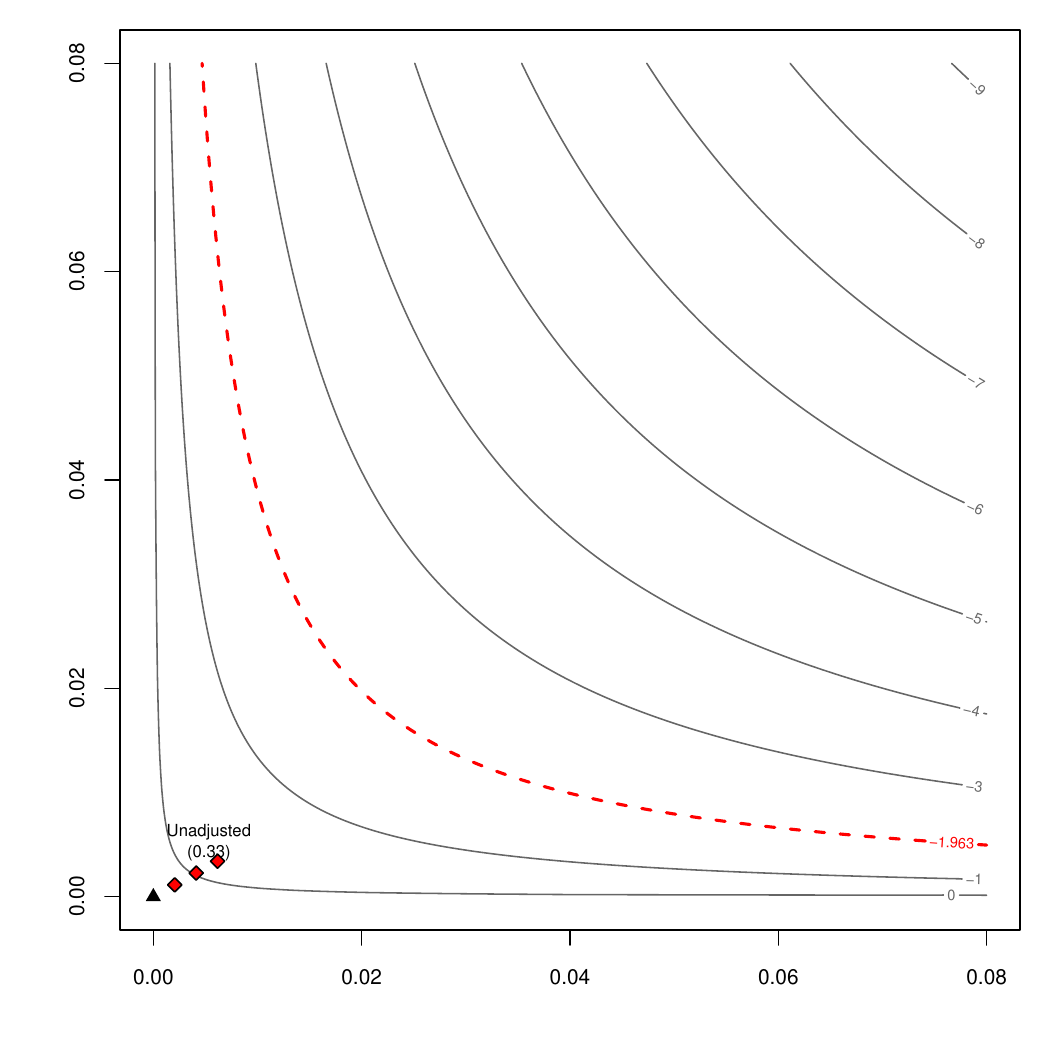}};
        \node[below=of img3,yshift=1.5cm] (img5)  {\includegraphics[scale=0.45]{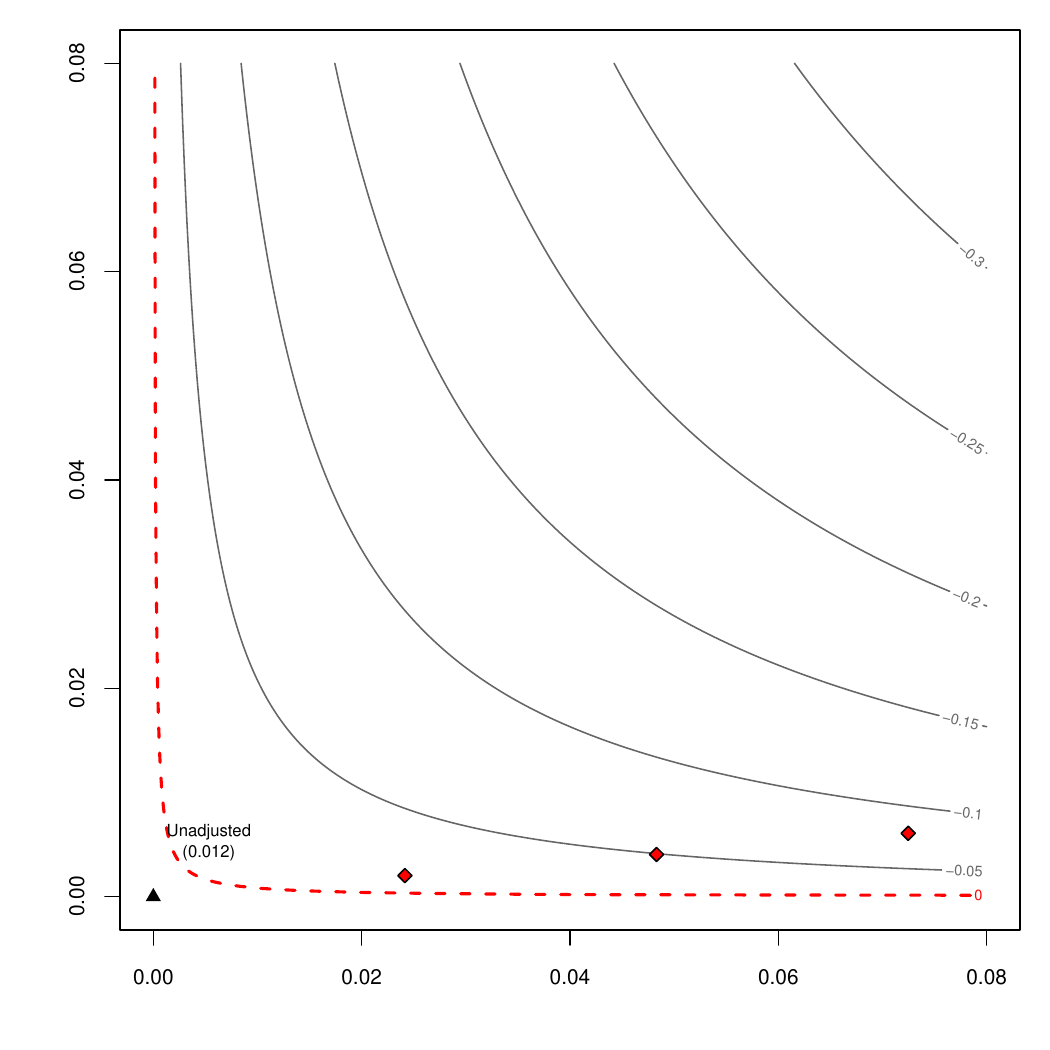}};
        \node[below=of img5, node distance=0cm, yshift=1.3cm,font=\color{black}] {\tiny Partial $R^2$ of confounder(s) with the treatment};
        \node[left=of img5, node distance=0cm, rotate=90, anchor=center,yshift=-1cm,font=\color{black}] {\tiny Partial $R^2$ of confounder(s) with the outcome};
        \node[left=of img5, node distance=0cm, rotate=90, anchor=center,yshift=0cm,font=\color{black}] {No. of past failed cycles};
        \node[below=of img4,yshift=1.5cm] (img6)  {\includegraphics[scale=0.45]{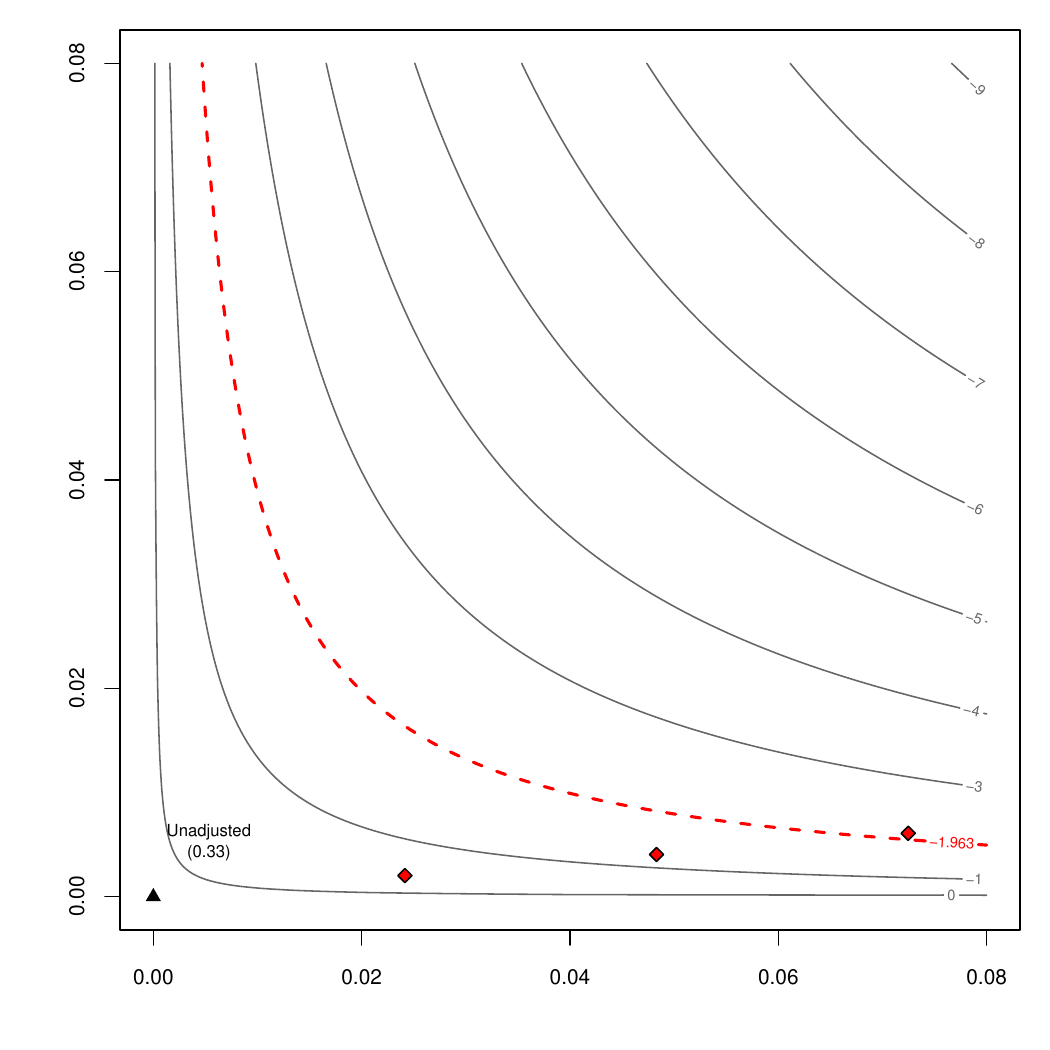}};
        \node[below=of img6, node distance=0cm, yshift=1.3cm,font=\color{black}] {\tiny Partial $R^2$ of confounder(s) with the treatment};
    \end{tikzpicture}
    \caption{Sensitivity of APGAR5 ATE estimate to unobserved confounding}
    \label{fig:sensemakr_plot_APGAR5}
\end{figure}

\begin{figure}[htb]
    \centering
    \thispagestyle{empty}
    \begin{tikzpicture}
        \node (img1)  {\includegraphics[scale=0.45]{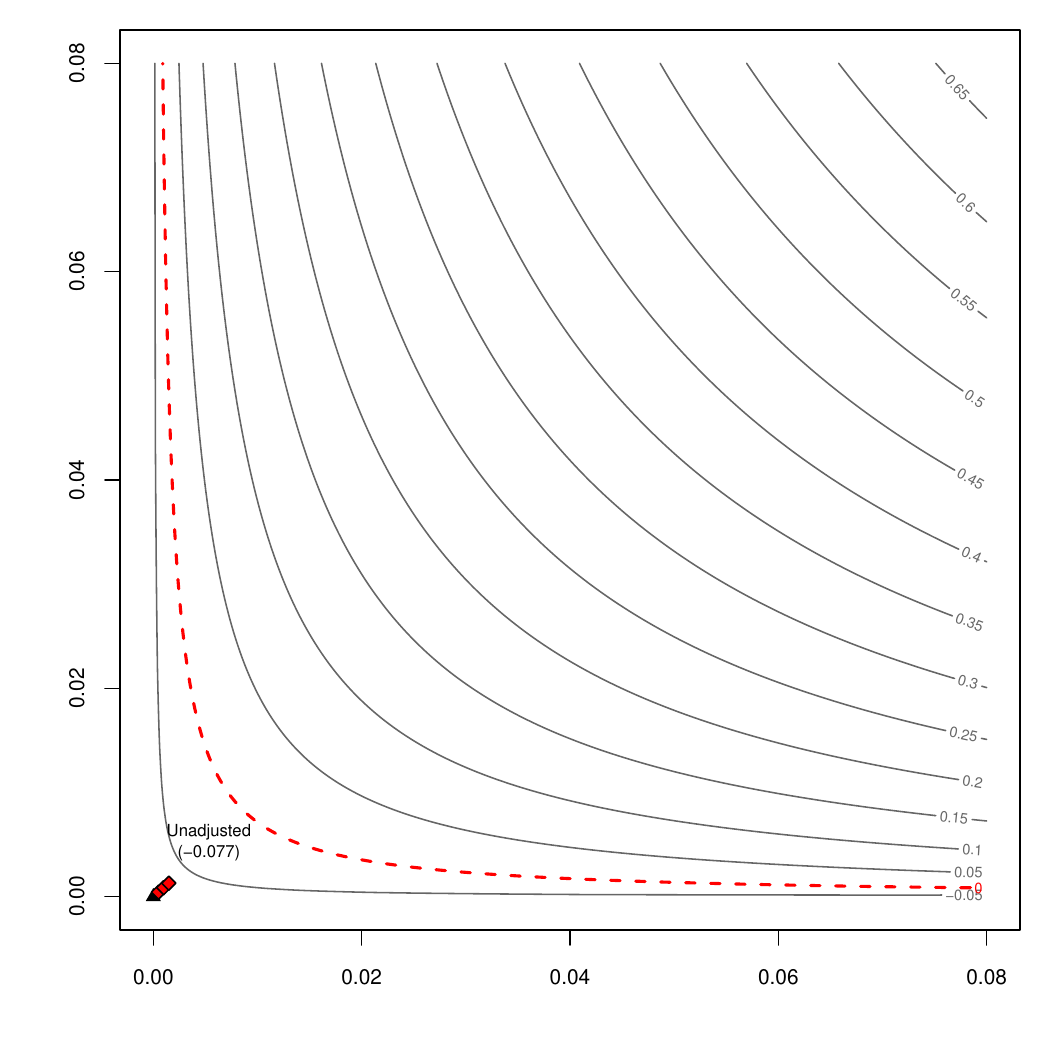}};
        \node[above=of img1, node distance=0cm, yshift=-1cm,font=\color{black}] {coefficient};
        \node[left=of img1, node distance=0cm, rotate=90, anchor=center,yshift=-1cm,font=\color{black}] {\tiny Partial $R^2$ of confounder(s) with the outcome};
        \node[left=of img1, node distance=0cm, rotate=90, anchor=center,yshift=0cm,font=\color{black}] {Maternal age at birth};
        \node[right=of img1,yshift=0cm, xshift=-1cm] (img2)  {\includegraphics[scale=0.45]{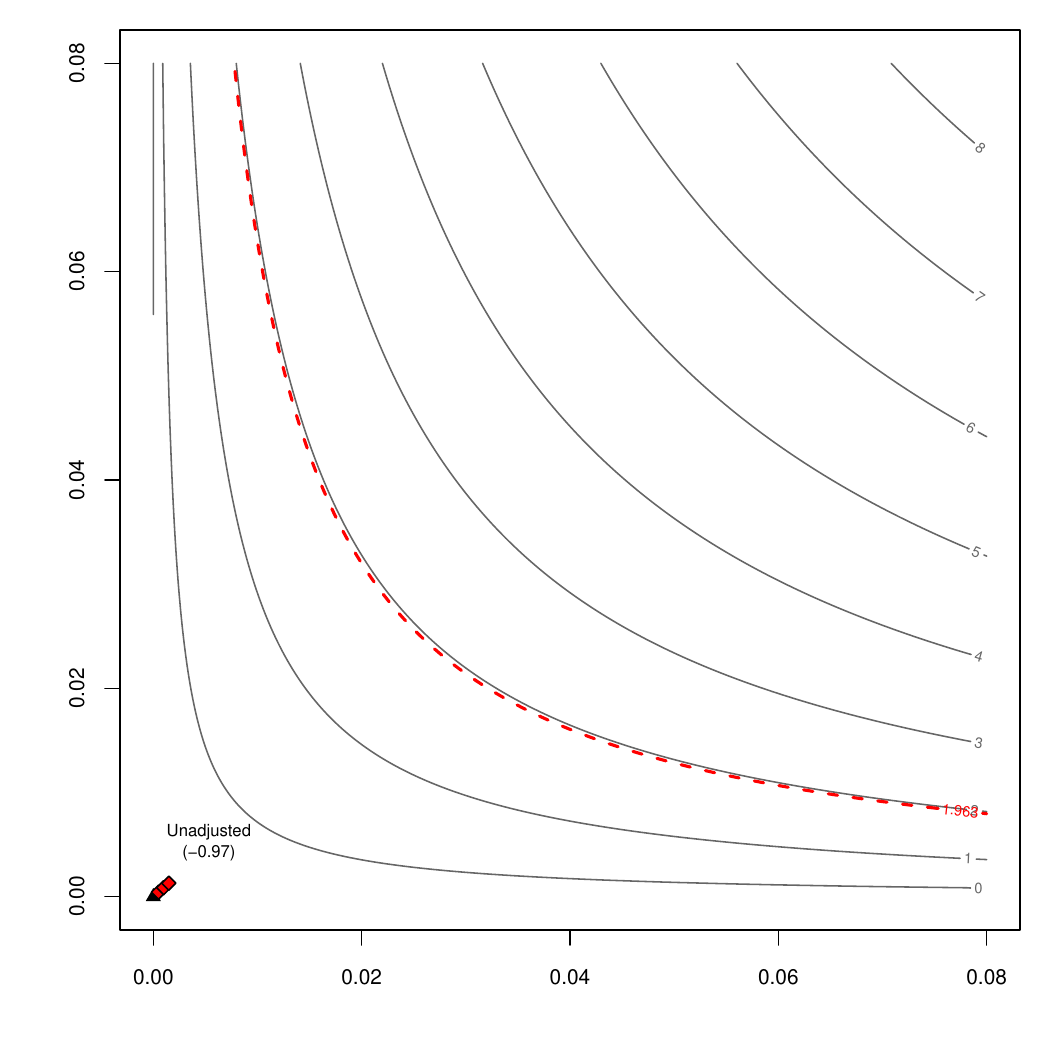}};
        \node[above=of img2, node distance=0cm, yshift=-1cm,font=\color{black}] {t-value};
        \node[below=of img1,yshift=1.5cm] (img3)  {\includegraphics[scale=0.45]{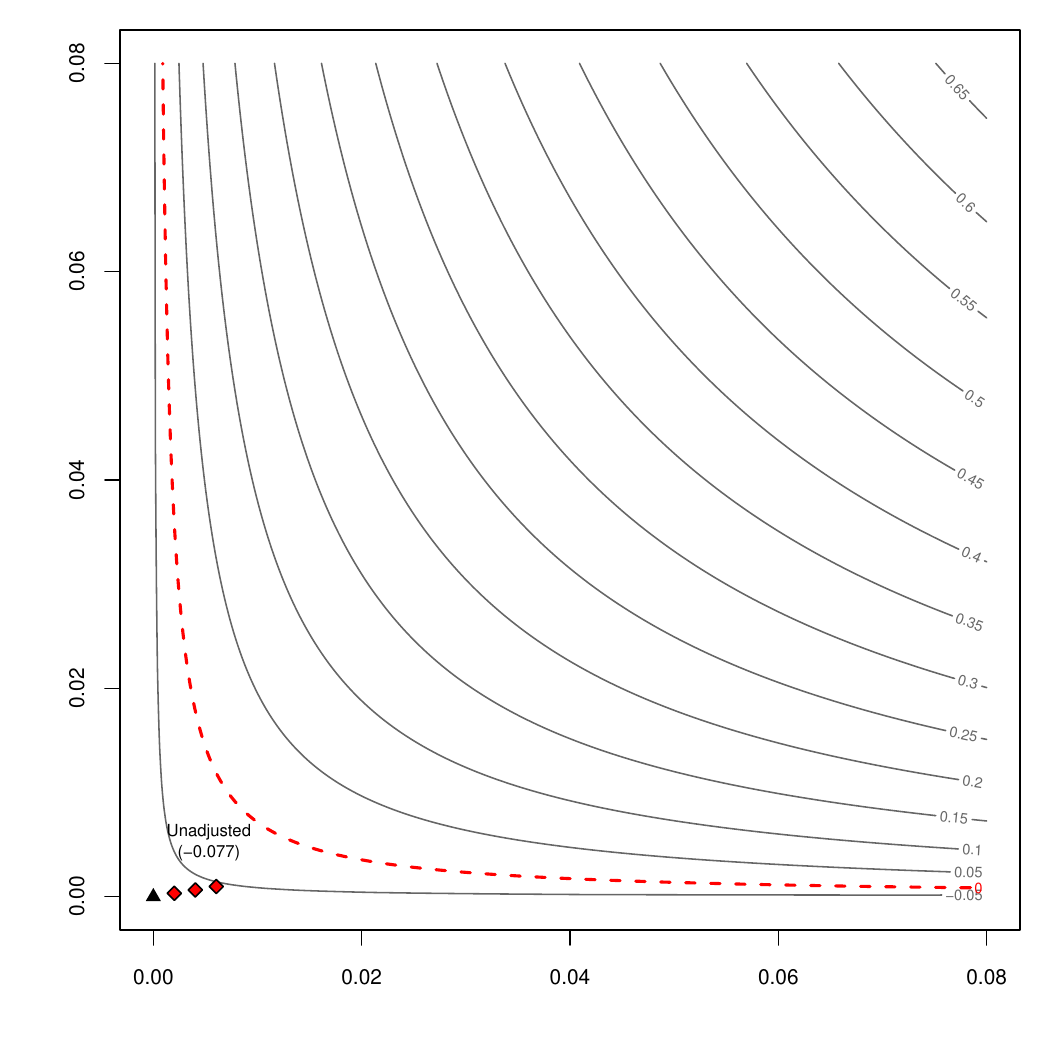}};
        \node[left=of img3, node distance=0cm, rotate=90, anchor=center,yshift=-1cm,font=\color{black}] {\tiny Partial $R^2$ of confounder(s) with the outcome};
        \node[left=of img3, node distance=0cm, rotate=90, anchor=center,yshift=0cm,font=\color{black}] {No. of past pregnancies};
        \node[below=of img2,yshift=1.5cm] (img4)  {\includegraphics[scale=0.45]{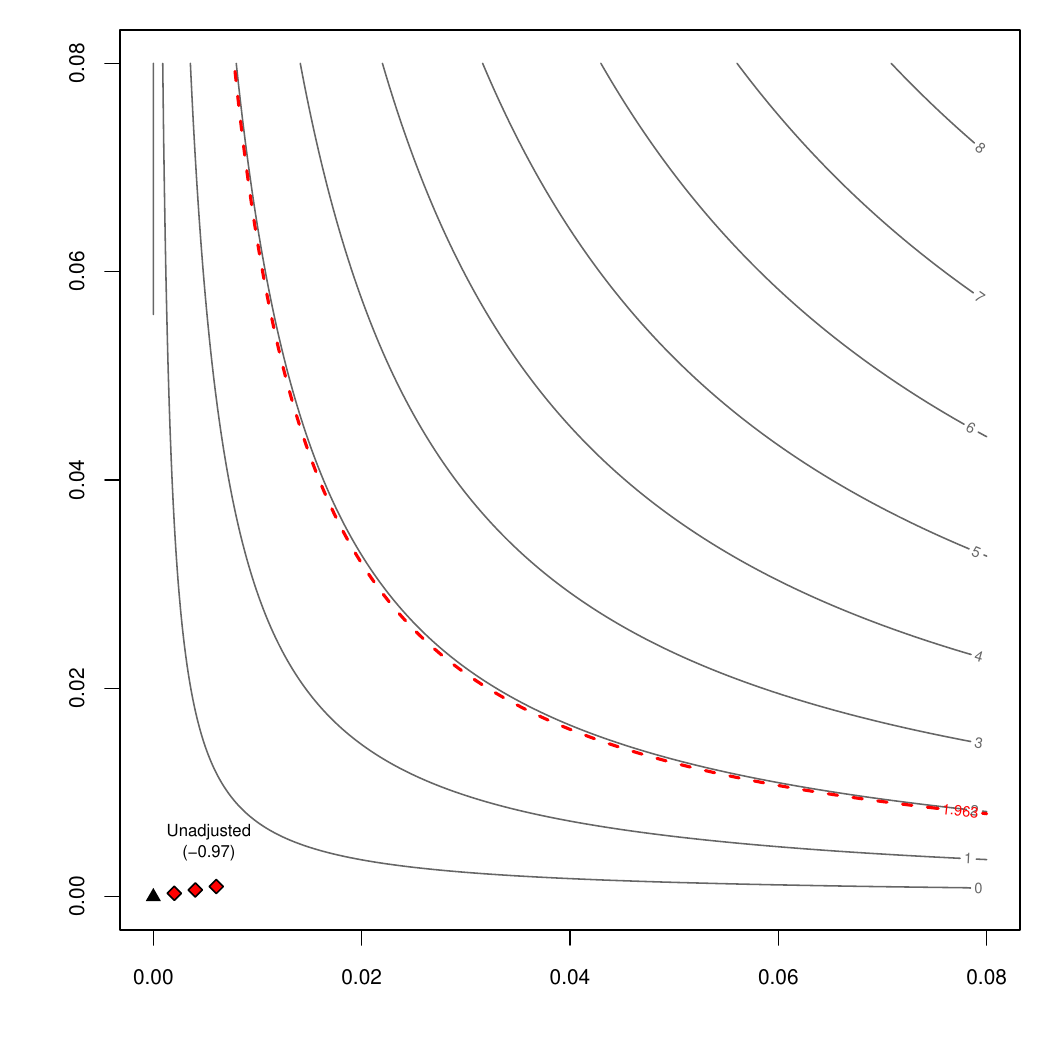}};
        \node[below=of img3,yshift=1.5cm] (img5)  {\includegraphics[scale=0.45]{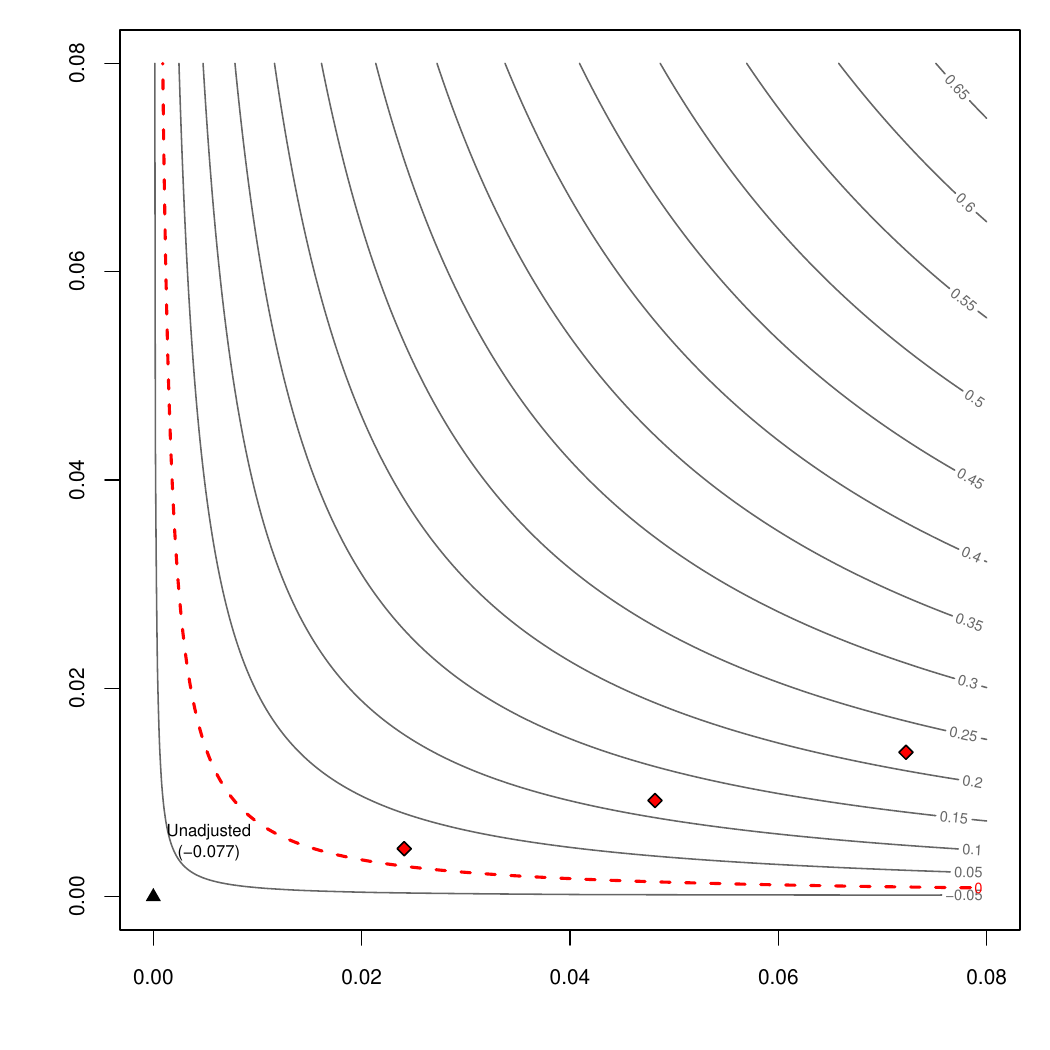}};
        \node[below=of img5, node distance=0cm, yshift=1.3cm,font=\color{black}] {\tiny Partial $R^2$ of confounder(s) with the treatment};
        \node[left=of img5, node distance=0cm, rotate=90, anchor=center,yshift=-1cm,font=\color{black}] {\tiny Partial $R^2$ of confounder(s) with the outcome};
        \node[left=of img5, node distance=0cm, rotate=90, anchor=center,yshift=0cm,font=\color{black}] {No. of past failed cycles};
        \node[below=of img4,yshift=1.5cm] (img6)  {\includegraphics[scale=0.45]{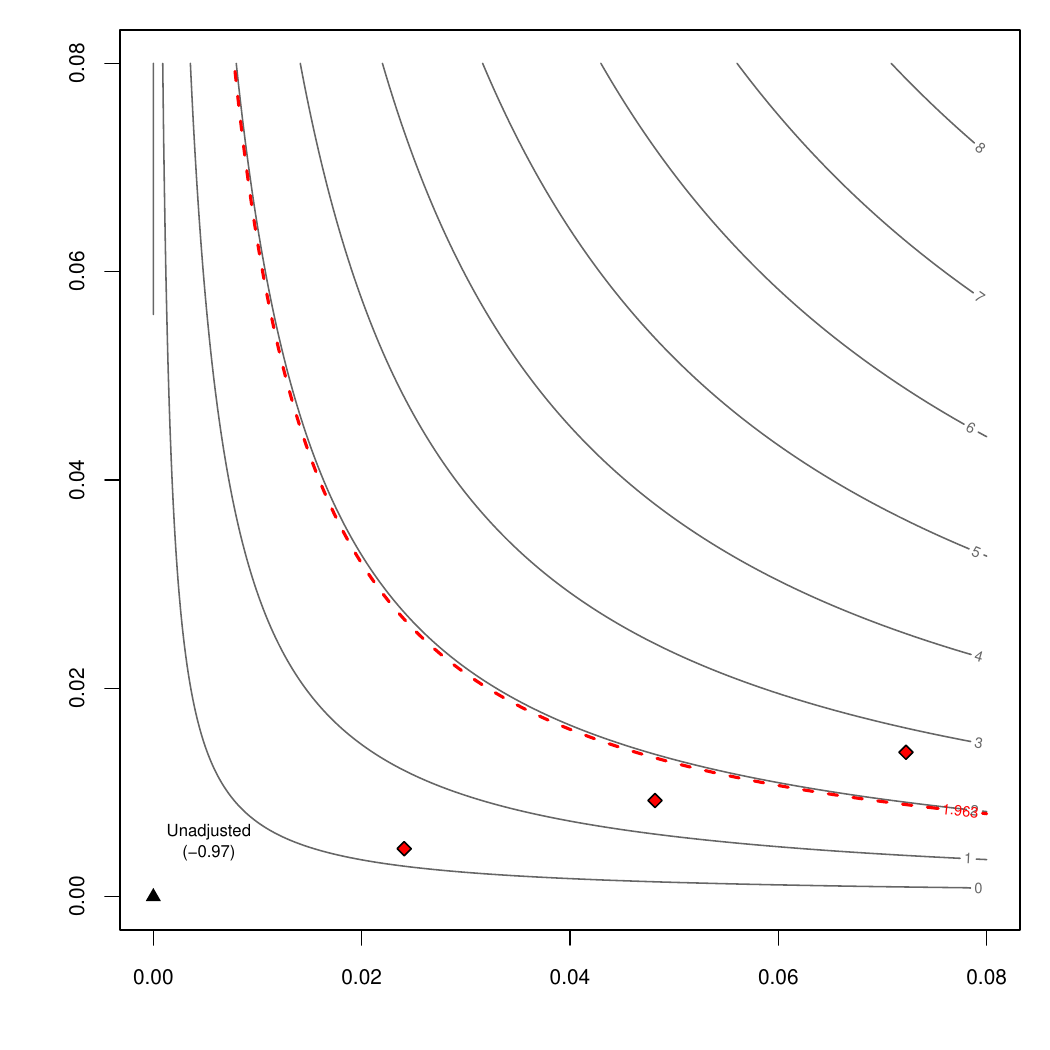}};
        \node[below=of img6, node distance=0cm, yshift=1.3cm,font=\color{black}] {\tiny Partial $R^2$ of confounder(s) with the treatment};
    \end{tikzpicture}
    \caption{Sensitivity of gestational age ATE estimate to unobserved confounding}
    \label{fig:sensemakr_plot_gestage}
\end{figure}

\begin{figure}[htb]
    \centering
    \thispagestyle{empty}
    \begin{tikzpicture}
        \node (img1)  {\includegraphics[scale=0.45]{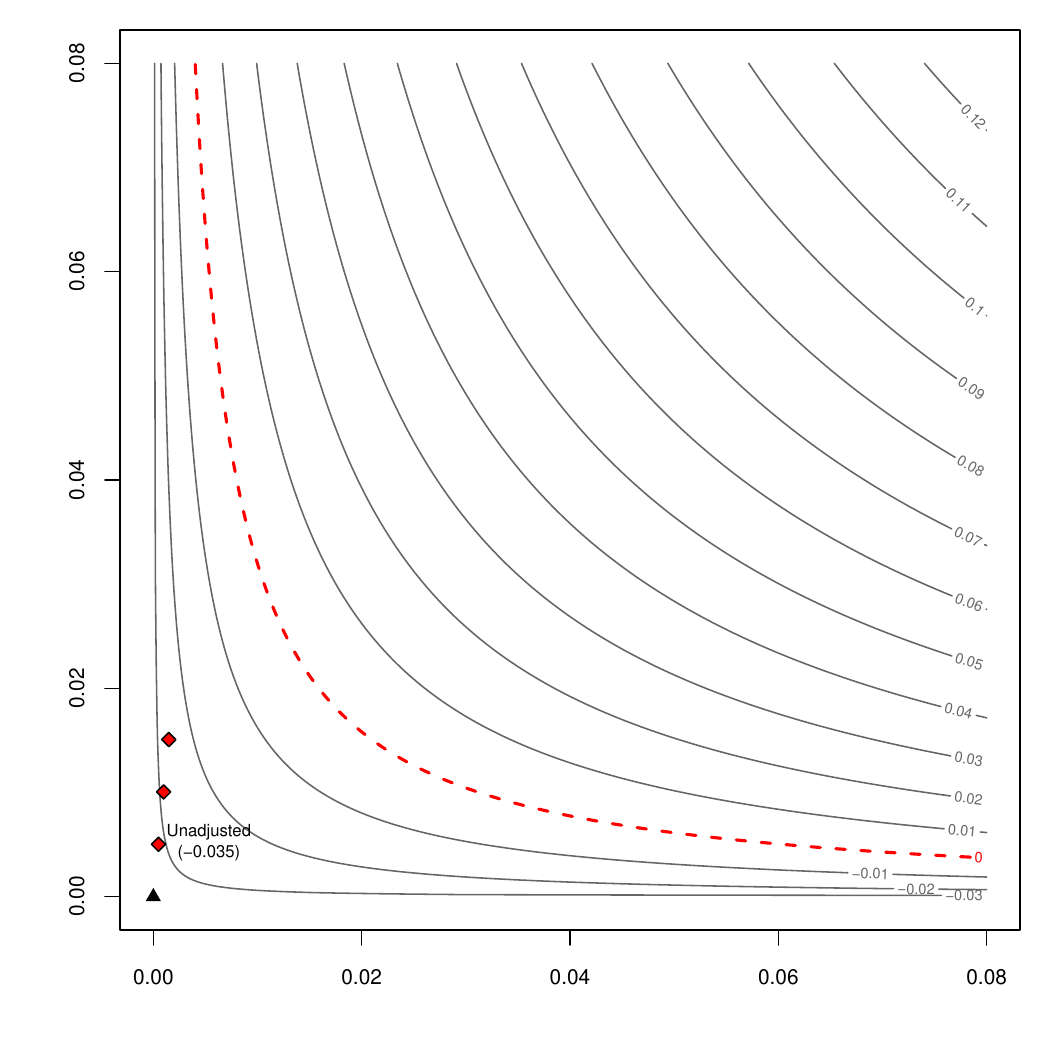}};
        \node[above=of img1, node distance=0cm, yshift=-1cm,font=\color{black}] {estimate};
        \node[left=of img1, node distance=0cm, rotate=90, anchor=center,yshift=-1cm,font=\color{black}] {\tiny Partial $R^2$ of confounder(s) with the outcome};
        \node[left=of img1, node distance=0cm, rotate=90, anchor=center,yshift=0cm,font=\color{black}] {Maternal age at birth};
        \node[right=of img1,yshift=0cm, xshift=-1cm] (img2)  {\includegraphics[scale=0.45]{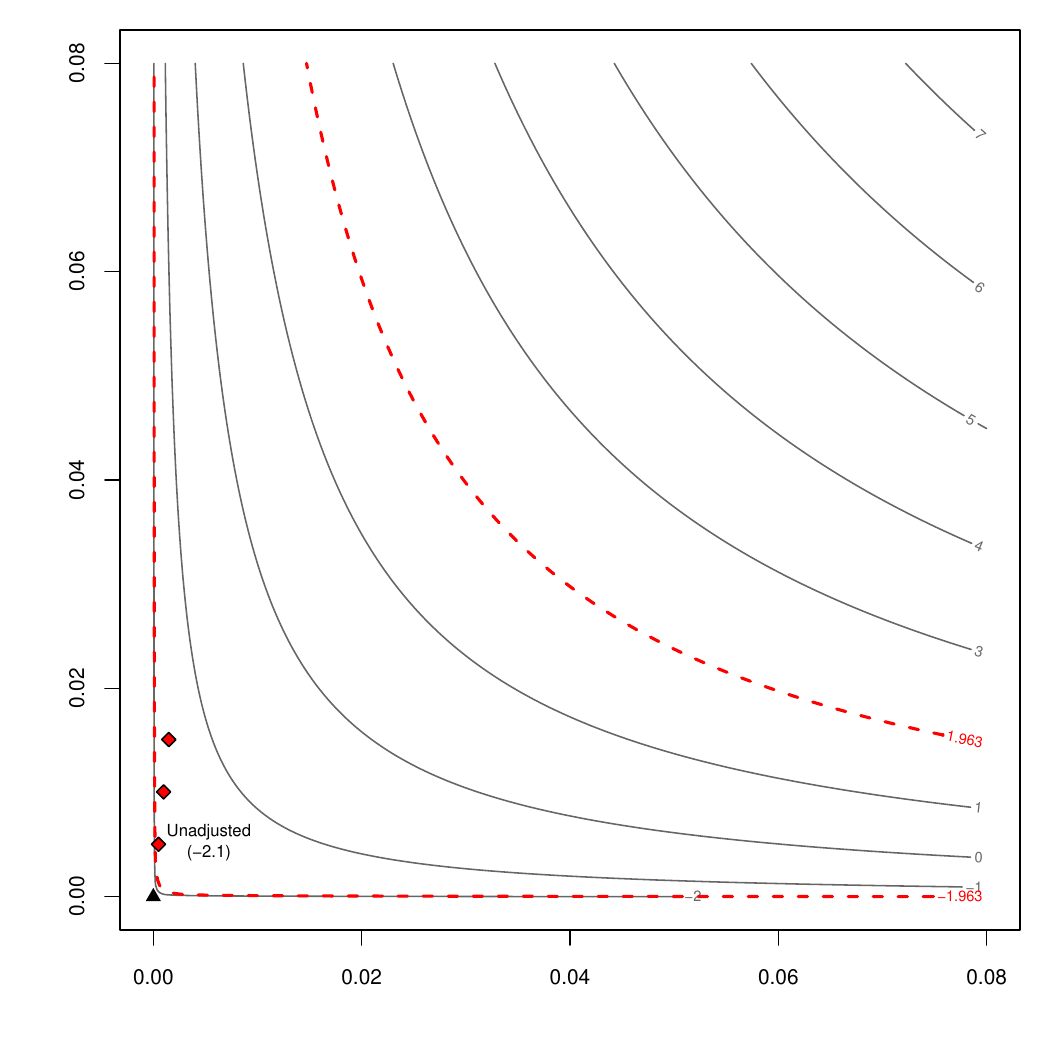}};
        \node[above=of img2, node distance=0cm, yshift=-1cm,font=\color{black}] {t-value};
        \node[below=of img1,yshift=1.5cm] (img3)  {\includegraphics[scale=0.45]{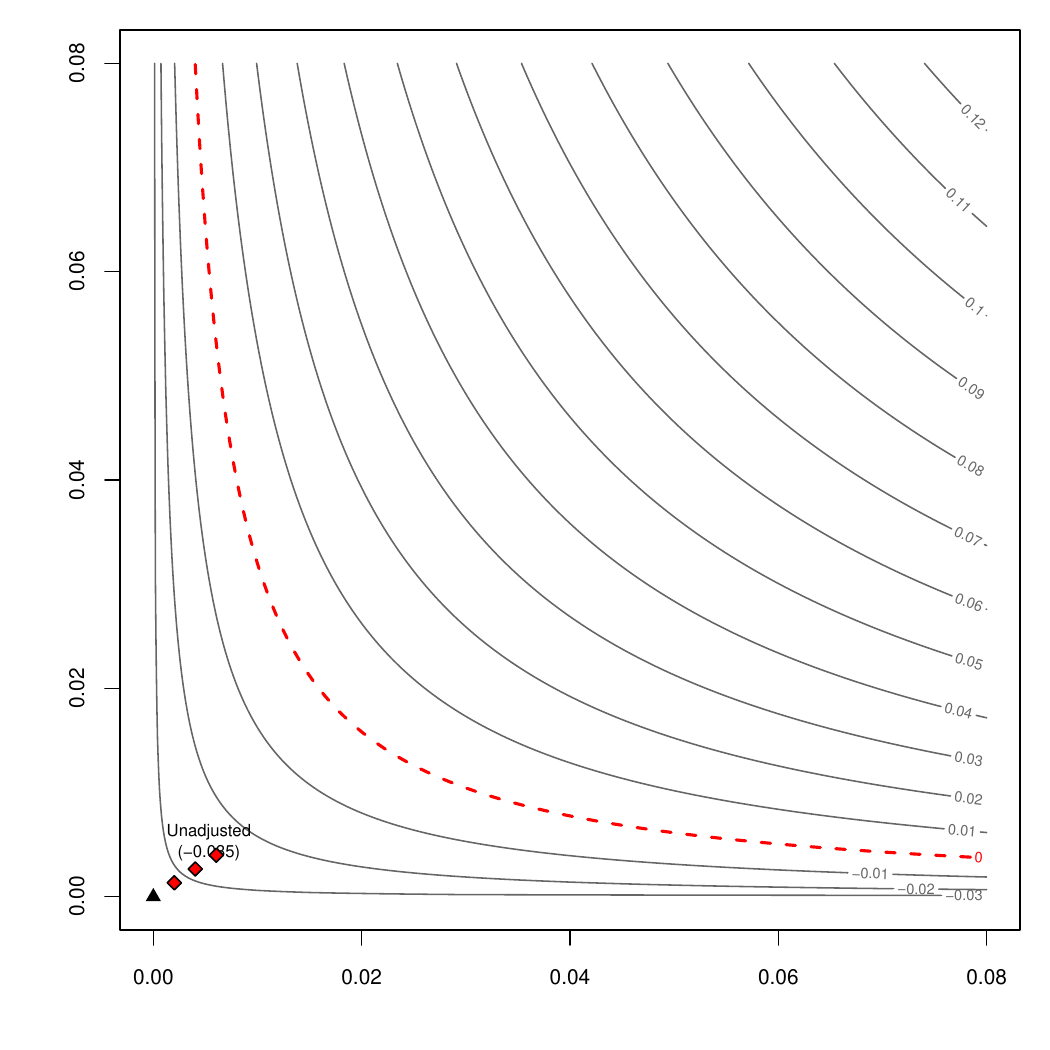}};
        \node[left=of img3, node distance=0cm, rotate=90, anchor=center,yshift=-1cm,font=\color{black}] {\tiny Partial $R^2$ of confounder(s) with the outcome};
        \node[left=of img3, node distance=0cm, rotate=90, anchor=center,yshift=0cm,font=\color{black}] {No. of past pregnancies};
        \node[below=of img2,yshift=1.5cm] (img4)  {\includegraphics[scale=0.45]{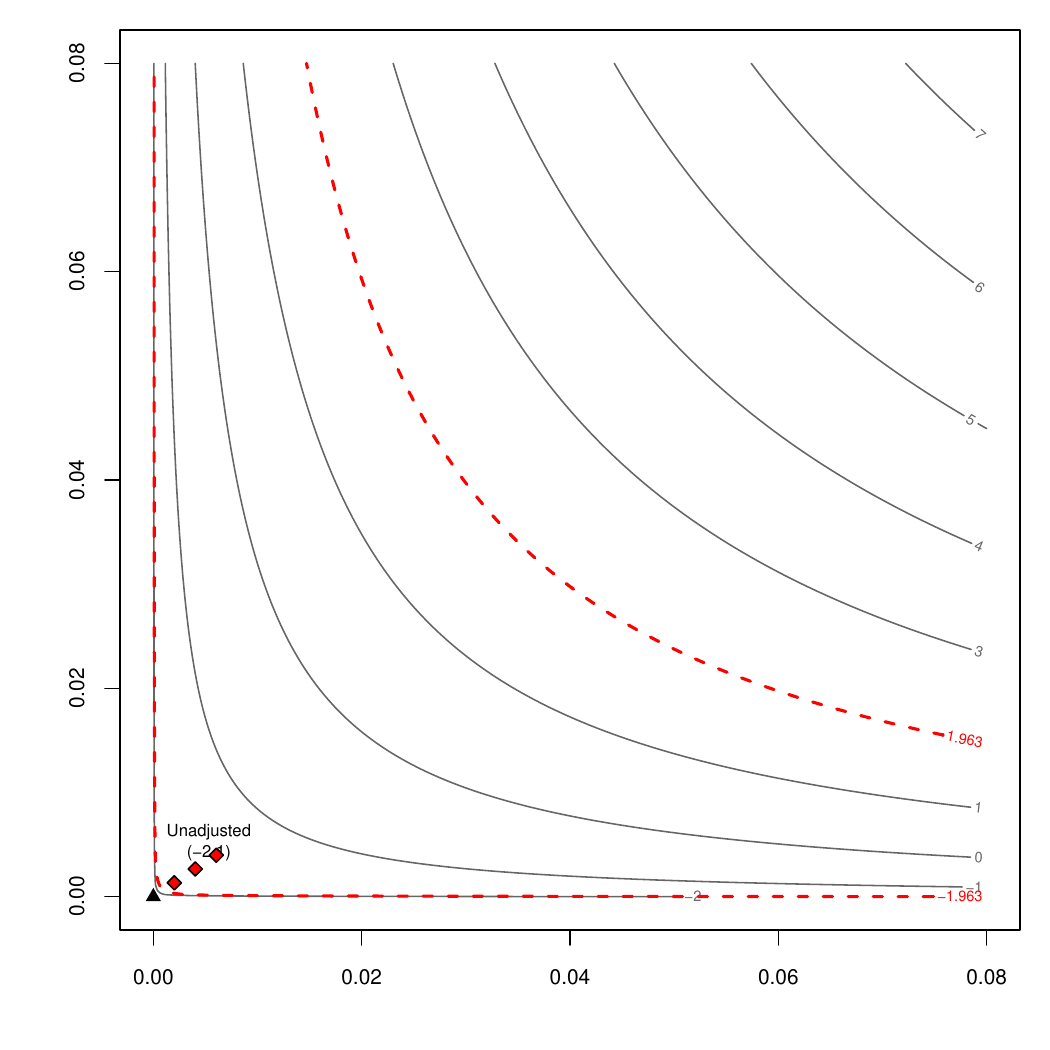}};
        \node[below=of img3,yshift=1.5cm] (img5)  {\includegraphics[scale=0.45]{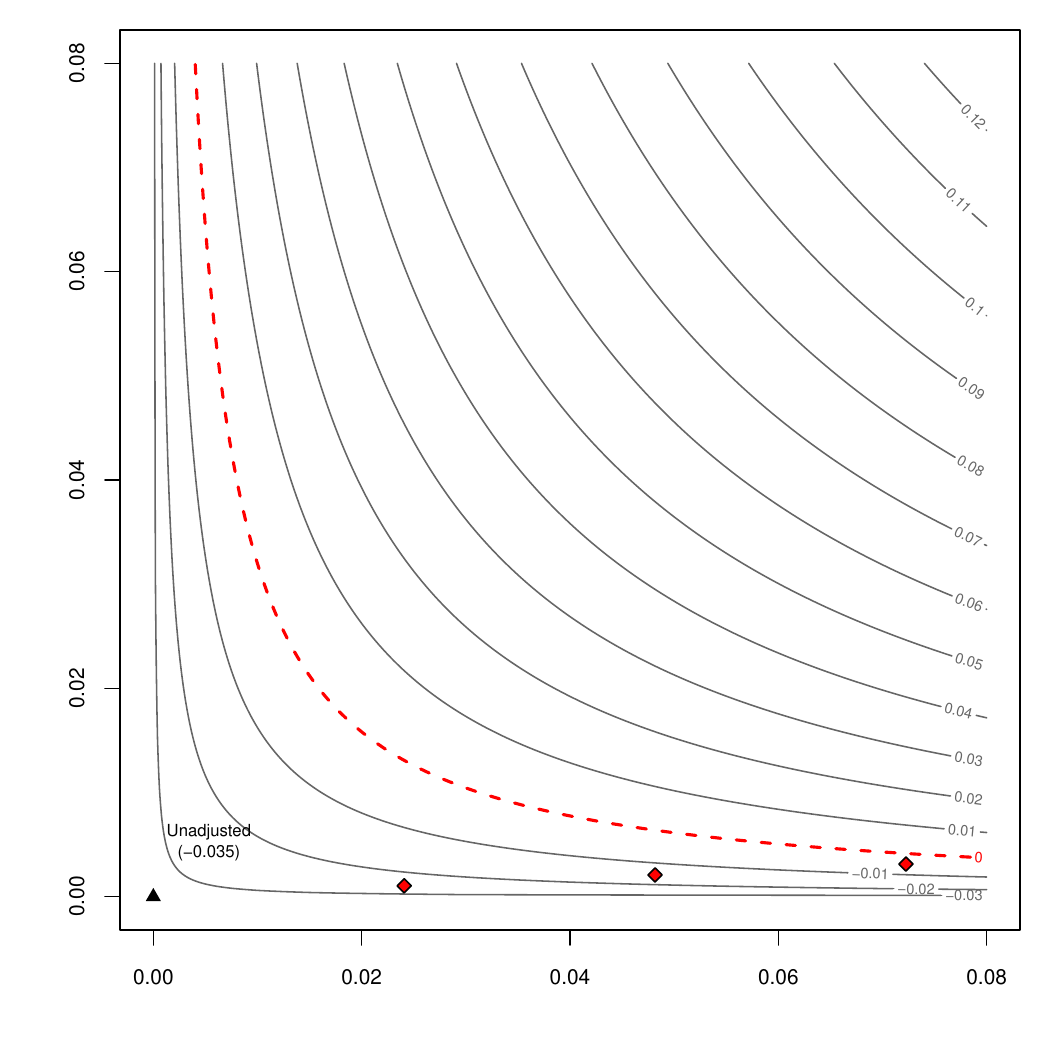}};
        \node[below=of img5, node distance=0cm, yshift=1.3cm,font=\color{black}] {\tiny Partial $R^2$ of confounder(s) with the treatment};
        \node[left=of img5, node distance=0cm, rotate=90, anchor=center,yshift=-1cm,font=\color{black}] {\tiny Partial $R^2$ of confounder(s) with the outcome};
        \node[left=of img5, node distance=0cm, rotate=90, anchor=center,yshift=0cm,font=\color{black}] {No. of past failed cycles};
        \node[below=of img4,yshift=1.5cm] (img6)  {\includegraphics[scale=0.45]{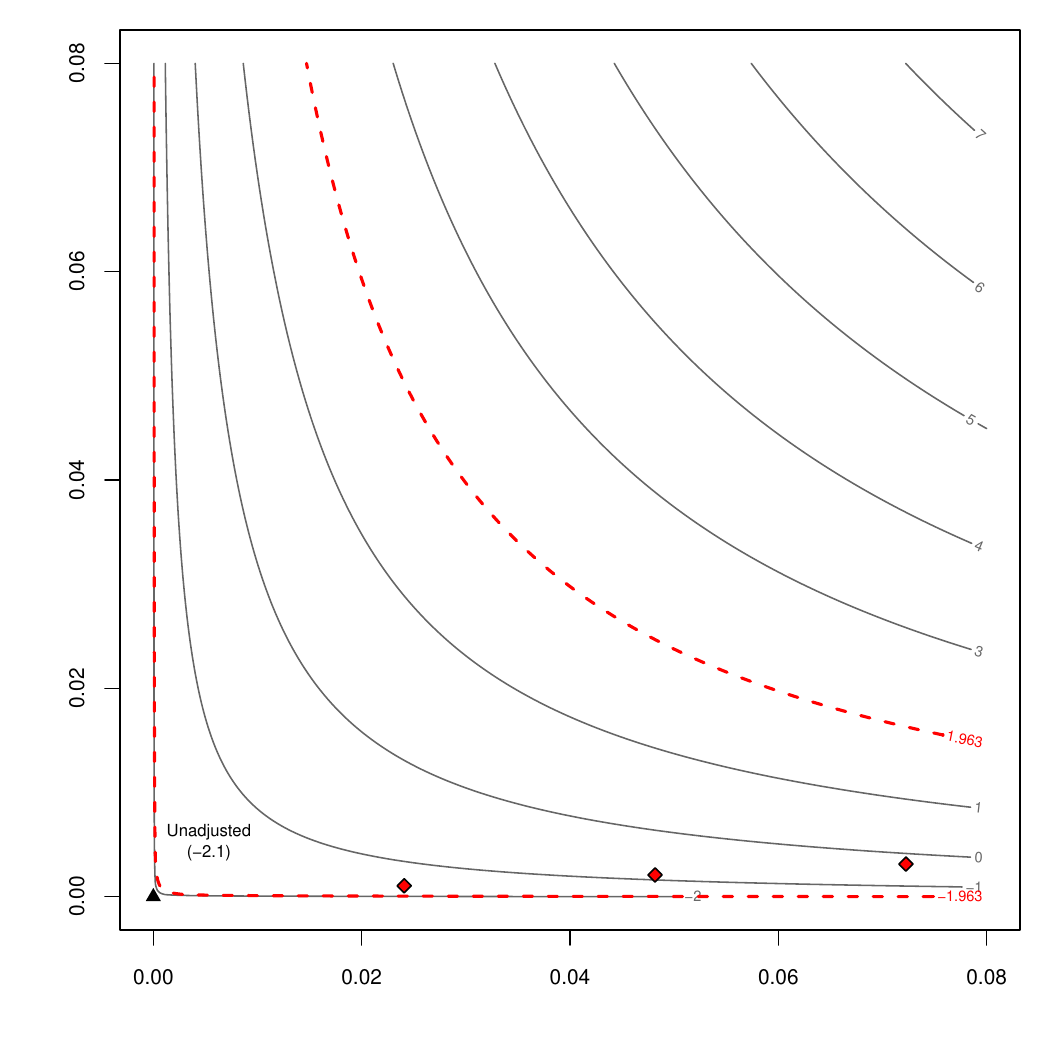}};
        \node[below=of img6, node distance=0cm, yshift=1.3cm,font=\color{black}] {\tiny Partial $R^2$ of confounder(s) with the treatment};
    \end{tikzpicture}
    \caption{Sensitivity of c-section ATE estimate to unobserved confounding}
    \label{fig:sensemakr_plot_csection}
\end{figure}
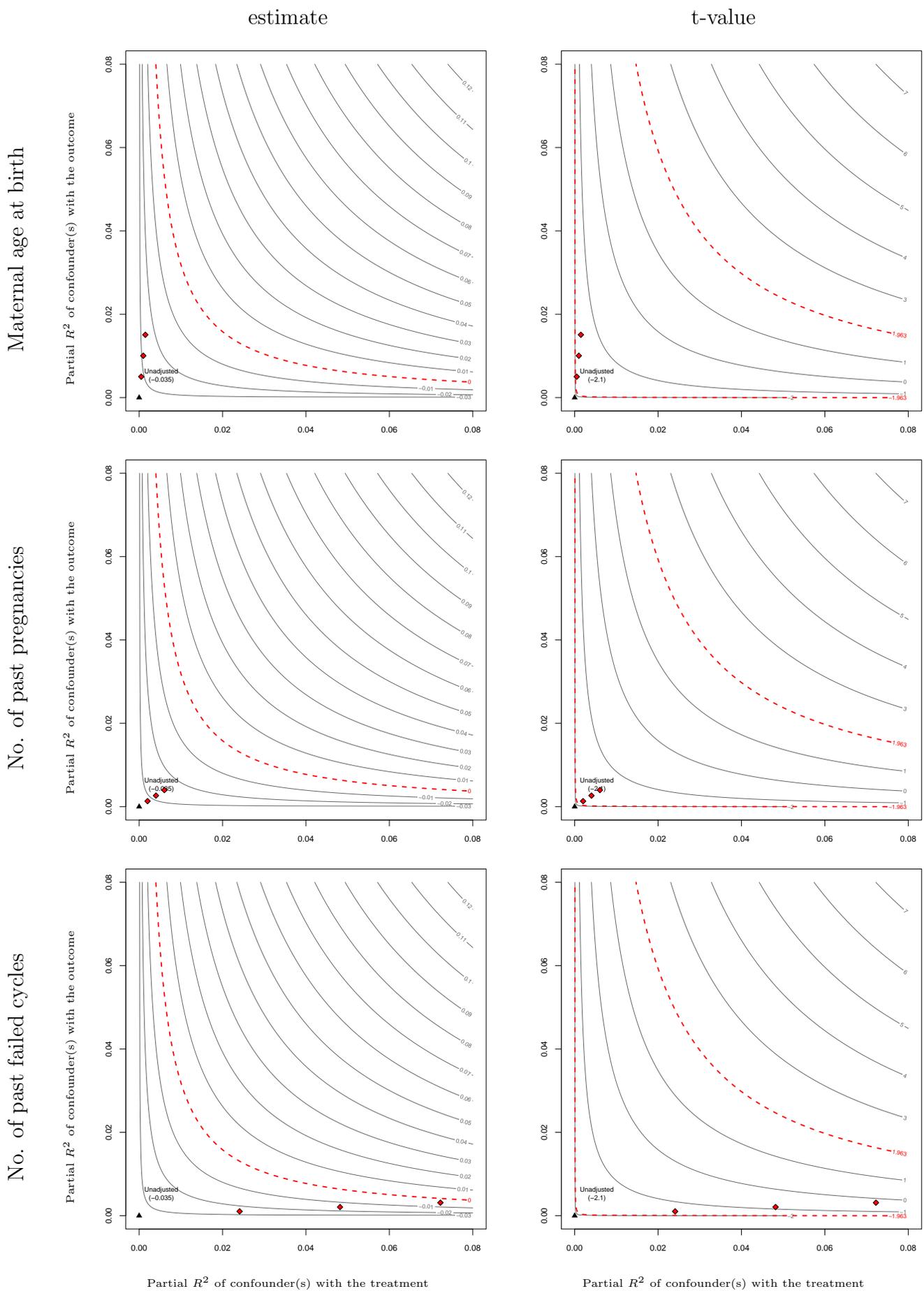

\begin{table}

    \caption{\label{tab:tab:anz_12_month_miss_table}Test for observations missing at random.}
    \centering
    \begin{threeparttable}
        \begin{tabular}[t]{lllll}
            \toprule
                                               & No miss. & With miss. & Diff.   & p-value \\
            \midrule
            Age                                & 34.9     & 35.09      & -0.19   & 0.0030  \\
                                               & (4.56)   & (4.51)     & (0.065) &         \\
            No. of past unsuccessful cycles    & 1.77     & 1.79       & -0.027  & 0.4092  \\
                                               & (2.24)   & (2.33)     & (0.033) &         \\
            Maternal year of birth             & 1977.82  & 1979.09    & -1.269  & 0.0000  \\
                                               & (4.73)   & (5.1)      & (0.07)  &         \\
            SEIFA                              & 3.49     & 3.46       & 0.026   & 0.1999  \\
                                               & (1.41)   & (1.39)     & (0.02)  &         \\
            Remoteness: major city             & 0.89     & 0.9        & -0.003  & 0.5283  \\
                                               & (0.31)   & (0.31)     & (0.004) &         \\
            Remoteness: inner regional         & 0.09     & 0.09       & 0.001   & 0.7256  \\
                                               & (0.28)   & (0.28)     & (0.004) &         \\
            Remoteness: outer regional         & 0.02     & 0.02       & 0.001   & 0.6701  \\
                                               & (0.13)   & (0.13)     & (0.002) &         \\
            Remoteness: remote and very remote & 0        & 0          & 0.001   & 0.3599  \\
                                               & (0.05)   & (0.04)     & (0.001) &         \\
            Mother born in Australia           & 0.66     & 0.65       & 0.012   & 0.0771  \\
                                               & (0.47)   & (0.48)     & (0.007) &         \\
            No. of past pregnancies            & 0.38     & 0.42       & -0.033  & 0.0006  \\
                                               & (0.63)   & (0.7)      & (0.01)  &         \\
            Diabetes mellitus                  & 0.01     & 0.01       & 0.006   & 0.0019  \\
                                               & (0.11)   & (0.08)     & (0.002) &         \\
            Gestational diabetes               & 0.09     & 0.08       & 0.013   & 0.0367  \\
                                               & (0.29)   & (0.27)     & (0.006) &         \\
            Chronic hypertension               & 0.01     & 0.01       & -0.001  & 0.5130  \\
                                               & (0.1)    & (0.1)      & (0.001) &         \\
            Pre-eclampsia                      & 0.02     & 0.02       & -0.002  & 0.2391  \\
                                               & (0.14)   & (0.15)     & (0.002) &         \\
            Gestational hypertension           & 0.05     & 0.05       & 0.004   & 0.1804  \\
                                               & (0.22)   & (0.21)     & (0.003) &         \\
            Smoked during pregnancy            & 0.01     & 0.01       & 0.003   & 0.0696  \\
                                               & (0.11)   & (0.1)      & (0.002) &         \\
            Endometriosis                      & 0.11     & 0.1        & 0.013   & 0.0032  \\
                                               & (0.32)   & (0.3)      & (0.004) &         \\
            Other cause of infertility         & 0.37     & 0.34       & 0.023   & 0.0006  \\
                                               & (0.48)   & (0.48)     & (0.007) &         \\
            Male infertility                   & 0.42     & 0.31       & 0.114   & 0.0000  \\
                                               & (0.49)   & (0.46)     & (0.007) &         \\
            Unexplained infertility            & 0.29     & 0.29       & 0.004   & 0.5408  \\
                                               & (0.46)   & (0.45)     & (0.006) &         \\
            \bottomrule
        \end{tabular}
        \begin{tablenotes}
            \item \textit{Notes: } This table compares the mean values of each
            variable in the population without missing observations and in that
            with at least one missing observation.
        \end{tablenotes}
    \end{threeparttable}
\end{table}

\restoregeometry

\end{document}